\documentclass{aa}
\pdfoutput=1 

\usepackage{xcolor}
\usepackage{txfonts} 
\usepackage{physics} 
\usepackage{graphicx} 
\usepackage[alsoload=astro]{siunitx} 
\usepackage{subcaption}
\usepackage{multirow}
\usepackage{tabularx}

\usepackage{hyperref} 
\hypersetup{
colorlinks   = true, 
urlcolor     = blue, 
linkcolor    = blue, 
citecolor   = blue 
}

\usepackage[textwidth=3.5cm]{todonotes}
\usepackage{blindtext}

\makeatletter
\renewcommand*\aa@pageof{, page \thepage{} of \pageref*{LastPage}}
\makeatother

\begin{document} 
\newcommand{\Map}{\mathcal{M}}
\newcommand{\Nap}{\mathcal{N}}
\newcommand{\Mperp}{\Map_{\perp}}

\newcommand{\NMap}{\expval{\Nap\Map}}
\newcommand{\NNMap}{\expval{\Nap\Nap\Map}}
\newcommand{\NNMperp}{\expval{\Nap\Nap\Mperp}}
\newcommand{\NMapMap}{\expval{\Nap\Map^2}}

\newcommand{\NNM}[2]{\expval{\Nap^{#1}\Nap^{#2}\Map}}

\newcommand{\G}{\mathcal{G}}
\newcommand{\Gtilde}{\tilde{\mathcal{G}}}
\newcommand{\I}{\mathrm{i}}
\newcommand{\varthetavec}{\vb*{\vartheta}}
\newcommand{\thetavec}{\vb*{\theta}}
\newcommand{\alphavec}{\vb*{\alpha}}
\newcommand{\ellvec}{\vb*{\ell}}
\newcommand{\yvec}{\vb*{y}}
\newcommand{\kparallel}{\vb*{k}^{\|}}
\newcommand{\kperp}{k^{\perp}}
\newcommand{\SigCrit}{\Sigma_{\textrm{crit}}}
\newcommand{\SigCritAve}{\bar{\Sigma}_{\textrm{crit}}}
\newcommand{\bggk}{\mathit{b}_{\mathrm{gg}\kappa}}
\newcommand{\Bggd}{\mathit{B}_{\mathrm{gg}\delta}}
\newcommand{\Bggdab}{\Bggd^{ab}}
\newcommand{\Bggdabh}[1]{{}_{#1}\Bggd^{ab}}
\newcommand{\Bggdaah}[1]{{}_{#1}\Bggd^{aa}}
\newcommand{\bggkab}{\bggk^{ab}}
\newcommand{\dirac}{\delta_\textrm{D}}
\newcommand{\kronecker}[2]{\delta^\textrm{K}_{#1 #2}}
\newcommand{\Msun}{{M}_{\odot}}
\newcommand{\SNR}{\text{SNR}}

\newcommand{\Ns}[1]{N_{\mathrm{sat}}^{#1}}
\newcommand{\Nc}[1]{N_{\mathrm{cen}}^{#1}}
\newcommand{\Nss}[2]{N_{\mathrm{sat}, #2}^{#1}}
\newcommand{\Ncc}[2]{N_{\mathrm{cen}, #2}^{#1}}

\newcommand{\rmg}{_\mathrm{g}}

\newcommand{\pma}[2]{_{-#1}^{+#2}}

\title{KiDS+VIKING+GAMA: Halo occupation distributions and correlations of satellite numbers with a new halo model of the galaxy-matter bispectrum for galaxy-galaxy-galaxy lensing}

   \author{Laila Linke\inst{1}
          \and Patrick Simon \inst{1} \and Peter Schneider \inst{1} \and Daniel J. Farrow \inst{2,} \inst{3} \and Jens R\"odiger \inst{1} \and Angus H. Wright \inst{4}
          }

   \institute{Argelander-Institut für Astronomie, Rheinische Friedrich-Wilhems Universität Bonn,  Auf dem Hügel 71, 53121 Bonn\\
              \email{llinke@astro.uni-bonn.de}
              \and Max-Planck-Institut f{\"u}r extraterrestriche Physik, Giessenbachstrasse 
1, 85748 Garching, Germany
\and Universit{\"a}ts-Sternwarte, Fakult{\"a}t f{\"u}r Physik, Ludwig-Maximilians-Universit{\"a}t M{\"u}nchen, Scheinerstr. 1, 81679 M{\"u}nchen, 
Germany
              \and 	Ruhr University Bochum, Faculty of Physics and Astronomy, Astronomical Institute (AIRUB), German Centre for Cosmological Lensing, 44780 Bochum, Germany}

\titlerunning{KiDS+VIKING+GAMA: HODs with galaxy-galaxy-galaxy-lensing}
\authorrunning{L. Linke et al.}             

   \date{}

\abstract{Halo models and halo occupation distributions (HODs) are important tools to model the distribution of galaxies and matter.}{ We present and assess a new method for constraining the parameters of HODs using the mean gravitational lensing shear around galaxy pairs, so-called galaxy-galaxy-galaxy-lensing (G3L). In contrast to galaxy-galaxy-lensing, G3L is also sensitive to the correlations between the per-halo numbers of galaxies from different populations. We employ our G3L halo model to probe these correlations and test the default hypothesis that they are negligible.}  {We derive a halo model for G3L and validate it with realistic mock data from the Millennium Simulation and a semi-analytic galaxy model. Then, we analyse public data from the Kilo-Degree Survey (KiDS), the VISTA Infrared Kilo-Degree Galaxy Survey (VIKING) and data from the Galaxy And Mass Assembly Survey (GAMA) to infer the HODs of galaxies at $z<0.5$ in five different stellar mass bins between $10^{8.5}h^{-2}\Msun$ and $10^{11.5}h^{-2}\Msun$ and two colours (red and blue),  as well as correlations between satellite numbers.}{The analysis accurately recovers the true HODs in the simulated data for all galaxy samples within the $68\%$ credibility range. The model best-fits agree with the observed G3L signal on the $95\%$ confidence level. The inferred HODs vary significantly with colour and stellar mass. In particular red galaxies prefer more massive halos $\gtrsim 10^{12} \Msun$, while blue galaxies are present in halos $\gtrsim 10^{11}\Msun$. There is strong evidence ($>3\sigma$) for a high correlation, increasing with halo mass, between the numbers of red and blue satellites and between galaxies with stellar masses below $10^{10} \Msun$.}{Our G3L halo model accurately constrains galaxy HODs for lensing surveys of up to $10^3\,\mathrm{deg}^2$ and redshift below $0.5$ probed here. Analyses of future surveys may need to include non-Poisson variances of satellite numbers or a revised model for central galaxies. Correlations between satellite numbers are ubiquitous between various galaxy samples and are relevant for halos with masses $\gtrsim 10^{13} \Msun$, that is, of galaxy group scale and more massive. Possible causes of these correlations are the selection of similar galaxies in different samples, the survey flux limit, or physical mechanisms like a fixed ratio between the satellite numbers of distinct populations. The decorrelation for halos with smaller masses is probably an effect of shot noise by low-occupancy halos. The inferred HODs can be used to complement galaxy-galaxy-lensing or galaxy clustering HOD studies or as input to cosmological analyses and improved mock galaxy catalogues.}
 

   \maketitle
%

\section{Introduction}
\label{sec:intro}
Accurate models of the distribution of galaxies inside the cosmic large-scale structure are crucial to understanding the physics of galaxy evolution and inferring cosmological parameters from galaxy surveys. Popular frameworks for analytically describing the galaxy and matter distribution are halo models (e.g., \citealp{Cooray2002, Scoccimarro2001, Kravtsov2004, Zheng2005}). Their key ingredient is the halo occupation distribution (HOD), which gives the expected number of galaxies inside a halo of a given mass \citep{Berlind2002}. Here we present a new method to accurately infer galaxy HODs with a halo model for galaxy-galaxy-galaxy lensing (G3L) -- the mean gravitational lensing shear around galaxy pairs \citep{Schneider2005}. We demonstrate that with this higher-order statistic we can obtain the HODs of various galaxy samples selected, for example, by their colour, in current galaxy surveys. Additionally, in contrast to other lensing methods, G3L can probe the correlation of per-halo numbers of satellite galaxies from different populations. In a first application, we infer the HODs and correlations of various galaxy selections in the Kilo-Degree Survey (KiDS; \citealp{Kuijken2015}), the VISTA Infrared Kilo-degree Galaxy survey (VIKING; \citealp{Edge2013, Venemans2015}) and the Galaxy And Mass Assembly survey (GAMA; \citealp{Driver2009, Liske2015}).

Halo models are the backbone of many analytical expressions for the statistics of the large-scale structure. They postulate that matter is distributed in virialized halos, and galaxies only exist within these halos \citep{White1978}. For known halo density profiles and HODs, they can predict all statistics of the galaxy and matter distributions \citep{Cooray2002}. Therefore, with halo models, we can infer HODs from the measured galaxy and matter statistics, such as galaxy clustering \citep{Zehavi2011, Ishikawa2021} and the galaxy-matter power spectrum \citep{Mandelbaum2006, Clampitt2016, Dvornik2018}.

Since halo models are analytic, calculating their predictions is faster and easier than more complex approaches such as semi-analytic galaxy models (SAMs; e.g., \citealp{Henriques2015}) or hydrodynamical simulations \citep{Vogelsberger2020}. Even though they rely on simple assumptions, they can accurately describe second-order statistics of galaxies and matter. For example, halo model predictions for the galaxy-galaxy two-point correlation function agree well with observations \citep{Zehavi2011}. \citet{Mead2015} shows that halo models can also describe the non-linear matter power spectrum in $N$-body simulations with $5\%$ accuracy.

Although halo models are prevalent for two-point statistics, it is unclear whether they are sufficiently accurate for modelling higher-order statistics, such as the galaxy- and matter bispectrum. These higher-order statistics contain complementary information to the two-point statistics \citep{Berlind2002} so it is worthwhile to expand halo models to include them to improve and cross-validate constraints from second-order statistics. Here we extend halo models to measurements of G3L \citep{Simon2008, Simon2013, Linke2020b}, a third-order statistic.

The G3L signal is induced on a background galaxy (the `source’) by the weak gravitational lensing of matter around a pair of foreground galaxies (the `lenses’). It directly depends on the galaxy-matter bispectrum integrated over the spread of lenses along the line-of-sight \citep{Schneider2005}. Unlike galaxy-galaxy lensing, G3L is sensitive to the mean number of galaxy pairs inside halos and, therefore, the correlation of halo satellite numbers. If the galaxies in a lens pair belong to different samples, the G3L signal is higher for positively correlated satellite numbers and lower if they are anti-correlated. While the default assumption in the literature is that satellite numbers are uncorrelated \citep{Scranton2001, Scranton2002, Zehavi2005}, some galaxy clustering studies suggest a correlation between galaxy populations, such as red and blue galaxies (\citealp{Zehavi2011, Ross2009, Wang2007, Simon2009}).

We challenge the default assumption by constructing a halo model for the galaxy-galaxy-matter bispectrum and the G3L signal, inferring the HODs and satellite correlations for various galaxy samples. Our model is based on the approaches by \cite{Zheng2007}, \cite{Zehavi2011}, \cite{Clampitt2017}, but goes further by including the correlation between the satellite numbers of different galaxy samples. We validate the model and our inference procedure with G3L estimates in a simulated lensing survey based on the Millennium Simulation (MS; \citealp{Springel2005}) populated with SAM galaxies by \citet{Henriques2015}. As a first real-data application, we analyse G3L measurements in the overlap region of KiDS, VIKING, and GAMA, with lenses selected from GAMA and sources from KiDS+VIKING \citep{Linke2020b}. We infer the HODs of galaxies in five different stellar mass bins between $10^{8.5}$ to $10^{11.5}\,h^{-2}\, M_{\odot}$ and two colours (red and blue), as well as the cross-correlations between the per-halo satellite numbers for these galaxy samples. 

This paper is structured as follows: In Sect.~\ref{sec:theoryG3L}, we give an overview of the basics of G3L. We present our halo model for G3L in Sect.~\ref{sec:halomodel}. The simulated and observed data sets are described in Sect.~\ref{sec:data}. Section~\ref{sec:method} describes the estimator for the G3L signal and our analysis of the measurements. We present our results in Sect.~\ref{sec:results} and discuss them in Sect.~\ref{sec:discussion}.

In the analysis of the MS, we use the cosmological parameters of this simulation, namely $\Omega_\mathrm{m}=0.25$, $\Omega_\mathrm{b}=0.045$, $H_0=\SI{73}{\km \per \second \per \mega \parsec}$,  and $\sigma_8=0.9$. For the analysis of the observations, we use the parameters from the \citet{Planck2018}, namely $\Omega_\mathrm{m}=0.315$, $\Omega_\mathrm{b}=0.049$, $H_0=\SI{67.4}{\km \per \second \per \mega \parsec}$,  and $\sigma_8=0.811$. Throughout we assume a flat cosmology with $\Omega_\Lambda = 1- \Omega_\mathrm{m}$.

\section{Theory of galaxy-galaxy-galaxy lensing}
\label{sec:theoryG3L}
G3L is a weak gravitational lensing effect (see, e.g., \citealp{Bartelmann2001} for a review on weak lensing), first described by \citet[SW05 herafter]{Schneider2005}. There are two types of this effect: The lensing of background galaxy pairs by matter around individual foreground galaxies (lens-shear-shear G3L) and the lensing of individual background galaxies by matter around foreground galaxy pairs (lens-lens-shear G3L). We concentrate on the latter. 

\begin{figure}
	\centering
	\resizebox{\hsize}{!}{\includegraphics{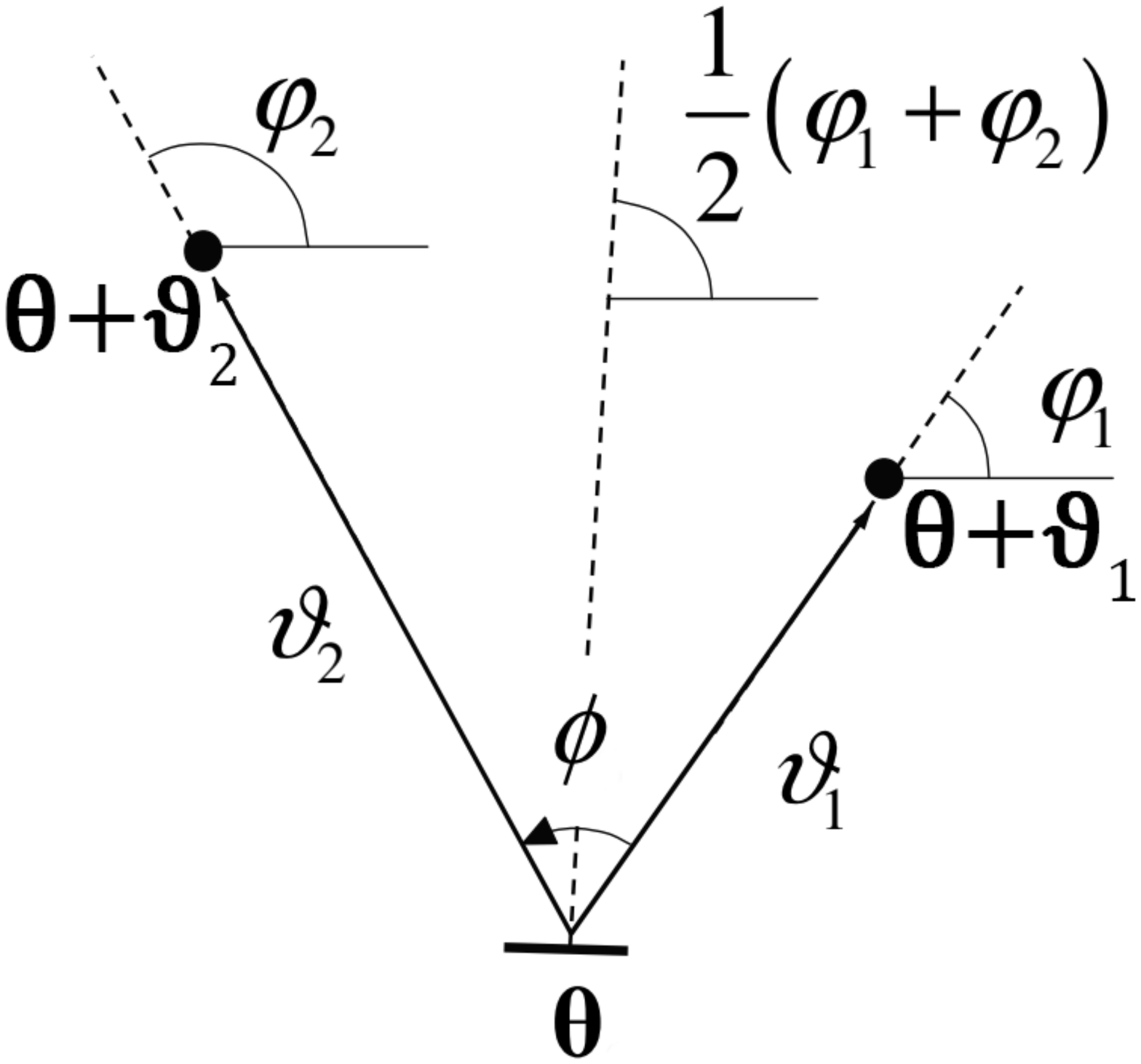}}
	\caption{Geometry of a G3L configuration with one source and two lens galaxies on the sky; adapted from \citet{Schneider2005}. Lens galaxies are at angular positions $\vec{\theta}_1 = \thetavec + \varthetavec_1$ and $\vec{\theta}_2 = \thetavec+\varthetavec_2$ on the sky; the source galaxy is at $\vec{\theta}$. The angle between the source-lens connections is the opening angle $\phi$.}
	\label{fig:G3L}
\end{figure}

Figure~\ref{fig:G3L} shows the geometric configuration for a lens-lens-shear G3L system on the sky. A background galaxy (`source'), located at angular position $\vec{\theta}$, is gravitationally lensed by matter around two foreground galaxies (`lenses'), located at $\vec{\theta} + \varthetavec_1$ and $\vec{\theta} + \varthetavec_2$. Due to the lensing, the source experiences a tangential shear $\gamma_\mathrm{t}$, which is measured with respect to the bisector of the angle $\phi$ between the source-lens connections. The main observables for G3L are the three-point correlation function $\Gtilde^{ab}$ and the closely related aperture statistics $\NNM{a}{b}$, where $a$ and $b$ denote the considered galaxy populations. We describe the observables in Sects.~\ref{sec:theoryG3L:Gtilde} and \ref{sec:theoryG3L:NNM}. To model $\Gtilde^{ab}$ and $\NNM{a}{b}$, we first introduce the galaxy-galaxy-matter bispectrum and its relation to the matter and galaxy number density distributions.

\subsection{Projected fields and the galaxy-galaxy-matter bispectrum}
\label{sec:theoryG3L:projections}
The distribution of matter and galaxies is defined by the density $\rho(\vec{x}, z)$ and discrete galaxy number density $n\rmg^{a}(\vec{x}, z)$ at comoving position $\vec{x}$ and redshift $z$. The subscript $a$ refers to the `sample' of the galaxies, chosen with the same selection function, for example early- or late-type galaxies. Fluctuations in the densities $\rho$ and $n\rmg^a$ are the matter- and galaxy number density contrast, $\delta(\vec{x}, z)$ and $\delta^a\rmg(\vec{x}, z)$, defined as
\begin{equation}
	\label{eq:definition delta}
	\delta(\vec{x},z)=\frac{\rho(\vec{x},z)-\bar{\rho}(z)}{\bar{\rho}(z)}\;,
\end{equation}
and
\begin{equation}
	\label{eq:definition delta_g}
	\delta^{a}\rmg(\vec{x},z)=\frac{n\rmg^{a}(\vec{x},z)-\overline{n\rmg^a}(z)}{\overline{n\rmg^a}(z)}\;,
\end{equation}
where bars denote ensemble averages.

Since G3L involves the correlation of galaxy pairs with the matter field, it is natural to derive its model from the galaxy-galaxy-matter bispectrum $\Bggdab(\vec{k}_1, \vec{k}_2; \vec{k}_3, z)$ at comoving modes $\vec{k}_1$, $\vec{k}_2$, and $\vec{k}_3$ and redshift $z$. This bispectrum is defined by
\begin{align}
	\label{eq:defintion bispectrum}
	& \expval{\hat{\delta}\rmg^{a}(\vec{k}_1, z)\,\hat{\delta}\rmg^{b}(\vec{k}_2, z)\, \hat{\delta}(\vec{k}_3, z)}\; \\
	&\notag=(2\pi)^3\, \dirac(\vec{k}_1+\vec{k}_2+\vec{k}_3)\,	\Bggdab (\vec{k}_1, \vec{k}_2; \vec{k}_3, z)
\end{align}
where hats denote Fourier transforms for which we use the convention
\begin{equation}
	\hat{\delta}(\vec{k})=\int \dd[3]{x}\; \delta(\vec{x})\, \exp(-{\rm i}\vec{k}\cdot\vec{x})\;.
\end{equation}
The bispectrum is defined only for $\vec{k}_3=-\vec{k}_1-\vec{k}_2$, so we abbreviate
\begin{equation}
	\Bggdab(\vec{k}_1, \vec{k}_2; -\vec{k}_1-\vec{k}_2, z) =: \Bggdab(\vec{k}_1, \vec{k}_2, z)\;.
\end{equation}

As a gravitational lensing effect, G3L depends mainly on the projections $\kappa$ and $\kappa^a\rmg$ of $\delta$ and $\delta^a\rmg$ along the line-of-sight. In a flat universe the lensing convergence is 
\begin{equation}
	\kappa(\thetavec)= \frac{3H_0^2\, \Omega_\mathrm{m}}{2c^2}\, \int_0^\infty \dd{\chi}\; q(\chi)\, \chi\, \frac{\delta\Big(\vec{x}(\thetavec, \chi), z(\chi)\Big)}{a(\chi)}\;,
\end{equation}
where $\chi$ is the comoving distance, $z(\chi)$ is the redshift at $\chi$, $\vb*{x}(\thetavec, \chi)= (\chi\,\thetavec, \chi)$, and $q(\chi)$ is an integral over the probability distribution function
$p_\mathrm{s}(\chi)$ of source galaxies with comoving distance $\chi$, given as
\begin{equation}
	q(\chi)=\int_{\chi}^{\infty} \dd{\chi^\prime}\; p_\mathrm{s}(\chi^\prime)\, \frac{\chi^\prime-\chi}{\chi^\prime}\;.
\end{equation}
The projected galaxy number density contrast is given by
\begin{equation}
	\kappa\rmg^a(\thetavec) = \int_0^\infty \dd{\chi}\; p^a(\chi)\, \delta\rmg^a\Big(\vec{x}(\thetavec, \chi), z(\chi)\Big)\;,
\end{equation}
where $p^a(\chi)$ is the probability distribution function of lens galaxies from sample $a$ in comoving distance. This distribution strongly depends on the selection criteria for sample $a$. With $p^a(\chi)$ we also define the projected galaxy number density $N\rmg^a$ as
\begin{equation}
	N\rmg^a(\thetavec) = \int_0^\infty \dd{\chi}\; p^a(\chi)\, n\rmg^a\Big(\vec{\chi}(\thetavec, \chi), z(\chi)\Big)\;.
\end{equation}

To arrive at a G3L signal, we convert the galaxy-galaxy-matter bispectrum $\Bggdab$ to its projected counterpart $\bggkab$, defined by
\begin{equation}
	\expval{\hat{\kappa}^a\rmg(\ellvec_1)\, \hat{\kappa}^b\rmg(\ellvec_2)\, \hat{\kappa}(\ellvec_3)} = (2\pi)^2\, \dirac\left(\ellvec_1 + \ellvec_2 + \ellvec_3\right)\,\bggkab(\ellvec_1, \ellvec_2, \ellvec_3)\;,
\end{equation}
where hats again denote Fourier transforms. Similar as for $\Bggdab$, we abbreviate
\begin{equation}
	\bggkab(\ellvec_1, \ellvec_2, -\ellvec_1-\ellvec_2) =: \bggkab(\ellvec_1, \ellvec_2)\;.
\end{equation}
As shown in \citet{Kaiser1992}, under the assumptions of the Limber equation, the projected bispectrum is
\begin{align}
	\label{limber}
	\bggkab(\ellvec_1, \ellvec_2)& = \frac{3H_0^2\, \Omega_\mathrm{m}}{2c^2} \int \dd{\chi} \frac{q(\chi)\, p^a(\chi)\, p^b(\chi)}{\chi^3\, a(\chi)}\\
	&\notag\quad\times \Bggdab\Big(\ellvec_1, \ellvec_2, z(\chi)\Big),
\end{align}
which we use to model the observables of G3L: the correlation function $\Gtilde^{ab}$ and the aperture statistics $\NNM{a}{b}$. Equation \eqref{limber} requires that the product $q(\chi)\,p^a(\chi)\,p^b(\chi)$ in the integrand does not vary strongly on scales smaller than the typical correlation length (i.e., a few Mpc) of the galaxy distribution  \citep{Bartelmann2001}.

\subsection{G3L correlation function}
\label{sec:theoryG3L:Gtilde}
The G3L correlation function $\Gtilde^{ab}$ correlates the projected lens galaxy number densities $N_\mathrm{g}^a$ and $N_\mathrm{g}^b$ of two lens galaxy samples $a$ and $b$ with the tangential shear $\gamma_\mathrm{t}(\thetavec)$ of the source galaxies,
\begin{equation}
	\label{eq:Gtilde}
	\tilde{\mathcal{G}}^{ab}(\vartheta_1, \vartheta_2, \phi)=\frac{1}{\overline{N_\mathrm{g}^a}\,\overline{N_\mathrm{g}^b}}\,\expval{ N_\mathrm{g}^{a}(\vec{\theta}+\vec{\vartheta}_1)\, N_\mathrm{g}^b(\vec{\theta}+\vec{\vartheta}_2) \, \gamma_\textrm{t}(\vec{\theta})}\;.
\end{equation}
Due to the statistical homogeneity and isotropy of the matter and galaxy fields, $\Gtilde^{ab}$ only depends on $\vartheta_1=|\vec{\vartheta}_1|$ and $\vartheta_2=|\vec{\vartheta}_2|$ of the lens-source separations, and the angle $\phi$ between $\varthetavec_1$ and $\varthetavec_2$. For $a\neq b$, the two lenses in each G3L configuration are from two different samples (mixed lens pairs);  otherwise, they are from the same sample (unmixed lens pairs).

As shown in SW05, $\Gtilde^{ab}$ can be calculated from the projected galaxy-galaxy-matter bispectrum $\bggkab$ (c.f. their Equation 40), by integrating $\bggkab(\ellvec_1, \ellvec_2)$, multiplied with a complex kernel function containing the second-order Bessel function. Consequently, $\Gtilde^{ab}$ can be modelled from the bispectrum and directly compared to measurements. However, it is numerically preferable to consider the G3L aperture statistics, a linear transform of $\Gtilde$. We describe them in the following subsection.

\subsection{G3L aperture statistics}
\label{sec:theoryG3L:NNM}
Aperture statistics are moments of aperture masses $\Map_\theta(\varthetavec)$ and aperture number counts $\Nap_\theta^p(\varthetavec)$, 
\begin{equation}
	\label{eq:DefinitionApertureMass}
	\Map_{\theta}(\varthetavec)    = \int \dd[2]{\vartheta'} U_\theta(|\varthetavec-\varthetavec'|)\,\kappa(\varthetavec') 
\end{equation} 
and
\begin{equation}
	\Nap_{\theta}^a(\varthetavec)    = \frac{1}{\overline{N\rmg^a}} \int \dd[2]{\vartheta'} U_\theta(|\varthetavec-\varthetavec'|)\,N\rmg^a(\varthetavec')\;,
\end{equation} 
with a filter function $U_\theta$ of aperture scale radius $\theta$. The filter function $U_\theta$ has to be compensated, i.e., $\int \dd{\vartheta}\, \vartheta\, U_\theta(\vartheta) = 0$. For G3L, the relevant aperture statistics are $\NNM{a}{b}$, given by
\begin{align}
	\label{eq:Definition NNMap}
	&    \NNM{a}{b}(\theta_1, \theta_2, \theta_3) = \expval{\Nap_{\theta_1}^a(\varthetavec)\,\Nap_{\theta_2}^b(\varthetavec)\, \Map_{\theta_3}(\varthetavec)} \;.
\end{align}

The aperture statistics are related to the bispectrum $\bggkab$ by
\begin{align}
	\label{eq:NNMapFromBggkappa}
	&\NNM{a}{b} (\theta_1, \theta_2, \theta_3) \\
	&\notag= \int \frac{\dd[2]{\ell_1}}{(2\pi)^2} \int \frac{\dd[2]{\ell_2}}{(2\pi)^2} \; \hat{U}_{\theta_1}(\ell_1)\, \hat{U}_{\theta_2}(\ell_2)\, \hat{U}_{\theta_3}(|\vb*{\ell}_1+\vb*{\ell}_2|)\, \bggkab(\ellvec_1, \ellvec_2)\;,
\end{align}
where $\hat{U}_\theta$ is the Fourier transform of $U_\theta$. We choose the exponential filter function by \citet{Crittenden2002},
\begin{equation}
	\label{eq:exponentialFilterFunction}
	U_\theta(\vartheta)=\frac{1}{2\pi\theta} \, \left(1-\frac{\vartheta^2}{2\theta^2}\right)\, \exp(-\frac{\vartheta^2}{2\theta^2})\;,
\end{equation}
which is commonly used for studies of higher-order aperture statistics (e.g. \citealp{SchneiderKilbinger2005}, \citealp{Jarvis2004}, \citealp{Simon2009}, \citealp{Saghiha2017}) due to its favourable analytical properties. In particular, SW05 show that for this filter function the correlation function $\Gtilde^{ab}$ can be connected analytically to $\NNM{a}{b}$ through
\begin{align}
	\label{eq:NNMapFromGtilde}
	&\notag    \NNM{a}{b}(\theta_1, \theta_2, \theta_3)  \\
	&= \int_{0}^{\infty} \dd{\vartheta_1} \, \vartheta_1 \int_{0}^{\infty} \dd{\vartheta_2} \, \vartheta_2
	\int_{0}^{2\pi} \dd{\phi} \, \Gtilde^{ab}(\vartheta_1, \vartheta_2, \phi)\,\\
	&\notag\qquad\times \mathcal{A}_{\Nap\Nap\Map}(\vartheta_1, \vartheta_2, \phi \mid \theta_1, \theta_2, \theta_3)\;,
\end{align}
with the kernel function $\mathcal{A}_{\Nap\Nap\Map}(\vartheta_1, \vartheta_2, \phi \mid \theta_1, \theta_2, \theta_3)$ given in the appendix of SW05. 

As alluded to before, there are practical advantages to modelling $\NNM{a}{b}$ instead of $\Gtilde^{ab}$. First, evaluating Eq.~\eqref{eq:NNMapFromBggkappa} is numerically more stable than calculating $\Gtilde$ from the bispectrum because the filter function $U_\theta$ is localized and non-oscillating. Second, in contrast to $\Gtilde$, the aperture statistics do not
depend on the galaxy-matter two-point correlation function. Therefore, they do not need a model of the galaxy-galaxy-lensing signal $\expval{\gamma_\mathrm{t}}$. Third, a  $\NNM{a}{b}$ data vector is a condensed summary statistic, where a few aperture radii ($\sim$ tens) contain a similar amount of information as $\Gtilde$ over hundreds of bins. 

Consequently, we use the aperture statistics $\NNM{a}{b}$ as the primary observable and model it with Eq.~\eqref{eq:NNMapFromBggkappa}, based on a halo-model based bispectrum $\bggkab$. To measure the aperture statistics, we estimate  $\Gtilde^{ab}$ (see Sect.~\ref{sec:method:measurement}) and convert it to aperture statistics with Eq.~\eqref{eq:NNMapFromGtilde}. We focus on the equilateral statistics, i.e., $\theta_1=\theta_2=\theta_3$, to reduce the size of the data vector and use the shorthand
\begin{equation}
	\NNM{a}{b}(\theta):=\NNM{a}{b}(\theta, \theta, \theta)\;,
\end{equation}
for convenience. We describe the `ingredients' of our halo model and the derivation of the model bispectrum in the next section.

\section{G3L halo model}
\label{sec:halomodel}
In this section, we derive the galaxy-galaxy-matter bispectrum based on the halo model assumption, from which we can predict the G3L aperture statistics with Eq.~\eqref{eq:NNMapFromBggkappa}. For this derivation, we require several ingredients: the linear matter power spectrum and critical density contrast, the halo density profile, the halo mass function (HMF), the halo bias, the spatial distribution of galaxies within a halo, and the first- and second-order moments of the HOD. We will give details on these ingredients next.

\subsection{Ingredients}
\label{sec:halomodel:ingredients}

Halo models assume that all matter and galaxies are within self-bound, virialized halos. Halos form in regions at redshift $z$ where the linear density contrast of matter exceeds the critical density contrast $\delta_\mathrm{c}(z)$. They reach the virial density $\rho_\mathrm{vir}=\Delta(z)\,\bar{\rho}$, where $\bar{\rho}$ is the cosmic mean density of matter and $\Delta(z)$ is the fractional overdensity within the virialized region.

The first ingredients are the linear matter power spectrum $P_\mathrm{lin}$ and the critical density contrast $\delta_\mathrm{c}$. We use the $P_\mathrm{lin}$ by \citet{Eisenstein1998} that includes baryonic effects, and the fitting formula by \citet{Nakamura1997} for the critical density contrast, 
\begin{align}
	\delta_\mathrm{c}(z)& = \frac{3}{20}\,(12\pi)^{2/3} \left[ 1 + 0.012299 \log_{10}\left(1+ \frac{\Omega_\mathrm{m}^{-1} -1}{(1+z)^3}\right)\right]\;,
\end{align}
which was derived for flat $\Lambda$CDM universes. For $\Omega_\mathrm{m}=1$, it reduces to the value for an Einstein--de Sitter Universe, $\delta_\mathrm{c}\simeq 1.69$.

Second, we require the density profile $\rho(|\vec{r}-\vec{r}_0|\,|\,m, z)$ of a halo with mass $m$ and redshift $z$, centred at $\vec{r}_0$. We assume halos follow (truncated) Navarro-Frenk-White profiles (NFW; \citealt{Navarro1996}), given as
\begin{align}
	\rho(r \,|\, m, z) &= \frac{200\,\bar{\rho}(z)}{3 \frac{r}{r_{200}}\left[\frac{1}{c(m,z)} + \frac{r}{r_{200}}\right]^2}\\
	&\notag \quad \times \frac{1+c(m,z)}{\ln\left(1+c(m, z)\right)[1+c(m,z)] - c } \\
	&= m\, u(r\,|\,m, z)\;,
\end{align}
where $r_{200}$ is the radius of a sphere around the halo centre, in which the mean density \footnote{In their original derivation, \citet{Navarro1996} defined the mass by a sphere of enclosed mean density $200\,\rho_\mathrm{crit}$. However, the fitting function for the concentration by \citet{Bullock2001}, employed here, defines $m$ in terms of the mean cosmic density $\bar{\rho}$.} is 200 times the Universes mean density, $m$ is the mass inside this sphere, $c$ is the halo concentration parameter and $u$ is the normalized density profile.

We model the mass and redshift dependence of $c$ with the fitting formula by \citet{Bullock2001},
\begin{equation}
	c(m,z) = \frac{c_0}{1+z}\left(\frac{m}{m_\star(z)}\right)^{-\alpha}\;,
\end{equation}
with $c_0=9$ and $\alpha=0.13$. The mass $m_\star$ is that enclosed by a sphere of radius $r_\star$, 
\begin{equation}
	m_\star(z) = \frac{4\pi}{3}\,\bar{\rho}(z) r_\star^3(z)\;,
\end{equation}
where $r_\star$ is the scale at which the standard deviation $\sigma(r, z)$ of linear density fluctuations is equal to the critical overdensity $\delta_\mathrm{c}$, i.e., $\sigma(r_\star, z ) = \delta_\mathrm{c}(z)$.  The $\sigma(r, z)$ is given by the convolution of the linear matter power spectrum $P_\mathrm{lin}(k, z)$ with the Fourier transformed tophat filter $\hat{W}(x)$,
\begin{equation}
	\label{eq:defSigma(m)}	
	\sigma^2(r, z) = 2\pi \int \dd{k} \; k^2\, P_\mathrm{lin}(k, z)\, \hat{W}^2\Big(r\,k\Big)\;.
\end{equation}

The third group of ingredients are the HMF $n(m, z)\dd{m}$ and the halo bias. The HMF describes the comoving number density of dark matter halos with mass between $m$ and $m+\dd{m}$. We use the HMF by \citet{Sheth1999}, 
\begin{align}
	\label{eq:sheth-tormen-hmf}
	n(m, z)\, \dd{m}
	&= A\, \frac{\bar{\rho}}{m^2} \,\dv{\ln\nu}{\ln m}\, \dd{m}\\
	&\notag \quad \times \left[1 + \frac{1}{(q\,\nu)^{2p}}\right]\sqrt{\frac{(q\,\nu)^2}{2\pi}}\, \exp(-\frac{(q\nu)^2}{2}) \;,
\end{align}
where the parameters $A=0.322$, $p=0.3$, and $q=0.707$ were found in $N$-body simulations, and $\nu = \delta_\mathrm{c}/\sigma(r_m,z)$ for $r_m^3=3m\,(4\pi\bar{\rho})^{-1}$.

The halo bias quantifies the clustering of halos by the ratio of the halo density contrast, $\delta_\mathrm{h}$, and the matter density contrast, $\delta$. We assume a linear halo bias, 
\begin{align}
	\delta_\mathrm{h}(\vec{x}, z\,|\,m) &= b_1(m, z)\, \delta(\vec{x}, z),
\end{align}
neglecting terms higher than linear in $\delta$. Using the so-called peak-background split formalism \citep{Mo1996, Scoccimarro2001} the linear bias $b_1$ is
\begin{equation}
	b_1(m, z) = 1 + \frac{q\,\nu^2(z) -2}{\delta_\mathrm{c}(z)} + \frac{2p}{1+q^p\,\nu^{2p}(z)} \, \frac{1}{\delta_\mathrm{c}(z)}\;,
\end{equation}
with the $q$ and $p$ as in the Sheth-Tormen HMF. Assuming a deterministic halo bias, the halo power spectrum is then
\begin{equation}
	\label{eq:halo power spectrum}
	P_\mathrm{h}(k, z\,|\,m_1, m_2) = b_1(m_1, z)\, b_1(m_2, z)\, P_\mathrm{lin}(k, z)\;,
\end{equation}
and the halo bispectrum is
\begin{align}
	\label{eq:halo bispectrum}
	&B_\mathrm{h}(\vec{k}_1, \vec{k_2}, z \,|\,m_1, m_2, m_3) \\
	&\notag= b_1(m_1, z)\, b_1(m_2, z)\, b_1(m_3, z)\, B_\mathrm{lin}(\vec{k}_1, \vec{k}_2, z)
\end{align}
for the linear matter bispectrum by \cite{Bernardeau2002},
\begin{align}
	B_\mathrm{lin}(\vec{k}_1, \vec{k}_2, z) &= 2\, F(\vec{k}_1, \vec{k}_2)\, P(k_1, z)\, P(k_2, z)\\
	&\notag \quad + 2\, F(\vec{k}_1, \vec{k}_3)\, P(k_1, z)\, P(k_3, z)\\
	&\notag \quad + 2\, F(\vec{k}_2, \vec{k}_3)\, P(k_2, z)\, P(k_3, z)\;,
\end{align}
where $\vec{k}_3=-\vec{k}_1-\vec{k}_2$ and
\begin{equation}
	F(\vec{k}_1, \vec{k}_2) = \frac{5}{7} + \frac{2}{7}\, \frac{(\vec{k}_1\cdot\vec{k}_2)^2}{k_1^2\,k_2^2} + \frac{1}{2}\, \frac{\vec{k}_1\cdot \vec{k}_2}{k_1\,k_2}\left(\frac{k_1}{k_2} + \frac{k_2}{k_1}\right)\;.
\end{equation}

Fourth, for the average number density of satellite galaxies $a$ inside halos of mass $m$, $\expval{\Ns{a}}u^a\rmg(\vec{x}\,|\,m)$, we assume an NFW profile with concentration 
\begin{equation}
    \label{eq:galaxy concentration}
    c^a\rmg = f^a c\;.
\end{equation}
This concentration may differ from that of the halo matter if $f^a\ne1$. A similar parameter was, e.g., introduced by \citet{Cacciato2012} in galaxy clustering halo models, who show that $f^a$ affects galaxy clustering at scales below $1\,h^{-1}\, \rm Mpc$.

Finally, we express the first-- and second--order moments of the HODs by the model of \citet{Zheng2007}, where galaxies are split into centrals and satellites, each with their own expected number per halo, such that the expected number $\expval{N^a\,|\,m}$ of galaxies from sample $a$ in a halo of mass $m$ is
\begin{equation}
	\expval{N^a\,|\,m} = \expval{\Nc{a}\,|\,m} + \expval{\Ns{a}\,|\, m}\;.
\end{equation}
Each halo hosts at most one central galaxy, $\Nc{a}=1$, situated at the halo centre but can contain several satellite galaxies, $\Ns{a}$. With this split, the galaxy-galaxy-matter bispectrum requires the specification of the mean numbers $\expval{\Nc{a}\,|\,m}$ and $\expval{\Ns{a}\,|\,m}$ of central and satellite galaxies,
as well as the numbers of central and satellite pairs $\expval{\Nc{a}\Nc{b}\,|\,m}$, $\expval{\Ns{a}\Ns{b}\,|\,m}$,  for $a\neq b$, and $\expval{\Ns{a}(\Ns{a}-1)\,|\,m}$  for $a=b$.

The mean number $\expval{\Nc{a}\,|\,m}$ of central galaxies depends only on halo mass. For small halo masses, no galaxy formation occurs, so $\expval{\Nc{a}\,|\,m}=0$. Halos with masses above a certain threshold, though, will contain central galaxies, at most one per halo. Similar to \citet{Zheng2007}, we assume
\begin{equation}
	\label{eq:HOD_Ncen}
	\expval{\Nc{a}\,|\,m} = \frac{\alpha^{a}}{2}\,\left\lbrace 1+\erf\left[\frac{\log(m)-\log(M_\mathrm{th}^{a})}{\sigma^{a}}\right]\right\rbrace\;, 
\end{equation}
with the free parameters $\alpha^{a}$, $M_\mathrm{th}^{a}$, and $\sigma^{a}$. The mass $M_\mathrm{th}$ is the halo mass below which we do not expect halos to contain galaxies.  The parameter $\sigma^{a}$ determines the transition of $\expval{\Nc{a}\,|\,m}$ from 0 to $\alpha^a$. If $\sigma^{a}$ is small, the transition from $\expval{\Nc{a}\,|\,m}=0$ to $\expval{\Nc{a}\,|\,m}=\alpha^{a}$ occurs quickly, whereas the transition is slower for larger $\sigma^{a}$. The parameter $0\le\alpha^a\le1$, for the maximum of $\expval{N^a|m}$, gives the fraction of massive halos ($m\gg M_\mathrm{th}^a$) with a central galaxy from population $a$. Its inclusion is necessary because we are splitting galaxies into samples: only one sample can contain the central galaxy in a
halo at a time. Since no more than one central galaxy per halo is allowed, the sum of all $\alpha^a$ from disjunct samples can never exceed unity.

\begin{figure}
	\includegraphics[width=\linewidth, trim={0.5cm 0.5cm 0.5cm 0}, clip]{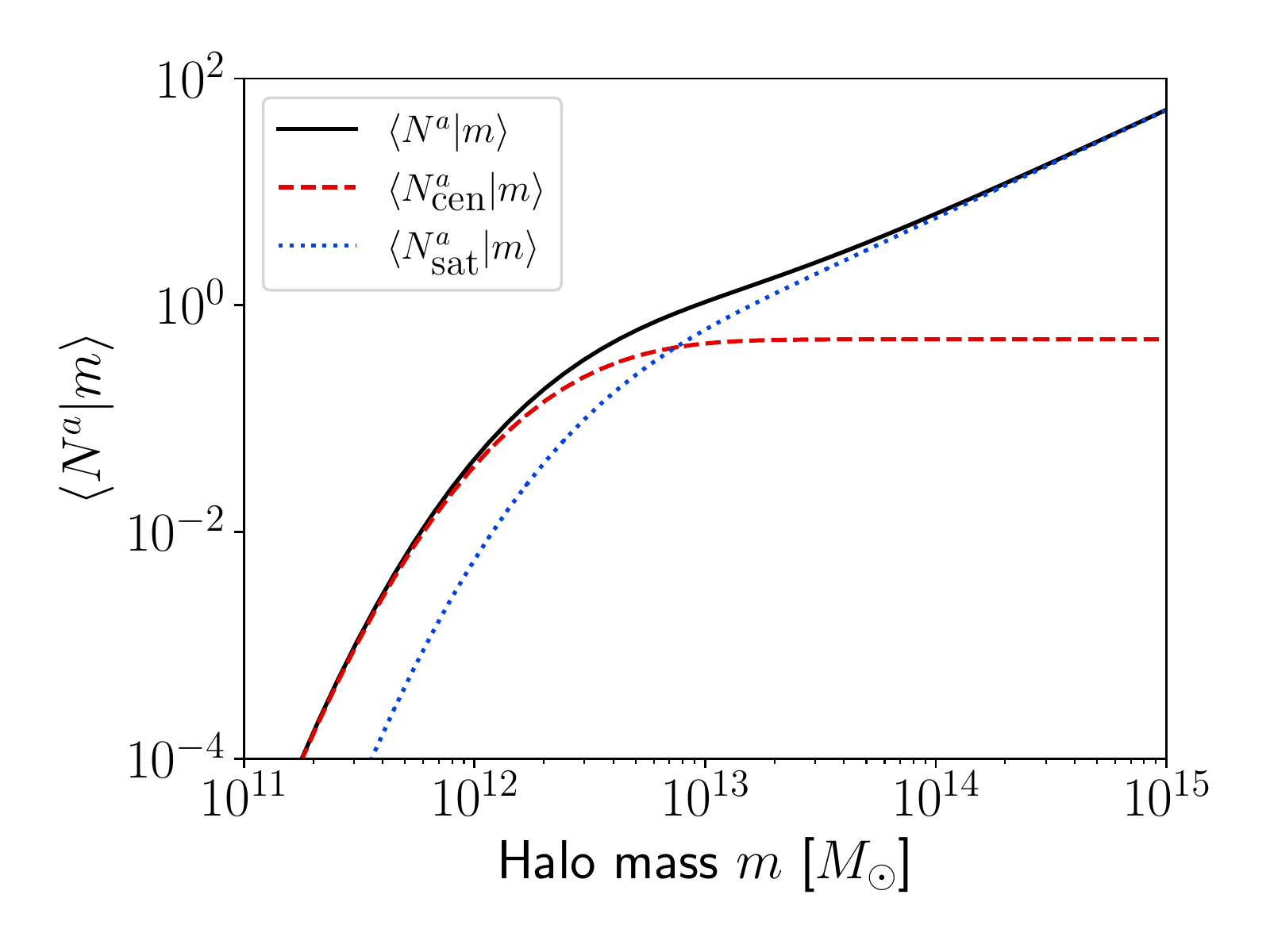}
	\caption{Mean per-halo numbers of galaxies for fiducial HOD parameters in Table~\ref{tab: hod params}. The solid black line shows the total galaxy number per halo, the dashed red line shows the fraction of halos with central galaxies, and the dotted blue line shows the number of satellite galaxies per halo.}
	\label{fig:HOD}
\end{figure}

For the mean number $\expval{\Ns{a}\,|\,m}$ of satellites, we assume, based on \citet[]{Zehavi2005},
\begin{equation}
	\label{eq:HOD_Nsat}
	\expval{\Ns{a}\,|\,m} = \frac{1}{2} \left\lbrace 1+\erf\left[\frac{\log(m)-\log(M_\mathrm{th}^{a})}{\sigma^{a}}\right]\right\rbrace \left(\frac{m}{M^{\prime a}}\right)^{\beta^{a}}\;,
\end{equation}
with the free parameters $M^{\prime a}$ and $\beta^{a}$. The satellite number therefore follows the central galaxy number for small halo masses and becomes a power law at high halo masses. To illustrate the dependence of the HOD terms on halo mass, Fig~\ref{fig:HOD} shows the expected number of satellite and central galaxies and the parameters in Table~\ref{tab: hod params}. The total number of galaxies depends strongly on the central galaxy distribution for low-mass halos. Satellite galaxies predominate in massive halos.

The galaxy-galaxy matter bispectrum depends not only on the expected numbers of centrals and satellites, but also on the expected number $\expval{\Ns{a}\, (\Ns{a}-1)\,|\,m}$ of unmixed satellite pairs, the expected number $\expval{\Ns{a}\,\Ns{b}\,|\,m}$ of mixed satellite pairs, i.e,  $a\neq b$, and the number $\expval{\Nc{a} \Ns{b}\,|\,m}$ of central-satellite pairs per halo of mass $m$. In principle, it also depends on the number $\expval{\Nc{a}\Nc{b}\,|\,m}$ of central pairs, but this number is 0 for all $m$, as each halo contains only one central galaxy.
We assume, as, e.g., in \cite{Cacciato2012}, that the occupation numbers of centrals and satellites are statistically independent, 
\begin{equation}
  \expval{\Nc{a}\, \Ns{b}\,|\, m} = \expval{\Nc{a}\,|\,m}\,\expval{\Ns{b}\,|\,m}\;.
\end{equation}
This common assumption entails that the mean number of type $b$ satellite galaxies in halos of mass $m$ is independent of whether a central galaxy $a$ is present or not. Furthermore, following \citet{Kravtsov2004}, we assume that satellite occupation numbers vary according to a Poisson statistic,
\begin{equation}
\label{eq:HOD_NsatNsat_same}
	\expval{\Ns{a}\left(\Ns{a}-1\right)\,|\,m}=\expval{\Ns{a}\,|\,m}^2\;.
\end{equation}
We test this assumption and its impact on the model accuracy with galaxies inserted via SAM in a cosmological simulation in Appendix \ref{sec:superPoissonian}.

Finally, we introduce as a new parameter the cross-correlation coefficient $r^{ab}$ of satellites, defined for mixed galaxy samples $a\neq b$ by
\begin{equation}
	\label{eq:HOD_NsatNsat}
	\expval{\Ns{a}\,\Ns{b}\,|\,m} = \expval{\Ns{a}}\, \expval{\Ns{b}} + r^{ab}(m)\, \sqrt{\expval{\Ns{a}\,|\,m} \expval{\Ns{b}\,|\,m}}\;.
\end{equation}
The coefficient $r^{ab}$ is negative if the numbers of satellite galaxies $a$ and $b$ are anti-correlated; and positive if they are positively correlated. If the satellite galaxies are distributed Poissonian, i.e.,  $\sigma^2(\Ns{a}\,|\,m) = \expval{\Ns{a}\,|\,m}$, then $r^{ab}$ is a Pearson correlation coefficient. As we show in Sect.~\ref{sec:results}, mock galaxies from a SAM imply a mass dependence of $r^{ab}(m)$, which we model by
\begin{equation}
\label{eq:model_r}
	r^{ab}(m)= A^{ab} \left(\frac{m}{10^{12}\,M_\odot}\right)^{\epsilon^{ab}}\;,
\end{equation}
with the free parameters $A^{ab}$ and $\epsilon^{ab}$. A value of $A^{ab}=0$ corresponds to uncorrelated galaxy samples. The cross-correlation between galaxy samples is independent of halo mass if $\epsilon=0$. Ignoring a mass dependence of $r^{ab}$ in a more simplistic approach, as, e.g., in \cite{Simon2009}, would result in a weighted average over the true $r^{ab}(m)$. The comparison of such an average value to the true $r^{ab}(m)$ in the simulation would not be straightforward, as the weighting per halo mass, given the survey characteristics, needs to be determined. A halo-mass dependent $r^{ab}(m)$, on the other hand, is easier to interpret, which is why we use the mass-dependent fitting formula above.

A decrease of $r^{ab}(m)$ towards small halo masses can be understood qualitatively as follows. The mean satellite number of galaxies from sample $a$ inside a halo decreases with halo mass until a regime is reached where we find at most one. Likewise, at small enough halo masses, we find at most one galaxy from a second sample $b$ (other galaxy populations may be present as well but are not of interest here). Then there are basically only four possibilities to populate a halo: (i) one galaxy from sample $a$, (ii) one galaxy from sample $b$, (iii) a pair of galaxies from $a$ and $b$, and (iv) no galaxies from $a$ or $b$. We shall denote the probabilities of these cases (i) to (iv) by $p_a$, $p_b$, $p_{ab}$, and $1-p_a-p_b-p_{ab}$, respectively. For this Poisson process with four outcomes, we find for the mean number of galaxies $a$ in a halo
\begin{align}
\label{eq:toycalculation_r}
  \langle \Ns{a} \rangle
  &= \langle \Ns{a}|(i)\rangle\,p_a+\langle \Ns{a}|(ii)\rangle\,p_b\\
  &\notag \quad +\langle \Ns{a}|(iii)\rangle\,p_{ab}+\langle \Ns{a}|(iv)\rangle\,(1-p_a-p_b-p_{ab})\\
  &= 1 \times p_a + 0 \times p_b + 1 \times p_{ab} + 0 \times (1-p_a-p_b-p_{ab})\\
  &=p_a+p_{ab}\;,
\end{align}
and $\langle \Ns{b}\rangle=p_b+p_{ab}$ for galaxies $b$. Similarly, the mean number of pairs is $\langle \Ns{a}\,\Ns{b}\rangle=p_{ab}$, and the variances are $\sigma^2(\Ns{a})=\langle (\Ns{a})^2\rangle-\langle \Ns{a}\rangle^2=(p_a+p_{ab})(1-p_a-p_{ab})$ and $\sigma^2(\Ns{b})=(p_b+p_{ab})(1-p_b-p_{ab})$. The Pearson cross-correlation coefficient of variations in the galaxy numbers consequently becomes
\begin{multline}
  r_\mathrm{pear}^{ab}=
  \frac{\langle N_a N_b\rangle-\langle N_a\rangle\,\langle N_b\rangle}
  {\sigma(\Ns{a})\,\sigma(\Ns{a})}\\
  =\frac{p_{ab}-(p_a+p_{ab})(p_b+p_{ab})}
  {\sqrt{(p_a+p_{ab})(p_b+p_{ab})(1-p_a-p_{ab})(1-p_b-p_{ab})}}\\
  \le\frac{(p_a-p_b)^2-2(p_a+p_b)+1}{1-(p_a-p_b)^2}\;.
\end{multline}
Therefore, at the extreme end, for halos too small to host more than one galaxy, i.e., $p_{ab}=0$, the value of $r_\mathrm{pear}^{ab}$ exactly converges to zero for $p_a,p_b\to0$ (that is $\langle \Ns{a}\rangle$, $\langle \Ns{b}\rangle\to0$). But, already in the intermediate regime, where $0\le p_{ab}\le1-p_a-p_b$ and $p_a,p_b\sim0.1$, the upper limit of $r_\mathrm{pear}^{ab}$ in the last line confines the correlation factor to smaller values: for example, $r^{ab}\le0.6$ for $p_a=p_b=0.1$, or $r^{ab}\le0.81$ for $p_a=0$ and $p_b=0.1$. Therefore, a decreasing $r_\mathrm{pear}^{ab}$ for $m\to0$ is a natural outcome of halos sparsely populated with galaxies $a$ and $b$. This also applies to $r^{ab}$ for sub-Poisson variances, expected at low halo masses, since $r^{ab}<r_\mathrm{pear}^{ab}$ in this case.

As stated above, $r^{ab}$ is a Pearson correlation coefficient only for Poisson variances of the satellite numbers, as assumed in our halo model. We show in Appendix \ref{sec:superPoissonian}, however, that this assumption is inaccurate for the SAMs, and possibly real galaxies, at high or low halo masses, with specifics depending on the galaxy selection. In extreme cases, variances may be an order of magnitude larger for $\langle N|m\rangle\gg1$, or smaller for $\langle N|m\rangle\ll1$. Consequently, $r^{ab}$ is at small halo masses systematically lower than the Pearson correlation coefficient ($r^{ab}<r_\mathrm{pear}^{ab}$) and higher for high halo masses ($r^{ab}>r_\mathrm{pear}^{ab}$). Nevertheless, true and reconstructed correlation parameters in this paper are comparable because we use the same $r^{ab}$ definition in both cases (and in the science verification). In addition, our following verification results show that the strict assumption of Poisson satellites still allows an accurate HOD reconstruction within the statistical errors of KV450$\times$GAMA data.

After describing our choices for the halo model parameters, we can model the galaxy-galaxy-matter bispectrum $\Bggd$. We derive the bispectrum in the following subsection.

\subsection{Galaxy-galaxy-matter bispectrum}
\label{sect:bispectrum}
Using Eqs.~\eqref{eq:definition delta}, \eqref{eq:definition delta_g}\, and \eqref{eq:defintion bispectrum}, the galaxy-galaxy-matter bispectrum is given by
\begin{align}
	\label{eq:bispectrum halo model}
	&\frac{1}{\bar{n}^a\rmg(z)\, \bar{n}^b\rmg(z)\, \bar{\rho}(z)}\expval{\hat{n}^a\rmg(\vec{k}_1, z) \, \hat{n}^b\rmg(\vec{k}_2, z) \, \hat{\rho}(\vec{k}_3, z)} =\\
	&\notag(2\pi)^3 \Bggdab(\vec{k}_1, \vec{k}_2, z) \, \dirac(\vec{k}_1+\vec{k}_2+\vec{k}_3) + \text{unconnected terms}  \;,
\end{align}
where $a$ and $b$ denote the galaxy samples, and `unconnected terms' are those proportional to $\dirac(\vec{k}_1)$, $\dirac(\vec{k}_2)$, or $\dirac(\vec{k}_3)$, which do not affect the bispectrum. The bispectrum can be divided into three terms: the 1-halo term $\Bggdabh{1}(\vec{k}_1, \vec{k}_2, z)$, the 2-halo term $\Bggdabh{2}(\vec{k}_1, \vec{k}_2, z)$, and the 3-halo term $\Bggdabh{3}(\vec{k}_1, \vec{k}_2, z)$, or together
\begin{align}
	\Bggdab(\vec{k}_1, \vec{k}_2, z) &= \Bggdabh{1}(\vec{k}_1, \vec{k}_2, z) + \Bggdabh{2}(\vec{k}_1, \vec{k}_2, z)\\
	&\notag\quad + \Bggdabh{3}(\vec{k}_1, \vec{k}_2, z)\;.
\end{align}
The 1-halo term depends on the correlation between galaxies numbers and matter density in the same halo. One part of the 2-halo term is caused by the correlation of galaxies in one halo with the matter in a different halo. Its second part is due to the correlation of galaxies and matter in one halo with galaxies in another halo. Correlations between matter and galaxies in three distinct halos cause the 3-halo term. 

To derive $\hat{\rho}$ and $\hat{n}$, we use that in the halo model the cosmic density field consists of $H$ halos with masses $\lbrace m_1,\dots,m_H \rbrace$ at positions $\lbrace \vec{x}_1,\dots,\vec{x}_H\rbrace$. With the normalized density profile $u$ the matter density field is
\begin{equation}
	\rho(\vec{x}, z) = \sum_{i=1}^H m_i\, u(\vec{x}-\vec{x}_i\,|\,m_i, z)\;,
\end{equation}
whose Fourier transform  is
\begin{equation}
	\label{eq:ft halo matter density}
	\hat{\rho}(\vec{k}, z) = \sum_{i=1}^H m_i\, \hat{u}(\vec{k}\,|\,m_i, z)\, \exp(-\I \vec{k}\cdot \vec{x}_i)\;.
\end{equation}

Galaxies are treated as discrete point particles. Each satellite galaxy $j$ from sample $a$ belongs to a halo centred at $\vec{x}_i$ and is at separation $\Delta \vec{x}^a_{ij}$ from the halo centre. Centrals are exactly at the halo centre. The number density of galaxies from sample $a$ is therefore
\begin{equation}
	\label{eq:galaxy number density}
	n\rmg^a(\vec{x}, z) = \sum_{i=1}^H \left[\Ncc{a}{i}\,\dirac(\vec{x}-\vec{x}_i) + \sum_{j=1}^{\Nss{a}{i}}\dirac(\vec{x} - \vec{x}_i - \Delta \vec{x}_{ij}) \right]\;,
\end{equation}
where $\Nss{a}{i}$ is the number of satellite galaxies and $\Ncc{a}{i}$ the number of central galaxies from sample $a$ inside halo $i$. The Fourier transform of the number density is
\begin{equation}
	\label{eq:ft galaxy number density}
	\hat{n}^a\rmg(\vec{k}, z) = \sum_{i=1}^{H} \left( \Ncc{a}{i}\, \mathrm{e}^{-\I \vec{k}\cdot \vec{x}_i} + \sum_{j=1}^{\Nss{a}{i}} \mathrm{e}^{-\I \vec{k}\cdot (\vec{x}_i + \Delta \vec{x}^a_{ij})}\right)\;.
\end{equation}

To derive the bispectrum we insert Eqs.~\eqref{eq:ft halo matter density} and ~\eqref{eq:ft galaxy number density} into Eq.~\eqref{eq:bispectrum halo model}, which leads to
\begin{align}
	\Bggdabh{1}(\vec{k}_1, \vec{k}_2, z) 
	&= \frac{1}{\bar{n}^a\rmg(z)\, \bar{n}^b\rmg(z)\, \bar{\rho}(z)}\, \int \dd{m}\;n(m)\, \\
	&\notag \quad \times m\,\hat{u}(\vec{k}_1+\vec{k}_2\,|\,m, z) \, G^{ab}(\vec{k}_1, \vec{k}_2\,|\,m, z)\;;
\end{align}
\begin{align}
	\Bggdabh{2}(\vec{k}_1, \vec{k}_2, z)
	&= \frac{1}{\bar{n}^a\rmg(z)\, \bar{n}^b\rmg(z)\, \bar{\rho}(z)} \int \dd{m_1} \int \dd{m_2}\; n(m_1)\, n(m_2)\,\\
	&\notag \quad \times \Big[ m_1\,\hat{u}(\vec{k}_1+\vec{k}_2\,|\,m_1, z)\, G^{ab}(\vec{k}_1, \vec{k}_2\,|\,m_2, z)\\
	&\notag \quad \quad \times  P_\mathrm{h}(|\vec{k}_1+\vec{k}_2|\,|\,m_1, m_2, z)\\
	&\notag \quad \quad + m_2\,\hat{u}(\vec{k}_1+\vec{k}_2\,|\,m_2, z)\,G^{a}(\vec{k}_1\,|\,m_1, z) \\
	&\notag \quad \quad \times  G^{b}(\vec{k}_2\,|\,m_2, z)\, P_\mathrm{h}({k}_1\,|\,m_1, m_2, z) \\
	&\notag \quad \quad +  m_2\,\hat{u}(\vec{k}_1+\vec{k}_2\,|\,m_2, z)\,G^{b}(\vec{k}_2\,|\,m_1, z) \\
	&\notag \quad \quad \times G^{a}(\vec{k}_1\,|\,m_2, z)\, P_\mathrm{h}({k}_2\,|\,m_1, m_2, z) \Big]\;;
\end{align}
and
\begin{align}
	&\Bggdabh{3}(\vec{k}_1, \vec{k}_2, z) \\
	&\notag=\frac{1}{\bar{n}^a\rmg(z)\, \bar{n}^b\rmg(z)\, \bar{\rho}(z)} \int \dd{m_1} \int \dd{m_2} \int \dd{m_3}\;n(m_1)\, n(m_2)\, n(m_3)\,\\
	&\notag\quad\times m_3\,\hat{u}(-\vec{k}_1-\vec{k}_2\,|\,m_3, z)\, G^{a}(\vec{k}_1\,|\,m_1, z)\,G^{b}(\vec{k}_2\,|\,m_2, z)\\
	&\notag\quad\times B_\mathrm{h}(\vec{k}_1, \vec{k}_2\,|\,m_1, m_2, m_3, z) \;,
\end{align}
with 
\begin{equation}
	G^{a}(\vec{k}\,|\,m, z) := \expval{\Nc{a}\,|\,m}+ \expval{\Ns{a}\,|\,m} \hat{u}^a_\mathrm{g}(\vec{k}\,|\,m, z)\;,
\end{equation}
and
\begin{align}
	G^{ab}(\vec{k}_1, \vec{k}_2\,|\,m, z) &:=	\expval{\Nc{a}\,(\Nc{b}-\kronecker{a}{b})\,|\,m}\\
	&\notag\quad + \expval{\Nc{a}\, \Ns{b}\,|\,m}\hat{u}_\mathrm{g}^b(\vec{k}_2\,|\,m, z)	\\
	&\notag\quad +\expval{\Nc{b}\, \Ns{a}\,|\,m}\hat{u}_\mathrm{g}^a(\vec{k}_1\,|\,m, z) \\
	&\notag\quad + \expval{\Ns{a}\, (\Ns{b}-\kronecker{a}{b})\,|\,m}\, \\
	&\notag\quad \quad \times \hat{u}_\mathrm{g}(\vec{k}_1\,|\,m, z)\, \hat{u}_\mathrm{g}(\vec{k}_2\,|\,m,z) \;,
\end{align}
where the Kronecker symbol $\kronecker{a}{b}$ is $1$ for $a=b$, and $0$ otherwise. We derive these equations in full in Appendix \ref{app:calculationBispectrum_halo model}. 

With the ingredients of the halo model from Sect.~\ref{sec:halomodel:ingredients}, the bispectrum is fully specified. By inserting it into Eq.~\eqref{limber} and \eqref{eq:NNMapFromBggkappa} we obtain the aperture statistics. Before fitting the model to simulated and actual G3L measurements, we discuss the parameter sensitivity of the model in the next section.

\subsection{Discussion of model parameters}
\label{sec:results:qual}
We qualitatively study the impact of the parameters on the theoretical G3L signal. For this, we vary each parameter inside the prior range, adopted for the likelihood
analysis below (Table~\ref{tab: hod params}), while keeping the other parameters fixed to `fiducial' values, i.e., to the centres of the parameter prior range. There are six free parameters for each galaxy sample and two additional parameters for the correlation of satellite numbers, hence in total $14$ parameters for the bispectrum of a combination of galaxy samples. We arrange these parameters inside the vector $\vec{p}$, 
\begin{equation}
	\vec{p}^\mathrm{T}=
	\left(\alpha^{a}\; \sigma^{a}\; M_\mathrm{th}^{a}\; \beta^{a}\; M^{\prime a}\, f^{a}\; \alpha^{b}\; \sigma^{b}\; M_\mathrm{th}^{b}\; \beta^{b}\; M^{\prime b} f^{b}\; A^{ab}\; \epsilon^{ab}\right)\;,
\end{equation}
and write the modelled aperture statistics for a set $\vec{p}$ of parameters and scale radius $\theta$ as $\NNMap(\theta, \vec{p})$.

The priors are aimed to be uninformative. For $\sigma$, $\log_{10}(M_\mathrm{th}^a/\Msun)$, $\beta^a$, and $\log_{10}(M^{\prime a}/\Msun)$, they are based on the galaxy-galaxy-lensing halo model by \citet{Clampitt2017}, whose prior ranges we doubled for the G3L analysis. For $\alpha$, the prior range contains all possible values between 0 and 1. For the parameters $f$, $A$, and $\epsilon$, which are unique to our G3L halo model, there are no reference values. Therefore, we chose arbitrary broad ranges centred on the values $f=0$ (indicating galaxy distributions perfectly following dark matter halos), $A=1$, and $\epsilon=1$ (indicating no correlations between galaxy HODs). To sanity-check the dependence on priors, we performed the analysis on the validation data also with prior ranges two times larger and found no difference in the final parameter constraints.

\begin{figure*}
	\centering
	\includegraphics[width=\linewidth, trim={3.5cm 0.5cm 4cm 3cm}, clip]{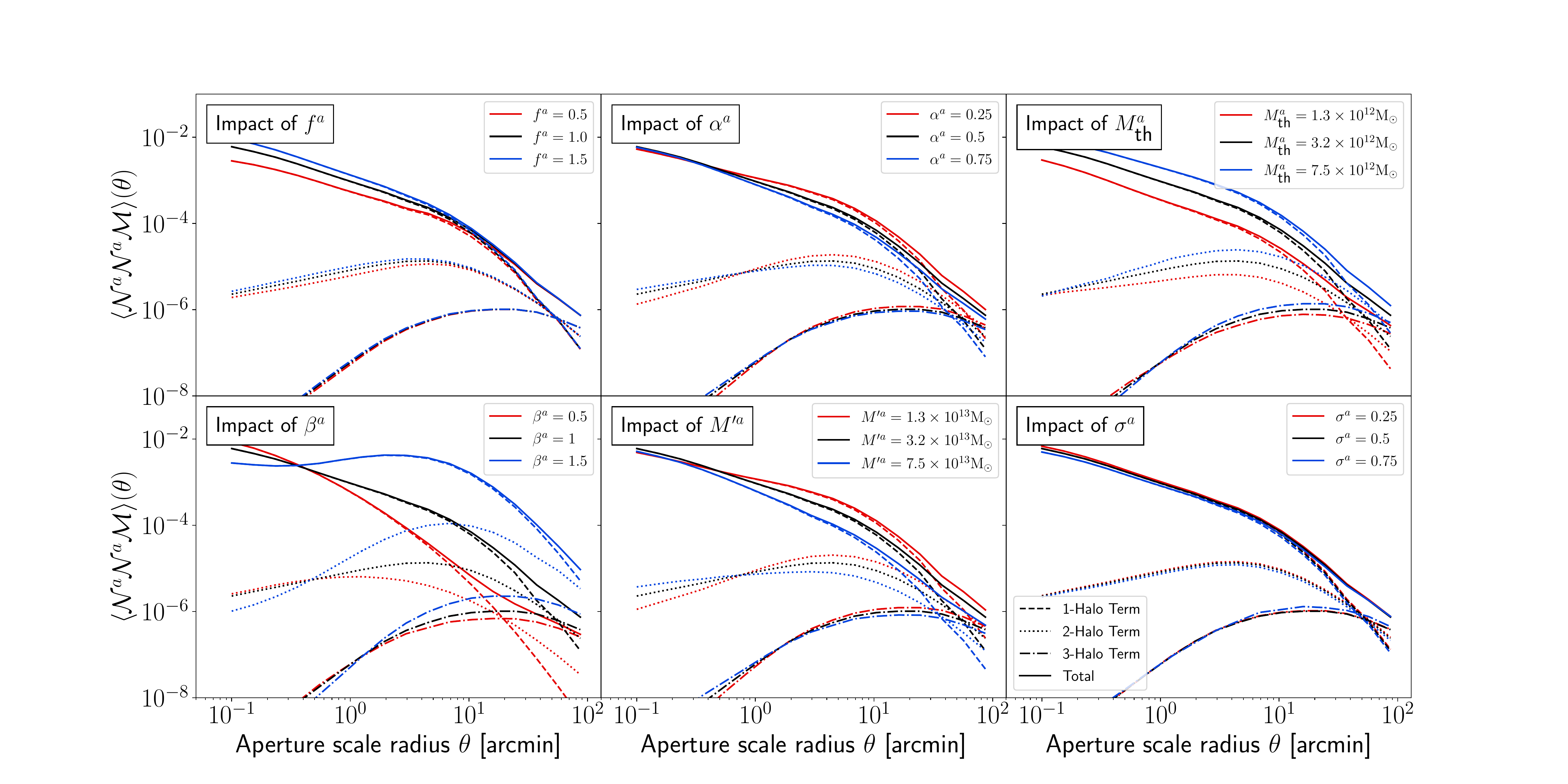}
	\caption[Impact of halo model parameters on $\NNM{a}{a}$ for unmixed lens pairs]{Impact of halo model parameters on $\NNM{a}{a}$ for unmixed lens pairs. In each panel, only one parameter is varied at a time. Solid lines indicate the total $\NNM{a}{a}$, while dashed lines show the 1-halo, dotted lines the 2-halo, and dash-dotted lines the 3-halo term.}
	\label{fig:ImpactParamsN2Map}
\end{figure*}

Figure \ref{fig:ImpactParamsN2Map} shows $\NNM{a}{a}$ for unmixed lens pairs, when varying $\alpha^a, \sigma^a, M_\mathrm{th}^a, M^{\prime a}$, and $f^a$. All parameters, except for $\sigma$, visibly impact the aperture statistics. However, we note that the trends seen in Fig.~\ref{fig:ImpactParamsN2Map} are for varying each parameter individually and do not show the correlation between the parameters. For example, the threshold mass $M_\mathrm{th}^a$ and the satellite mass scale $M^{' a}$ are tightly correlated: Increasing $M_\mathrm{th}^a$ while decreasing $M^{' a}$ could lead to the same $\NNM{a}{a}$.

\begin{table}
	\caption[Halo model parameters]{Fiducial values and flat priors of the halo model parameters.}
	\label{tab: hod params}
	\centering
	\begin{tabular}{cccc}
		\hline
		{Parameter} & {Fiducial} & {Prior Range} & Introduced in\\
		\hline
		$\alpha^a$ &  $0.5$ & $[0,1]$ & Eq.~\eqref{eq:HOD_Ncen}\\
		$\sigma^a$ &  $0.5$ & $[0.01,1]$ & Eq.~\eqref{eq:HOD_Ncen}\\
		$\log_{10}(M_\mathrm{th}^a/\Msun)$ &  $12.5$ & $[10, 15]$  & Eq.~\eqref{eq:HOD_Ncen}\\
		$\beta^a$  & $1$ & $[0.1,2]$  & Eq.~\eqref{eq:HOD_Nsat}\\
		$\log_{10}(M^{\prime a}/\Msun)$ & $13.5$& $[12,15]$   & Eq.~\eqref{eq:HOD_Nsat}\\
		$f^a$ & $1$ & $[0,2]$ & Eq.~\eqref{eq:galaxy concentration} \\
		$A^{ab}$ & $0$ & $[-1,1]\times10^{3}$ & Eq.~\eqref{eq:model_r}\\
		$\epsilon^{ab}$ & 0 & $[-5, 5]$ & Eq.~\eqref{eq:model_r}\\
		\hline
	\end{tabular}
\end{table}

The threshold halo mass $M^a_\mathrm{th}$ and the slope $\beta^a$ of the satellite HOD affect the signal the strongest on the scales we consider here. Increasing the threshold mass $M^a_\mathrm{th}$ from $3.2\times10^{12}\,\Msun$ to $7.5\times10^{12}\,\Msun$ roughly doubles the aperture statistics across the whole range of $\theta$ from $\ang[astroang]{;0.1;}$ to $\ang[astroang]{;100;}$ because $M^a_\mathrm{th}$ is closely connected to the galaxy bias. A higher $M^a_\mathrm{th}$ causes galaxies to reside in more massive halos, so the galaxy bias and the shear amplitude increase, which increases the amplitude of the G3L aperture statistics. The satellite parameter $M^{\prime a}$, which describes the mass scale where $\langle N_{\rm sat}|m\rangle\approx1$ if $M^a_\mathrm{th}\ll M^{\prime a}$, changes the signal amplitude for $\theta\gtrsim1^\prime$. A smaller $M^{\prime a}$ produces a higher amplitude by scaling up the number of satellites in a halo of a given mass.

The slope $\beta^a$ of $\expval{N^q_{\rm sat}|m}$ has the strongest impact on the statistics at $\theta \sim \ang[astroang]{;9;}$, an up to a 20-fold increase in amplitude. While a larger $\beta^a$ increases the signal on larger scales, at scales below $\ang[astroang]{;0.3;}$, it decreases the aperture statistics by up to $50\%$. These opposite trends have two reasons. First, a larger $\beta^a$ causes more massive halos to contain more satellite galaxies and pairs, giving the halos more weight in the second term in $G^a$ (Eq.~\ref{eq: def G_g}) and the last term in $G^{ab}$ (Eq.~\ref{eq: def G_gg}), compared to low-mass halos with fewer satellite pairs. This increases $\NNM{a}{a}$. Second, more massive halos are more extended and contribute predominantly at larger scales. Therefore, the signal increases at larger angular scales, but decreases at scales where lower-mass halos are more relevant. 

The concentration parameter $f^a$ affects the aperture statistics stronger at small scales than at large scales ($57\%$ at \ang[astroang]{;0.1;} and $40\%$ at \ang[astroang]{;1;}). A more concentrated galaxy profile (\mbox{$f^a>1$}) leads to more galaxies in the inner regions of each halo, which increases the galaxy density contrast $\delta\rmg$ at small scales. Consequently, the galaxy-galaxy-matter bispectrum and the aperture statistics increase at small scales. At scales above $\ang[astroang]{;11.5;}$, though, the change is less than $10 \%$.

\begin{figure}
	\centering
	\includegraphics[width=\linewidth, trim={0.5cm 0.5cm 0.5cm 0.5cm}, clip]{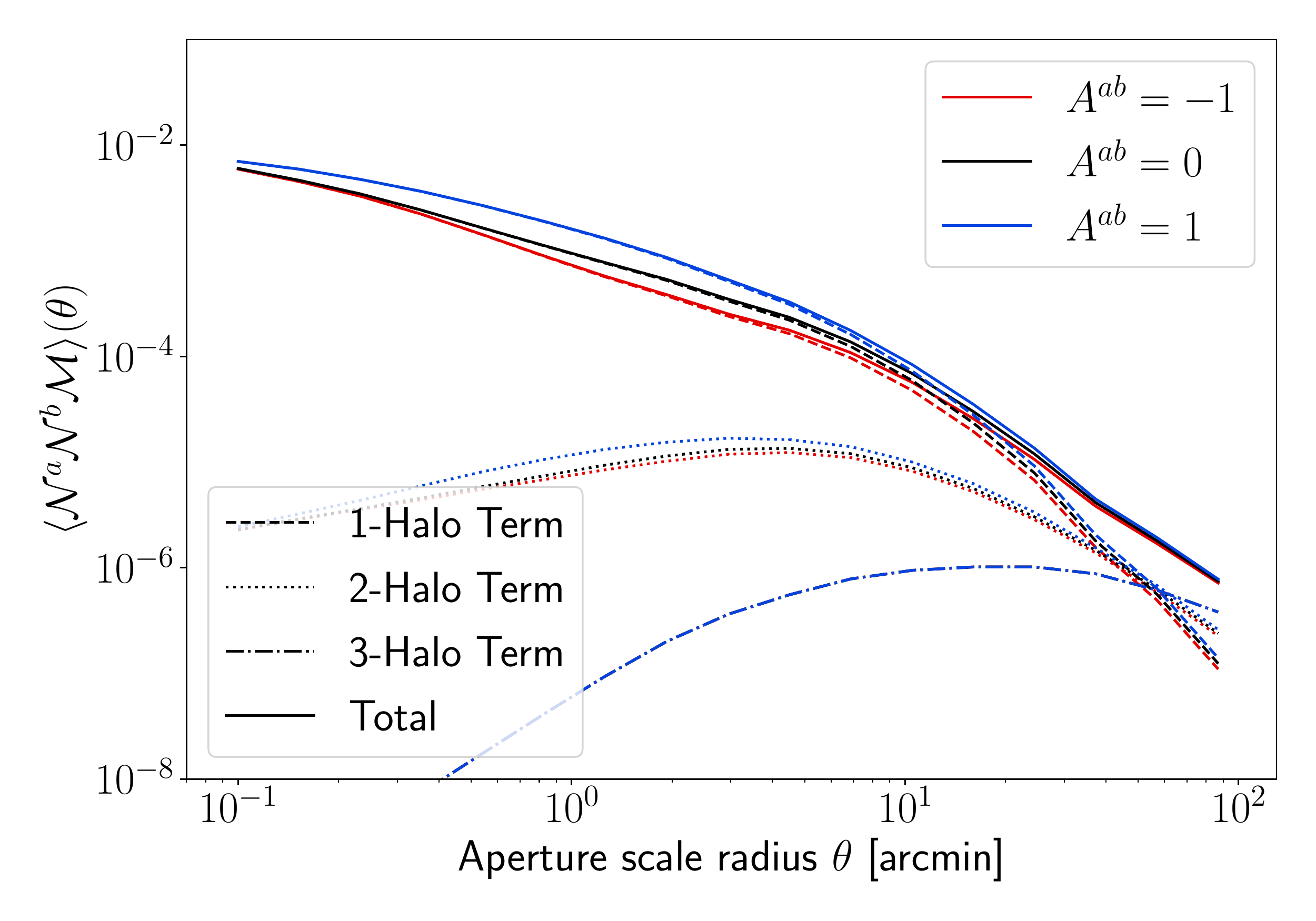}
	\caption[Impact of halo model parameters on the aperture statistics for mixed lens pairs]{Impact of the correlation of satellite numbers on the aperture statistics of mixed lens pairs. Satellite numbers are either fully correlated ($r^{ab}=1$, blue lines), uncorrelated ($r^{ab}=0$, black lines) or anti-correlated ($r^{ab}=-1$, red lines). Solid lines indicate the total aperture statistics, dashed lines the 1-halo, dotted lines the 2-halo, and dash-dotted lines the 3-halo term.}
	\label{fig:ImpactParamsNNMap}
\end{figure}

The sensitivity of the statistics for $r^{ab}$ is visible in Fig.~\ref{fig:ImpactParamsNNMap}, where we show $\NNM{a}{b}$ for mixed lens pairs for correlated satellite numbers ($r^{ab}=1$), uncorrelated numbers ($r^{ab}=0$), or anti-correlated numbers ($r^{ab}=-1$). For this example, we keep $r^{ab}$ constant with halo
mass, so $\epsilon^{ab}=0$, while $A^{ab}$ is varied between $A^{ab}\in\{0,\pm1\}$. The amplitude of $\NNM{a}{b}$ changes strongest for $\theta=\ang[astroang]{;1.1;}$ ($72\%$), where the signal is dominated by the 1-halo term containing $\expval{\Ns{a}\Ns{b}|m}$ and has a $r^{ab}$-dependence. The cross-correlation parameter has only a small effect (less than 10\% change) at scales above $\ang[astroang]{;30;}$ because these scales are dominated by the 3-halo term, independent of the per-halo number of satellite pairs.

In conclusion, the aperture statistics between \ang[astroang]{;0.1;} and \ang[astroang]{;100;} are most sensitive to the threshold halo mass $M^a_\mathrm{th}$ and the slope $\beta^a$. In contrast, we do not expect G3L to improve the $\sigma^a$ constraints beyond the prior range. To probe the cross-correlation parameter $r^{ab}$, the statistic $\NNM{a}{b}$ needs to be measured at scales below \ang[astroang]{;30;} and preferably around $\ang[astroang]{;1;}$. Both of these measurements are achieved for our survey data. 

\section{Data}
\label{sec:data}

Before describing the G3L measurement procedure and halo model fitting procedures, we briefly explain the simulated and observed data. 
These are the same data sets as in \citet{Linke2020b}: simulated (verification) data based on the MS and the SAM by \citet{Henriques2015}, and one composed of the overlap of KiDS, VIKING, and GAMA. As mentioned in Sect.~\ref{sec:intro}, the cosmology used to analyze the observation differs from the fiducial cosmology of the MS and is based on \citet{Planck2018}. This difference in cosmology should not impact our conclusions because we are not interested in a one-to-one comparison between the observation and simulation and merely use the mock data to validate our method. 

\subsection{\texorpdfstring{KV450 $\times$ GAMA}{KV450 x GAMA}}
\label{sec:data:obs}
We use observational data from the overlap of KiDS, VIKING, and GAMA, KV450 $\times$ GAMA for short, which is an area of approximately $\SI{180}{\deg\squared}$. KiDS \citep{Kuijken2015, deJong2015} and VIKING \citep{Edge2013, Venemans2015} are two photometric surveys covering approximately the same $\SI{1350}{\deg\squared}$ area, with KiDS observed in the optical with the VLT survey telescope, and VIKING observed in the near-infrared with the VISTA telescope. We use the public combined data release KV450, described in detail in \citet{Wright2019}. Galaxies were observed in the $u, g, r, i$ band for KiDS and $Z, Y, J, H, K_\mathrm{s}$ bands for VIKING with shape measurements performed in the $r$-band. This data set was processed with AstroWISE \citep{deJong2015} and THELI \citep{Erben2005, Schirmer2013} to give a catalogue of observed galaxies. The shapes of these galaxies were measured with \emph{lens}fit \citep{Miller2013, Kannawadi2019}. We use the KV450 galaxies with photometric redshifts between $0.5$ and $1.2$ (obtained by \citealt{Hildebrandt2020}) as sources for the G3L measurements.

Our lens galaxies are from GAMA \citep{Driver2009, Liske2015}, a spectroscopic survey conducted at the Anglo Australian Telescope. We use all galaxies listed in the table \verb*|distanceFramesv14| from the data management unit (DMU) \verb|LocalFlowCorrection| with a redshift quality parameter \verb|N_Q|$>2$ and with a spectroscopic redshift less than 0.5 to avoid overlap between lenses and sources. Each lens galaxy is assigned absolute magnitudes, restframe photometry, and stellar masses according to the table \verb*|stellarMassesLambdarv20| from the DMU \verb|StellarMasses| \citep{Taylor2011}. These were obtained with matched aperture photometry and the \verb|LAMBDAR| code \citep{Wright2016}, assuming the initial mass function by \citet{Chabrier2003}, stellar population synthesis according to \citet{Bruzual2003}, and dust extinction according to \citet{Calzetti2000}. To calculate the angular correlation function $\omega^{ab}$, we use the randoms in the table \verb*|randomsv02| from the DMU \verb|Randoms|, created by \citet{Farrow2015}.

Galaxies observed by GAMA are brighter than $r=19.8\,\mathrm{mag}$, rendering our lens galaxy sample flux-limited. We divide the GAMA galaxies into different samples: a `red' and a `blue' sample, using their restframe $(g-r)_0$ colour and their absolute magnitude $M_r$ in the $r$-band, and five samples defined by stellar masses. For this we use the same cuts as in \citet{Farrow2015} and \citet{Linke2020b}. An overview of these cuts is given in Table~\ref{tab:sub-samples}.

\begin{table}
	\caption{Selection criteria for lens samples}
	\label{tab:sub-samples}
	\centering
	\begin{tabular}{cc}
		\hline\hline
		Sample & {Selection Criterion} \\
		\hline
		m1 & $8.5<\log_{10}(M_*/(\Msun\, h^{-2}))\le9.5$ \\
		m2 & $9.5<\log_{10}(M_*/(\Msun\, h^{-2}))\le10$ \\
		m3 & $10<\log_{10}(M_*/(\Msun\, h^{-2}))\le10.5$ \\
		m4 & $10.5<\log_{10}(M_*/(\Msun\, h^{-2}))\le11$ \\
		m5 & $11<\log_{10}(M_*/(\Msun\, h^{-2}))\le11.5$\\
		\hline
		red & $(g-r)_0 + 0.03\, (M_r - 5 \log_{10}h+ 20.6) > 0.6135$  \\
		blue & $(g-r)_0 + 0.03\, (M_r - 5 \log_{10}h+ 20.6) \le 0.6135$ \\
		\hline
	\end{tabular}
	\tablefoot{Lenses are selected either according to their stellar mass $M_*$ or to their rest-frame $(g-r)_0$ colour and absolute $r$-band magnitude $M_r$ and need to have $r<19.8\,\mathrm{mag}$.}
\end{table}

\begin{figure*}
    \centering
    \includegraphics[width=0.49\linewidth, trim={0.5cm, 0.8cm, 0.5cm, 0.5cm}, clip]{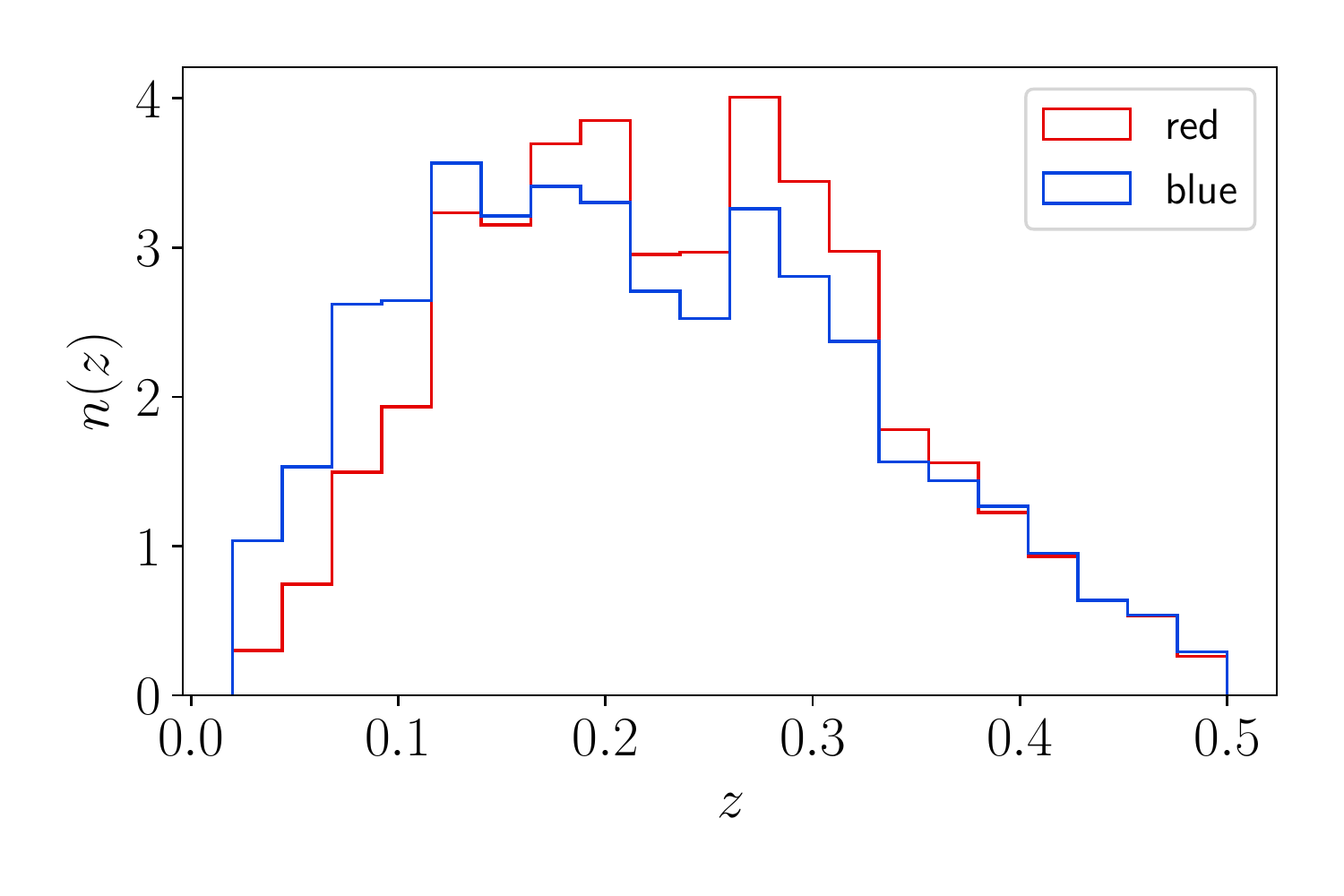}
    \includegraphics[width=0.49\linewidth, trim={0.5cm, 0.8cm, 0.5cm, 0.5cm}, clip]{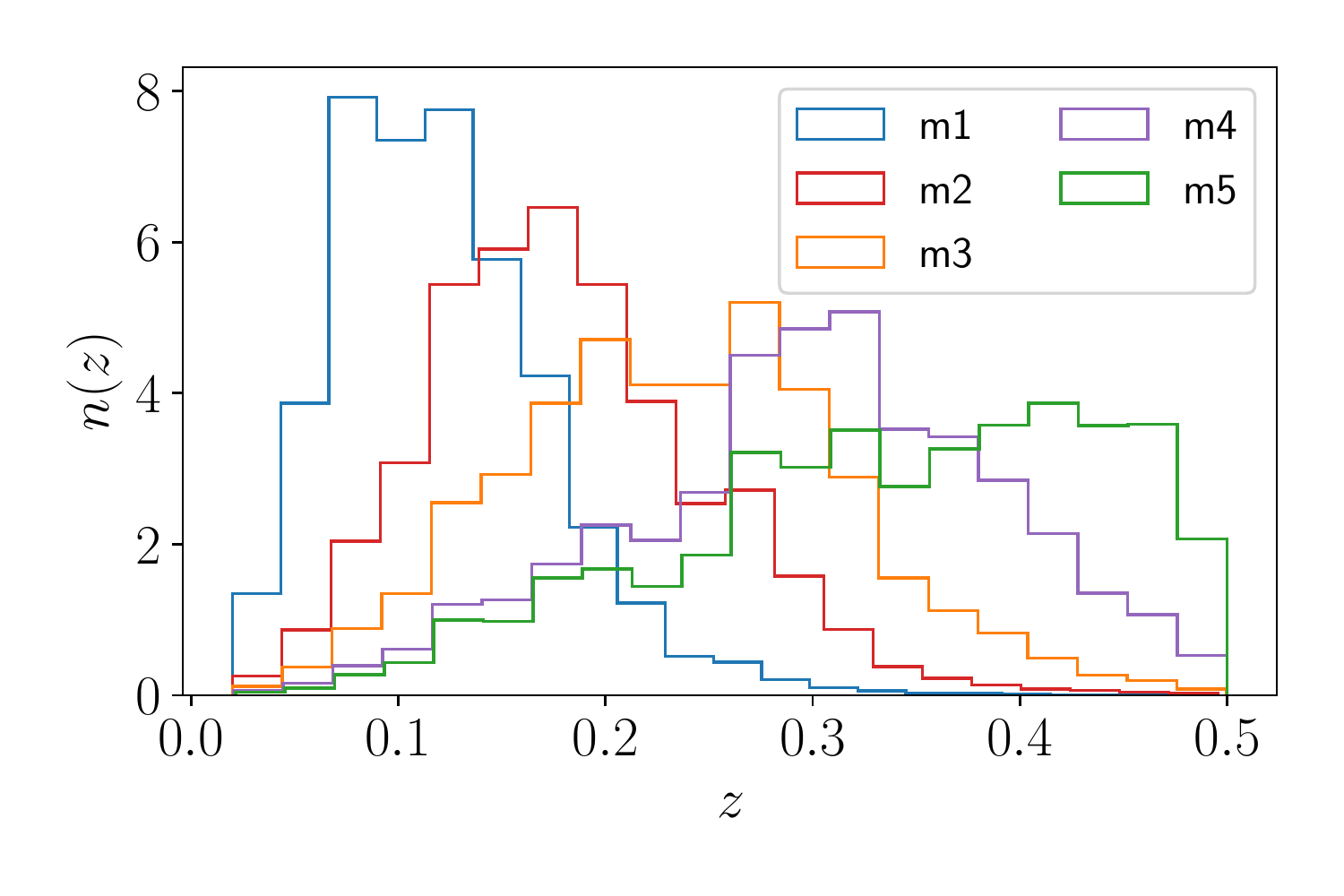}
    \caption{Normalized redshift distributions $n(z)$ of GAMA galaxies, selected by colour (left) and stellar mass (right).}
    \label{fig:nz}
\end{figure*}

These different samples have differing redshift distributions $n(z)$, shown in Fig.~\ref{fig:nz}. While the red and blue galaxies are distributed similarly, the $n(z)$ of the stellar-mass selected lenses differ strongly. Galaxies with lower stellar mass are predominantly observed at smaller redshifts; galaxies with higher stellar mass are found up to the limiting redshift $0.5$. These differences are caused by the flux limit of the survey; galaxies with lower stellar masses, typically fainter than more massive galaxies, are only visible at smaller redshifts. 

\begin{figure*}
\centering
\includegraphics[width=0.49\linewidth, trim={0.5cm, 0.8cm, 0.5cm, 0.5cm}, clip]{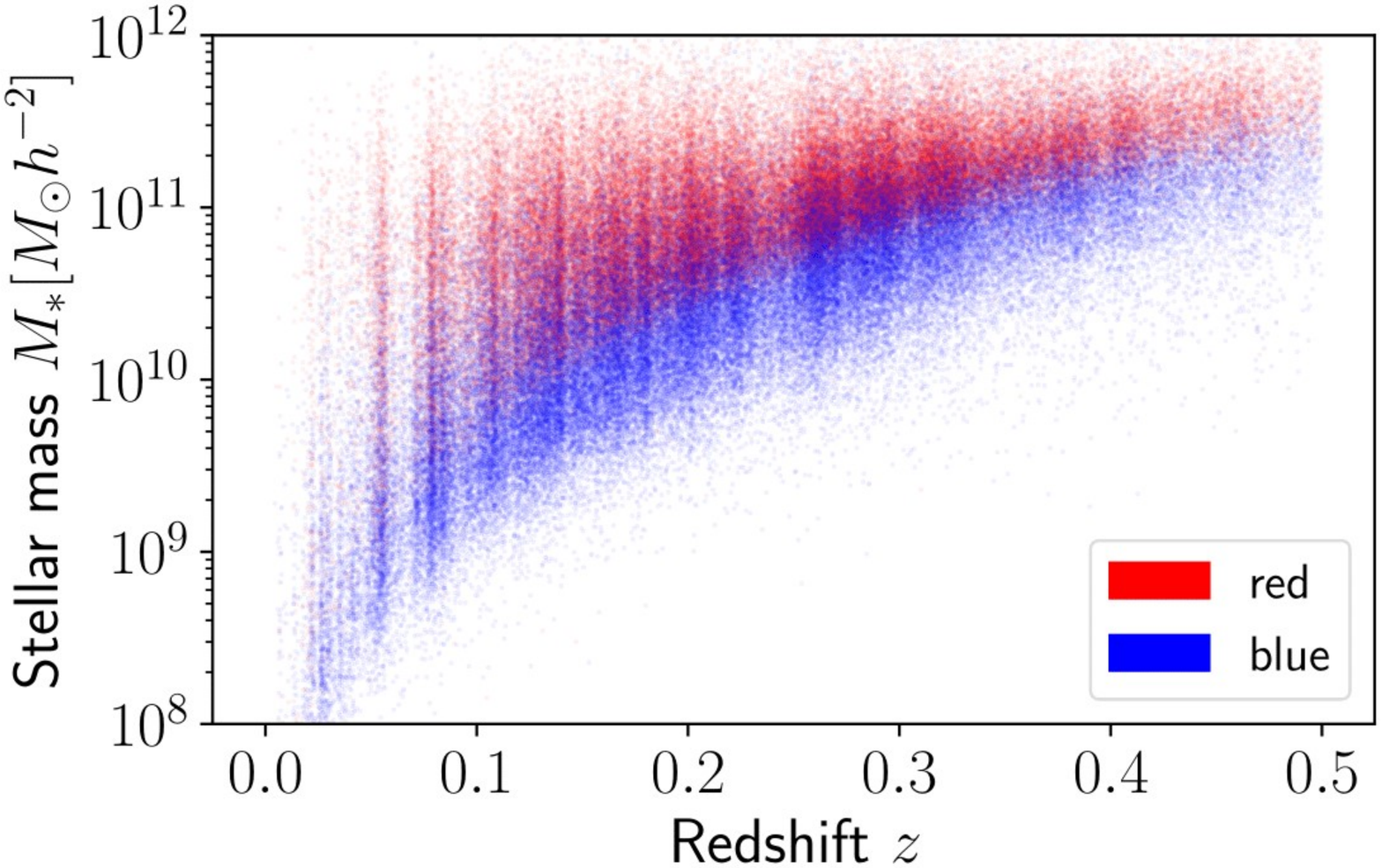}
\includegraphics[width=0.49\linewidth, trim={0.5cm, 0.8cm, 0.5cm, 0.5cm}, clip]{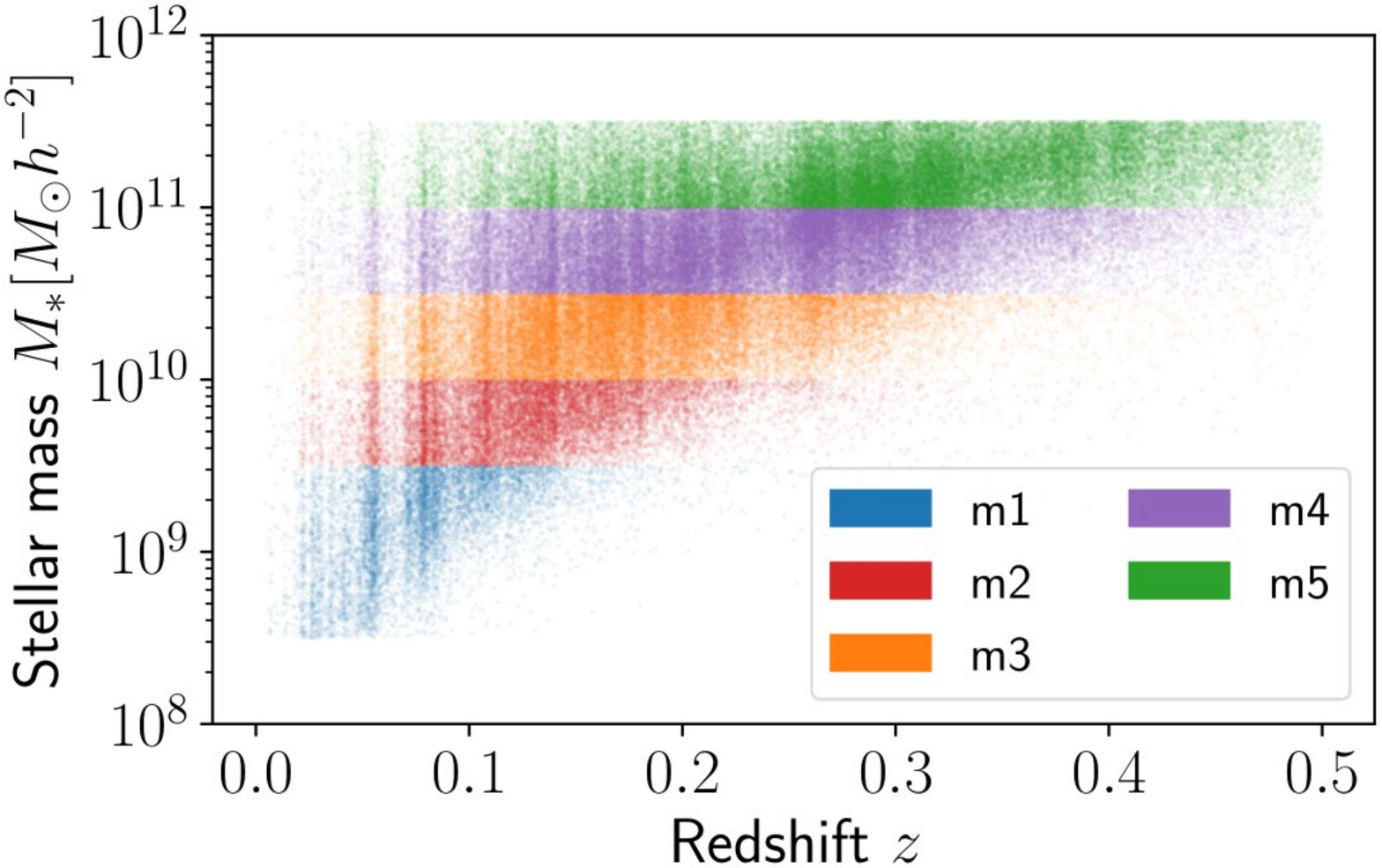}
\caption{Stellar-mass and redshift of GAMA galaxies, divided by colour (left) and stellar mass (right).}
\label{fig:bins}
\end{figure*}

The differences in the redshift distribution of galaxies from different samples are also visible in Fig.~\ref{fig:bins}, which shows the stellar mass $M_*$ and redshift $z$ of the GAMA galaxies. Observed galaxies at higher redshift have higher stellar masses on average, which is a direct consequence of the flux limit. Red and blue galaxies are distributed similarly with redshift, but differ in stellar mass: red galaxies tend to have higher stellar masses than blue galaxies.

\subsection{Millennium Simulation}
\label{sec:data:ms}
The simulation data are constructed from the MS \citep{Springel2005}, a dark-matter only cosmological $N$-body simulation. Its simulation box has a comoving side length of $500\, h^{-1}\si{\mega \parsec}$ and contains $2160^3$ particles with mass $8.6\times 10^8\, h^{-1}\, \Msun$ ($h=0.73$).

Maps of the lensing shear, $\gamma$, were created with the multiple-lens-plane raytracing algorithm by \citet{Hilbert2009}.  This algorithm generated shear maps of size $4\times 4\, \si{\deg \squared}$ on a regular grid of $4096\times4096\, \si{\deg \squared}$ for $64$ lines-of-sights (total area of $1024\,\rm deg^2$) at each redshift slice of the MS. We combine the shear redshift slices in a weighted average according to the observed KV450 galaxy redshift distribution, resulting in a combined shear without shear noise and intrinsic source alignment. The effective source density for the science verification corresponds to the pixel density, i.e., $291\, \mathrm{arcmin}^{-2}$.

Since all $64$ lines-of-sight originate from the same simulation, they are not independent at the largest scales. However, the lines-of-sight correspond to different observers placed in the simulation box such that the overlaps between the lines-of-sights are minimal, so we expect correlations predominantly at large scales, close to the simulation box size of $500\, h^{-1}\si{\mega \parsec}$. In contrast, the G3L signal is dominated by the 1-halo term, originating from small scales of approximately $1\, h^{-1}\si{\mega \parsec}$. Therefore, we treat the lines-of-sight as independent for our purposes. 

The lens galaxies for the simulated lenses \citet{Henriques2015} use the same initial mass function \citep{Chabrier2003} as assumed for the stellar mass estimates from GAMA, but a different stellar population model \citep{Maraston2005}. Nevertheless, as shown in \citet{Linke2020b}, the predictions of the MS combined with this SAM for the G3L signal agree with observations in KV450$\times$GAMA. 

To mimic the GAMA selection function, we use all lens galaxies with redshift less than $0.5$ and brighter than $r=19.8\, \mathrm{mag}$. We also divide the simulated lens galaxies by their colour and stellar masses by the same cuts as for the observed lens galaxies (Table~\ref{tab:sub-samples}).

\section{Methods}
\label{sec:method}
This section outlines our G3L estimators and the model fit to the data. The measurement procedure largely follows \citet{Simon2008} and \citet{Linke2020a} and is summarised in Sect.~\ref{sec:method:measurement}. The statistical analysis for the inference of HOD parameters from the G3L data is detailed in Sect.~\ref{sec:method:fitting}. For an assessment of the accuracy of the inferred HOD in a science verification with mock data, later on, we describe in Sect~\ref{sec:method:verification} the determination of the true HODs of the simulated galaxies. We make our codes publicly available at \texttt{github.com} \footnote{\url{https://github.com/llinke1/g3lhalo}}.

\subsection{Measuring G3L}
\label{sec:method:measurement}
We estimate the G3L aperture statistics from a catalogue of galaxy positions and source shapes for bins $B$ of similar triangles $(\vartheta_1,\vartheta_2,\phi)$ with the estimator $\Gtilde^{ab}_\mathrm{est}$ in \citet{Simon2008}. For $N^a$ lens galaxies from sample $a$, $N^b$ lens galaxies from sample $b$, and $N_\mathrm{s}$ source galaxies with complex ellipticities $\epsilon_k$, the estimate of $\Gtilde^{ab}(B)$ is the real part of
\begin{align}
	\label{eq:GtildeEst}
	&\notag \Gtilde_{\textrm{est}}^{ab}(B) \\
	&= \frac{\sum\limits_{i=1}^{N^a}\sum\limits_{j=1}^{N^b}\sum\limits_{k=1}^{N_\textrm{s}}(-1)\,w_k\, \epsilon_k\, \textrm{e}^{-\textrm{i}(\varphi_{ik} + \varphi_{jk})}\left[1 + \omega^{ab}(|\vec{\theta}_i - \vec{\theta}_j|)\right]\, \Delta_{ijk}(B)}{\sum\limits_{i,j=1}^{N_\textrm{d}}\sum\limits_{k=1}^{N_\textrm{s}}\, w_k\, \Delta_{ijk}(B)}\;,
\end{align}
where $\omega^{ab}(\theta)$ is the angular two-point correlation function of lens galaxies from samples $a$ and $b$ with lag $\theta$, and
\begin{equation}
	\Delta_{ijk}(B) = \begin{cases}
		1 &\textrm{for } (|\vec{\theta}_k - \vec{\theta}_i|, |\vec{\theta}_k - \vec{\theta}_j|, \phi_{ijk})\in B\\
		0 &\textrm{otherwise}
	\end{cases}\,.
\end{equation}
The angles $\varphi_{ik}$ and $\varphi_{jk}$ are the polar angles of the lens-source separation vectors $\vec{\theta}_i-\vec{\theta}_k$ and $\vec{\theta}_j-\vec{\theta}_k$ (corresponding to $\varphi_1$ and $\varphi_2$ in Fig.~\ref{fig:G3L}) and $\phi_{ijk}=\varphi_{ik}-\varphi_{jk}$ is the opening angle between the lens-source separation vectors (corresponding to $\phi$ in
Fig.~\ref{fig:G3L}). The weight $w_k$ of the source $k$ (set to $w_k\equiv1$ for our simulated data) measures the confidence in the shape measurements for the source, with higher weights indicating more precise ellipticities. The imaginary part of $\Gtilde_{\textrm{est}}^{ab}(B)$ is pure noise in the absence of systematic errors in the shear data \citep{Linke2020b}.

We give in the estimator equal weight to all lens pairs, which is in contrast to the recent analysis in \cite{Linke2020b}, where we weighted galaxy pairs based on the redshift difference between their members. The lack of weighting dilutes the signal-to-noise ratio of the G3L signal since non-physical pairs (widely separated along the line-of-sight) carry no signal but increase the noise. However, a model with weights requires abandoning the Limber approximation for the projection of the bispectrum in Eq.\eqref{limber}. The weighting introduces an additional factor into the integral, which depends on the line-of-sight distance between the lens galaxies in a pair. This factor, by definition, varies strongly on scales corresponding to the correlation length between galaxies. It is designed to give high weights to galaxy pairs within a correlation length while down-weighting galaxies outside this range. Consequently, the assumption that the integrand in Eq.\eqref{limber} is only slowly varying is no longer valid. With an alternative formalism for Eq.\eqref{limber} unclear at this point, we apply equal weights to our lens pairs for the scope of this work.

Our estimator for the two-point correlation function of the lens clustering, $\omega^{ab}(\theta)$, is that by \citet{Szapudi1998},
\begin{equation}
	\label{eq:szapudi&szalay}
	\omega^{ab}(\theta) = \frac{N_{\textrm{r}}^{a}\, N_{\textrm{r}}^{b}}{N_{\textrm{d}}^{a}\,N_{\textrm{d}}^{b}}\frac{{D_aD_b}(\theta)}{{R_aR_b}(\theta)} - \frac{N_{\textrm{r}}^{a}}{N_{\textrm{d}}^{a}}\frac{{D_aR_b}(\theta)}{{R_aR_b}(\theta)} - \frac{N_{\textrm{r}}^{b}}{N_{\textrm{d}}^{b}}\frac{{D_bR_a}(\theta)}{{R_aR_b}(\theta)} +1\;,
\end{equation}
for two lens samples $a$ and $b$ with $N_{\textrm{d}}^{a}$ and $N_{\textrm{d}}^{b}$ galaxies, and two `random samples'. These random samples contain $N_{\textrm{r}}^{a}$ and $N_{\textrm{r}}^{b}$ unclustered galaxies subject to the same selection functions as the lens samples. The ${D_aD_b}$, ${D_aR_b}$, ${D_bR_a}$,  and ${R_aR_b}$ are the pair counts of observed (D) and random galaxies (R).

Having obtained $\omega^{ab}$, we measure (on an approximate flat sky) $\Gtilde^{ab}$ with (\ref{eq:GtildeEst}) individually for tiles of size $\ang{1}\times\ang{1}$. For this, we divide the observational data into $N=189$ and the simulation data into $N=4\times 64 = 256$ tiles. The tiles allow us to project the galaxy positions and shear to Cartesian coordinates via an orthographic transformation and to quantify the uncertainty of the $\Gtilde^{ab}_{\rm est}$ with jackknife resampling on a tile by tile basis. We estimate $\Gtilde_i^{ab}$ for each tile $i$ on a regular grid of $128\times128\times128$ bins.  These bins are spaced logarithmically for $\vartheta_1$ and $\vartheta_2$, and linearly for $\phi$. We use $\vartheta_1, \vartheta_2 \in [\ang[astroang]{;0.15;}, \ang[astroang]{;85;}]$ and $\phi \in [0, 2\pi]$. For the total $\Gtilde_\mathrm{est}^{ab}$, individual tile estimates $\Gtilde_i^{ab}$ are averaged, weighted by the effective number of triplets per bin. 

Due to the finite number of galaxies, some of the bins will remain `empty', meaning that the data contains no lens-lens-source triplet fitting the configuration of the bin. Setting the correlation function in these bins to an arbitrary value biases the measurement, so we use the adaptive binning from \citet{Linke2020a}. This scheme uses Voronoi tesselation to redefine the bins, now $b_i$, such that no empty bins occur, effectively merging empty and `filled' bins. We found in \citet{Linke2020a} that estimating $\Gtilde$ in $128\times128\times128$ bins and then applying the tesselation leads to a measurement accuracy within $5\%$.

We convert a binned $\Gtilde_\mathrm{est}^{ab}$ to aperture statistics with a numerical approximation to Eq.~\eqref{eq:NNMapFromGtilde},
\begin{align}
	\NNM{a}{b}(\theta) &= \sum_{i=1}^{N_\mathrm{bin}} V(b_i)\, \Gtilde_\mathrm{est}^{ab}(b_i)\,\mathcal{A}_{\mathcal{NNM}}(b_i\,|\, \theta, \theta, \theta)\;,
\end{align}
where $N_\mathrm{bin}$ is the number of bins, $V(b_i)$ is the tesselation volume of bin $b_i$, and $\mathcal{A}_{\mathcal{NNM}}(b_i\,|\, \theta_1, \theta_2, \theta_3)$ is the kernel function of Eq.~\eqref{eq:NNMapFromGtilde}, evaluated at the center of bin $b_i$.

We estimate the covariance matrix of the estimates with jackknife resampling. For this, we combine the $\Gtilde^{ab}$ estimates for all but the $k$-th tile to the $k$-th jackknife sample $\Gtilde_{k, \mathrm{jn}}^{ab}$, which is then converted to the aperture statistics, Eq.~\eqref{eq:NNMapFromGtilde}, leading to $N$ samples $\NNM{a}{b}_k$. The estimate $\hat{C}$ of the $\NNM{a}{b}$ covariance matrix is then
\begin{align}
	\label{eq:covariancematrix}
	\hat{C}^{ab}_{ij}&= \frac{N}{N-1}\, \sum_{k=1}^{N}\left[\NNM{a}{b}_k(\theta_i)-\overline{\NNM{a}{b}_k}(\theta_i)\right]\\
	&\notag\quad \times \left[\NNM{a}{b}_k(\theta_j)-\overline{\NNM{a}{b}_k}(\theta_j)\right]\;,
\end{align}
where $\overline{\NNM{a}{b}_k}(\theta_i)$ is the average of all aperture statistics jackknife samples. The (mean-square) statistical uncertainty of the aperture statistics $\NNM{a}{b}(\theta_i)$ is \mbox{$\sigma_i=\sqrt{\hat{C}^{ab}_{ii}}$}.
The inverse of this covariance, needed for the likelihood analysis, is estimated with 
\begin{equation}
	\label{eq:inverseCovariancematrix}
	\left(C^{ab}\right)^{-1}_{ij} = \frac{N-N_\theta-2}{N-1} \left(\hat{C}^{ab}_{ij}\right)^{-1}\;,
\end{equation}
where $N_\theta$ is the number of aperture radii bins \citep{Hartlap2007, Anderson2003}. In our case $N_\theta = 3 \times 30 = 90$.

Formally, jackknife resampling assumes that all individual estimates of $\Gtilde$ are statistically independent. However, as all observed tiles originate from the same observation and are adjacent to each other, this assumption is not exactly valid. It is still probably a good approximation on scales smaller than the tile sizes. A possible bias in the empirical covariance is less pronounced for the simulation, as the tiles originate from $64$ (mostly) independent line-of-sights. 

\subsection{Fitting the halo model}
\label{sec:method:fitting}
We constrain the parameters of the G3L halo model by fitting it to measurements of both the auto-correlation aperture statistics, $\NNM{a}{a}(\theta)$ and $\NNM{b}{b}(\theta)$, and cross-correlation statistics, $\NNM{a}{b}(\theta)$, of two galaxy samples $a$ and $b$ for $N_\theta=30$ scale radii $\theta$ between \ang[astroang]{;0.1;} and \ang[astroang]{;30;}. To this end, we combine the measurements into a data vector of $3N_\theta$ elements,
\begin{equation}
\label{eq: order datavector}
	\vec{d}^{ab}=\begin{pmatrix}
		\NNM{a}{b}(\theta_1)\\ \vdots \\
		\NNM{a}{b}(\theta_{30})\\
		\NNM{a}{a}(\theta_1)\\ \vdots \\
		\NNM{a}{a}(\theta_{30})\\
		\NNM{b}{b}(\theta_1)\\ \vdots \\
		\NNM{b}{b}(\theta_{30})
	\end{pmatrix}\;.
\end{equation}
Likewise, we define the halo model vector $\vec{m}^{ab}(\vec{p})$ for each parameter set $\vec{p}$, which is obtained from Eqs. \eqref{eq:bispectrum halo model} and the bispectrum in Sect. \ref{sect:bispectrum},
\begin{equation}
	\vec{m}^{ab}(\vec{p})=\begin{pmatrix}
		\NNM{a}{b}(\theta_1\,|\,\vec{p})\\ \vdots \\
		\NNM{a}{b}(\theta_{30}\,|\,\vec{p})\\
		\NNM{a}{a}(\theta_1\,|\,\vec{p})\\ \vdots \\
		\NNM{a}{a}(\theta_{30}\,|\,\vec{p})\\
		\NNM{b}{b}(\theta_1\,|\,\vec{p})\\ \vdots \\
		\NNM{b}{b}(\theta_{30}\,|\,\vec{p})
	\end{pmatrix}\;.
\end{equation}

Optimal parameters $\vec{p}_\mathrm{opt}$ are those that minimise the goodness-of-fit
\begin{equation}
	\label{eq:chiSq_fit}
	\chi^2(\vec{p})=\left[\vec{d}^{ab}-\vec{m}^{ab}(\vec{p})\right]^\mathrm{T}\, {{C}^{-1}}^{ab}\, \left[\vec{d}^{ab}-\vec{m}^{ab}(\vec{p})\right]\;,
\end{equation}
determined by the Nelder-Mead algorithm \citep{Nelder1965} as implemented in the GNU Scientific Library \citep{Gough2009}. This algorithm is well-suited to multi-dimensional minimisation problems and requires only few (typically 1 or 2) function evaluations per iteration step. To avoid local minima, the algorithm is restarted multiple times at different, randomly chosen, initial parameter values.

To estimate the uncertainties on the best-fitting parameters $\vec{p}_\mathrm{opt}$ and to quickly sample the posterior distribution of the parameters $P(\vec{p}\,|\,\vec{d}^{ab})$, we approximate the posterior distribution with the importance function $q(\vec{p}\,|\,\vec{d}^{ab})$ following the importance sampling scheme \citep[e.g.,][]{Liu2004}. According to the Bayes theorem, the posterior density of $\vec{p}$ given the data is
\begin{equation}
	P(\vec{p}\,|\,\vec{d}^{ab}) \propto \mathcal{L}(\vec{d}^{ab}\,|\,\vec{p})\, {P}_\mathrm{prior}(\vec{p})\;,
\end{equation}
where ${P}_\mathrm{prior}(\vec{p})$ is the prior density of the parameters, as given in Table \ref{tab: hod params}, and $\mathcal{L}(\vec{d}^{ab}|\vec{p})$ is the likelihood of the data given the parameters $\vec{p}$. We assume Gaussian statistical errors in the data, i.e., the likelihood function is 
\begin{align}
	\mathcal{L}(\vec{d}^{ab}\,|\,\vec{p})\propto \exp\left[-\frac{1}{2}\, \chi^2(\vec{p})\right]\;,
\end{align}
with the $\chi^2$ as defined in Eq.~\eqref{eq:chiSq_fit}; the Bayesian evidence and thus the normalisation of $\cal{L}$ is not of interest here and will be ignored in what follows. The importance sampling function $q$ should be close to the posterior $P$ to achieve an efficient sampling. To find an appropriate $q$, we approximate $\mathcal{L}(\vec{d}^{ab}\,|\,\vec{p})$ in the proximity of the optimal parameters $\vec{p}_\mathrm{opt}$ by the Gaussian probability density 
\begin{equation}
	\tilde{\mathcal{L}}(\vec{d}^{ab}\,|\,\vec{p})\propto \exp[-\frac{1}{2}\, \left(\vec{p}-\vec{p}_\mathrm{opt}\right)^\mathrm{T}\, {{F}}(\vec{p}_\mathrm{opt})\left(\vec{p}-\vec{p}_\mathrm{opt}\right)]\;,
\end{equation}
where the matrix ${F}$ is the Fisher information, 
\begin{equation}
	{F}_{ij}(\vec{p})=\left(\pdv{\vec{m}^{ab}(\vec{p})}{p_i}\right)^{\rm T}\, {C}^{-1}\, \pdv{\vec{m}^{ab}(\vec{p})}{p_j} 
\end{equation}
\citep[e.g.,][]{Tegmark1997}, and choose $q$ as
\begin{equation}
	\label{eq:choosen q}
	q(\vec{p}\,|\,\vec{d}^{ab}) \propto \tilde{\mathcal{L}}(\vec{d}^{ab}\,|\,\vec{p})\, {P}_\mathrm{prior}(\vec{p})\;.
\end{equation}

We now draw $N_\mathrm{p}$ parameter sets ${\vec{p}}_i$ from $q$ which are then, by importance sampling, weighted to sample $P$. The allocated weights are
\begin{multline}
  w(\vec{p}_i\,|\,\vec{d}^{ab}) \propto \frac{P(\vec{p}_i\,|\,\vec{d}^{ab})}{q(\vec{p}_i\,|\,\vec{d})}\propto \frac{\mathcal{L}(\vec{d}^{ab}\,|\,\vec{p}_i)}{\tilde{\mathcal{L}}(\vec{d}^{ab}\,|\,\vec{p}_i)}\\
\propto  \exp[-\frac{1}{2}\,\chi^2(\vec{p}) + \frac{1}{2}\, \left(\vec{p}-\vec{p}_\mathrm{opt}\right)^\mathrm{T}\, {F}(\vec{p}_\mathrm{opt})\left(\vec{p}-\vec{p}_\mathrm{opt}\right) ]\;,
\end{multline}
normalized such that their sum is unity.

For each parameter $p^i$ we give the $\alpha$ credibility interval (CI) on the marginalized posterior $P(p^i | \vec{d}^{ab})$, defined by
\begin{equation}
    P(p^i | \vec{d}^{ab}) = \int \dd[14]p'\; P(\vec{p'}\,|\,\vec{d}^{ab})\; \dirac(p'^i - p^i)\;.
\end{equation}
We take the mode of the posterior as optimal parameter value $p_\mathrm{opt}^i$ and define the CI $I_\alpha$ of a single parameter as the interval around the optimal parameter value including $\alpha$ of the marginalized posterior, i.e.,
\begin{equation}
    \alpha = \int_{I_\alpha} \dd{p^i} P(p^i | \vec{d}^{ab})\;.
\end{equation}
To find this interval, we sort the sampling points by the distance $|p^i-p^i_\mathrm{opt}|$ from the optimal parameter value and take the first $N_{\rm p}$ points, for which the sum of their weights equals $\alpha$. The interval defined by these points is our estimate for the $\alpha$ credibility region of the optimal parameter.

\subsection{Science verification} 
\label{sec:method:verification}
We assess the accuracy of the model, estimators, and our statistical setup by comparing the inferred HODs to the true ones in mock data. In these mock data, each galaxy is associated with a dark matter halo, identified by its halo ID and virial halo mass from a friends-of-friends halo finder. To extract the true HOD, we count the number of galaxies of each sample $a$ in each halo and divide the halos into $50$ mass bins in the range between $10^{11}$ and $10^{15}\,h^{-2}\, \Msun$. For each mass bin, the number of satellite and central galaxies per halo is averaged, yielding the true average $\expval{N^a\,|\,m}=\expval{\Nc{a}\,|\,m}+\expval{\Ns{a}\,|\,m}$ in the data. We also verify the inference estimate of the correlation coefficient $r^{ab}(m)$, Eq.~\eqref{eq:HOD_NsatNsat}, by computing
\begin{equation}
  \label{eq:r(m)}
  r^{ab}(m) = \frac{\expval{\Ns{a}\Ns{b}|m}-\expval{\Ns{a}|m}\expval{\Ns{b}|m}}{\sqrt{\expval{\Ns{a}|m}\expval{\Ns{b}|m}}}
\end{equation}
from the halo catalogue and galaxies in the mock data. The uncertainties on $\expval{N^a\,|\,m}$ and $r^{ab}(m)$ are the standard errors on the mean over the $64$ lines-of-sights of the simulation.

\section{Results}
\label{sec:results}
In this section, we give the results of the science verification and the G3L analysis for KV450 $\times$ GAMA. We first present the results for lens samples defined by their colour in Sect.~\ref{sec:results:colour} and then for lens samples defined by their stellar mass in Sect.~\ref{sec:results:mass}. 

\subsection{Colour-selected lens samples}
\label{sec:results:colour}

\begin{figure*}
	\centering
	\begin{subfigure}{\linewidth}
		\includegraphics[width=\linewidth, trim={5cm 0 5cm 2cm}, clip]{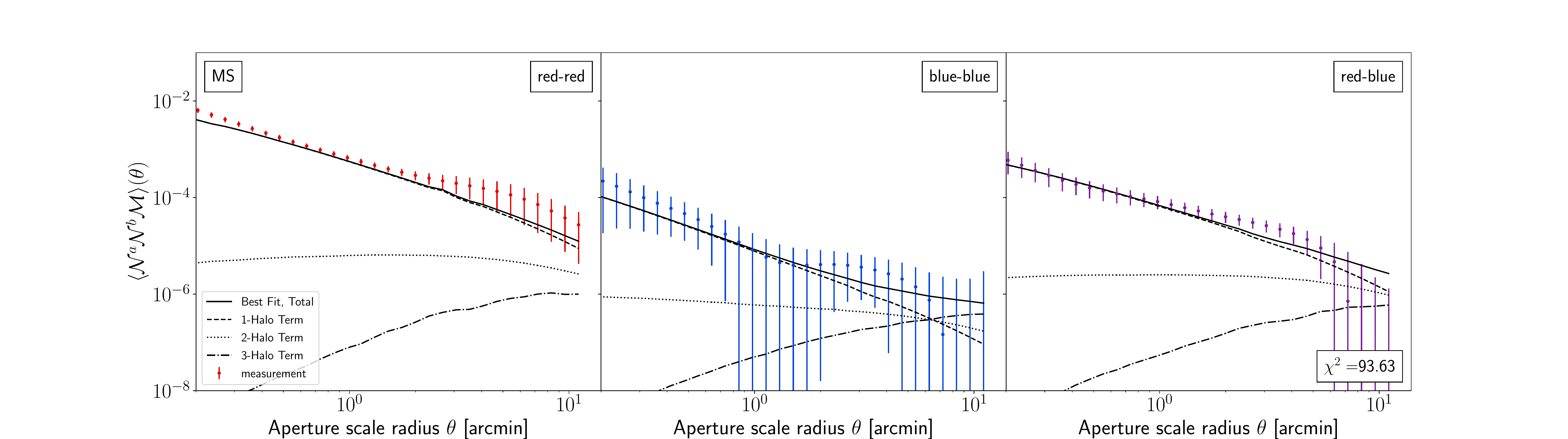}
		\caption{MS}
		\label{fig:FitResultsMR_colour}
	\end{subfigure}
	\begin{subfigure}{\linewidth}
		\includegraphics[width=\linewidth, trim={5cm 0 5cm 2cm}, clip]{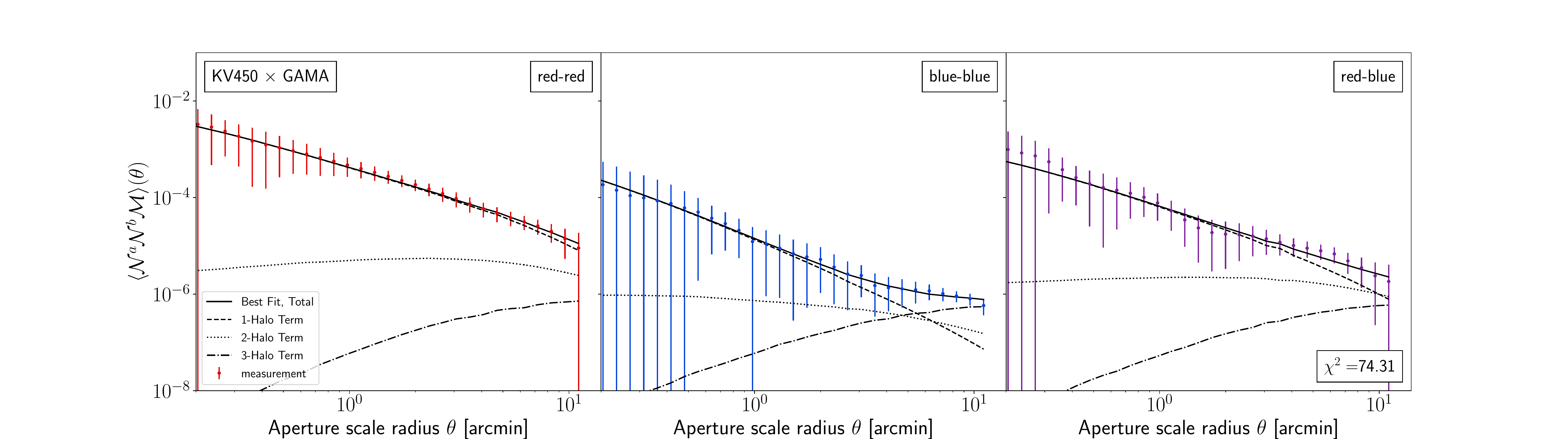}
		\caption{KV450 $\times$ GAMA}
		\label{fig:FitResultsKV450_colour}
	\end{subfigure}
	
	\caption{G3L measurement (points) and best-fitting halo model (lines) for red and blue galaxies in the MS (upper plot) and the KV450 $\times$ GAMA (lower plot). Solid lines indicate the total aperture statistics, dashed lines the 1-halo, dotted lines the 2-halo, and dash-dotted lines the 3-halo term of the fit. The left panels show the result for red-red galaxy pairs, the central panels for blue-blue galaxy pairs, and the right panels for red-blue mixed pairs.}
	\label{fig:FitResults_colour}
\end{figure*}

The G3L aperture statistics of simulated red and blue galaxies are shown in Fig. \ref{fig:FitResultsMR_colour}, along with the best-fit of the halo model and its decomposition into the 1-, 2-, and 3-halo terms. The parameter values corresponding to the best fit are given in the first two columns in Table \ref{tab: fitvalues_red_blue}. The goodness-of-fit is $\chi^2=93.63$ for $90-14=76$ degrees-of-freedom (dof), or $\chi^2/\mathrm{dof}=1.28$. This $\chi^2$ corresponds to a $p$-value of $0.083$, indicating no significant deviation between the fit and the simulated G3L signal within the $95\%$ confidence level (CL). 

\begin{table}[]
	\caption[Best fitting halo model parameters]{Best fitting values of halo model parameters for colour-selected lenses and $68\%$ credibility intervals (${\rm dof}=76$).}
	\label{tab: fitvalues_red_blue}
	\begin{tabular}{l|ll|ll}
		\hline
		& \multicolumn{2}{c|}{MS} & \multicolumn{2}{c}{KV450 $\times$ GAMA} \\
		{Parameter} 	& $a=\mathrm{red}$ 			& $b=\mathrm{blue}$			& $a=\mathrm{red}$ 		&$a=\mathrm{blue}$	 \\
		\hline
		$\alpha^a$ 		&  $0.47\pma{0.18}{0.26}$ 	& $0.10\pma{0.09}{0.10}$	& $0.34\pma{0.31}{0.29}$	& $0.13\pma{0.06}{0.42}$  \\[3pt]
		$\sigma^a$		&  $0.55\pma{0.47}{0.48}$	&  $0.47\pma{0.44}{0.54}$	& $0.52\pma{0.49}{0.46}$		& $0.47\pma{0.43}{0.53}$ \\[3pt]
		$M_\mathrm{th}^a[10^{11}\Msun]$ &  $23.0\pma{6.4}{5.0}$ 	& $1.19\pma{0.60}{0.64}$	& $15\pma{12}{67}$			& $1.4\pma{0.9}{7.2}$  \\[3pt]
		$\beta^a$ 		& $0.84\pma{0.09}{0.10}$ 	&  $0.73\pma{0.16}{0.10}$	& $0.88\pma{0.44}{0.32}$		& $0.55\pma{0.21}{0.16}$\\[3pt]
          $M^{\prime a} [10^{13}\Msun]$ 	& $5.8\pma{1.7}{3.4}$ 	& $32\pma{21}{20}$		& $3.6\pma{0.7}{3.5}$		& $20\pma{12}{30}$          \\[3pt]
		$f^a$ 			& $1.49\pma{0.53}{0.31}$ 	& $0.88\pma{0.45}{0.46}$	&	$1.27\pma{0.43}{0.46}$			& $0.83\pma{0.23}{0.56}$\\[3pt]
		$A^\text{red-blue} [10^{-2}]$ 			& \multicolumn{2}{c|}{$5.31\pma{0.92}{0.87}$}			&	\multicolumn{2}{c}{$1.62\pma{0.51}{0.62}$}		\\[3pt]
		$\epsilon^\text{red-blue}$ 			& \multicolumn{2}{c|}{$0.69\pma{0.04}{0.08}$}			&	\multicolumn{2}{c}{$0.99\pma{0.12}{0.20}$}		\\[3pt]
		\hline
		$\chi^2$ &\multicolumn{2}{c|}{93.62} & \multicolumn{2}{c}{74.31} \\
		\hline 
	\end{tabular}
\end{table}

\begin{figure}
	\centering
	\begin{subfigure}{\linewidth}
		\includegraphics[width=\linewidth, trim={0.5cm 0.5cm 0.5cm 0.5cm}, clip]{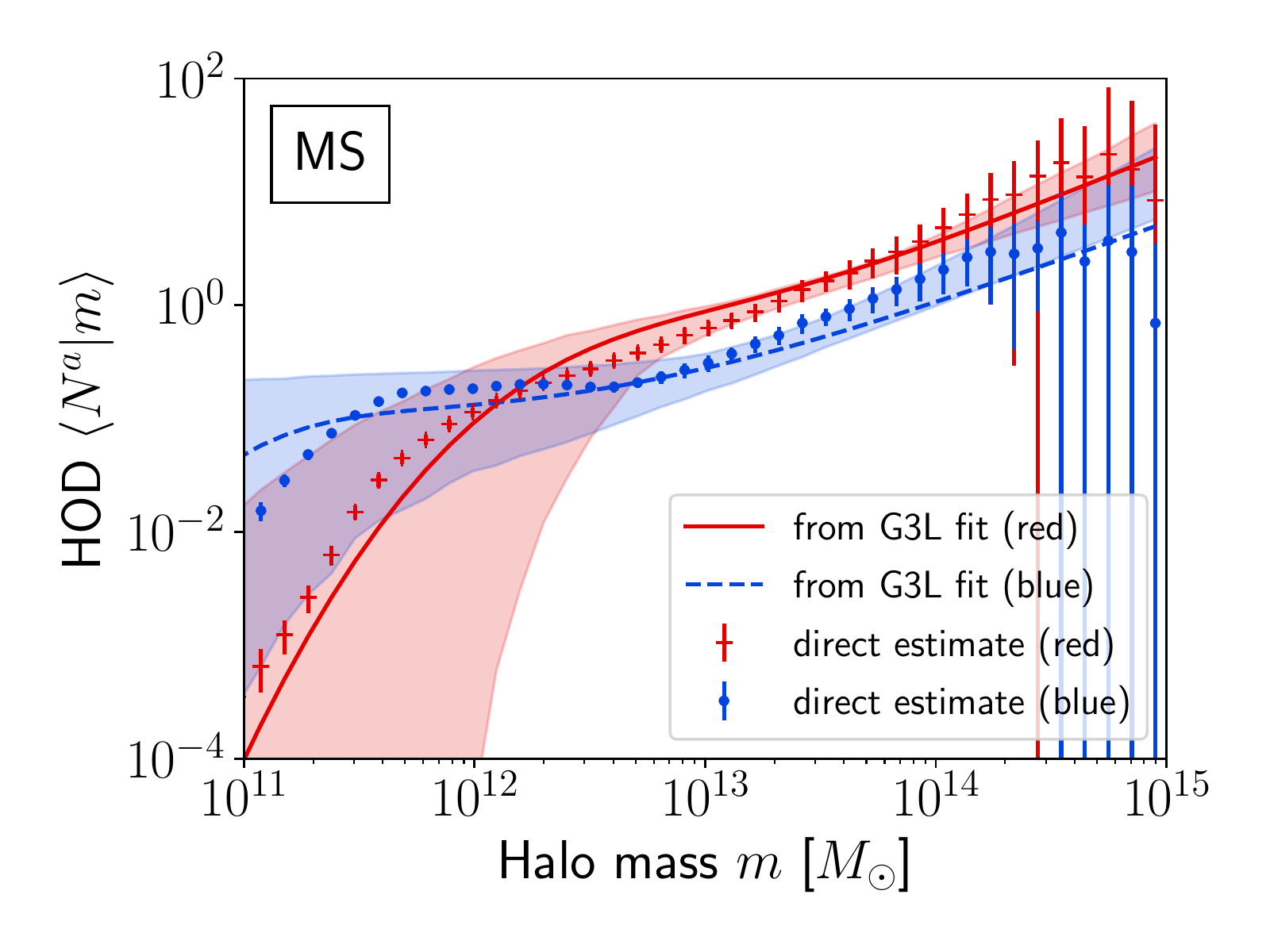}
		\caption{MS}
		\label{fig:HODs_MR_colour}
	\end{subfigure}
	\begin{subfigure}{\linewidth}
		\includegraphics[width=\linewidth, trim={0.5cm 0.5cm 0.5cm 0.5cm}, clip]{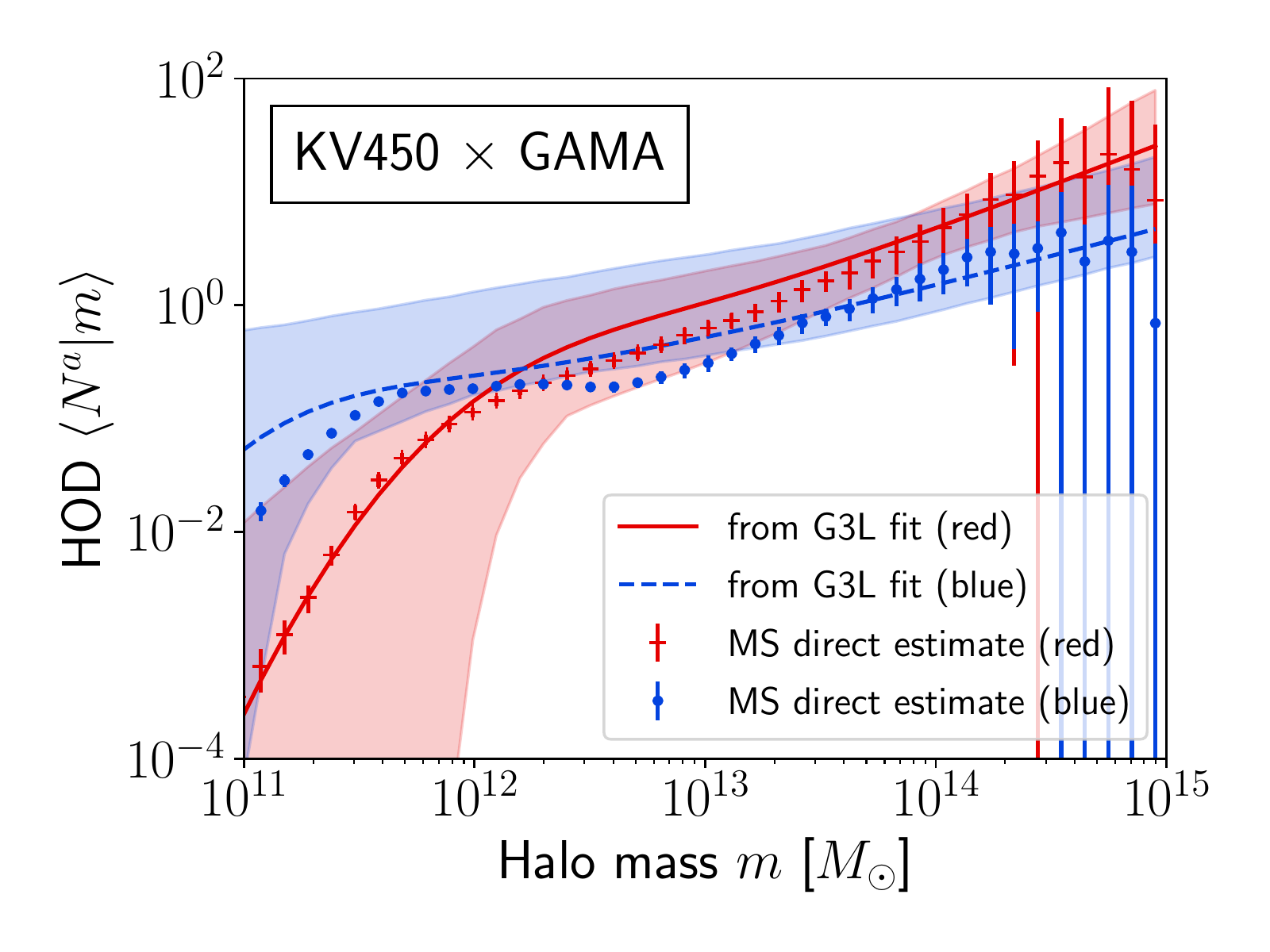}
		\caption{KV450 $\times$ GAMA}
		\label{fig:HODs_KV450_colour}
	\end{subfigure}
	\caption{Mean per-halo numbers of simulated galaxies (top) and observed galaxies (bottom) as function of halo mass. Red crosses (blue points) indicate the true HOD of simulated red (blue) galaxies, where the error bars are the standard deviation of the mean over the $64$ line-of-sights. The lines indicate the per-halo numbers inferred from the fit to the G3L signal for red (solid red) and blue galaxies (dashed blue). The shaded areas are the $68\%$ credibility areas of the halo model fit.}
	\label{fig:HODs_colour}
\end{figure}

\begin{figure}
	\includegraphics[width=\linewidth, trim={0.5cm 0.5cm 0.5cm 0.5cm}, clip]{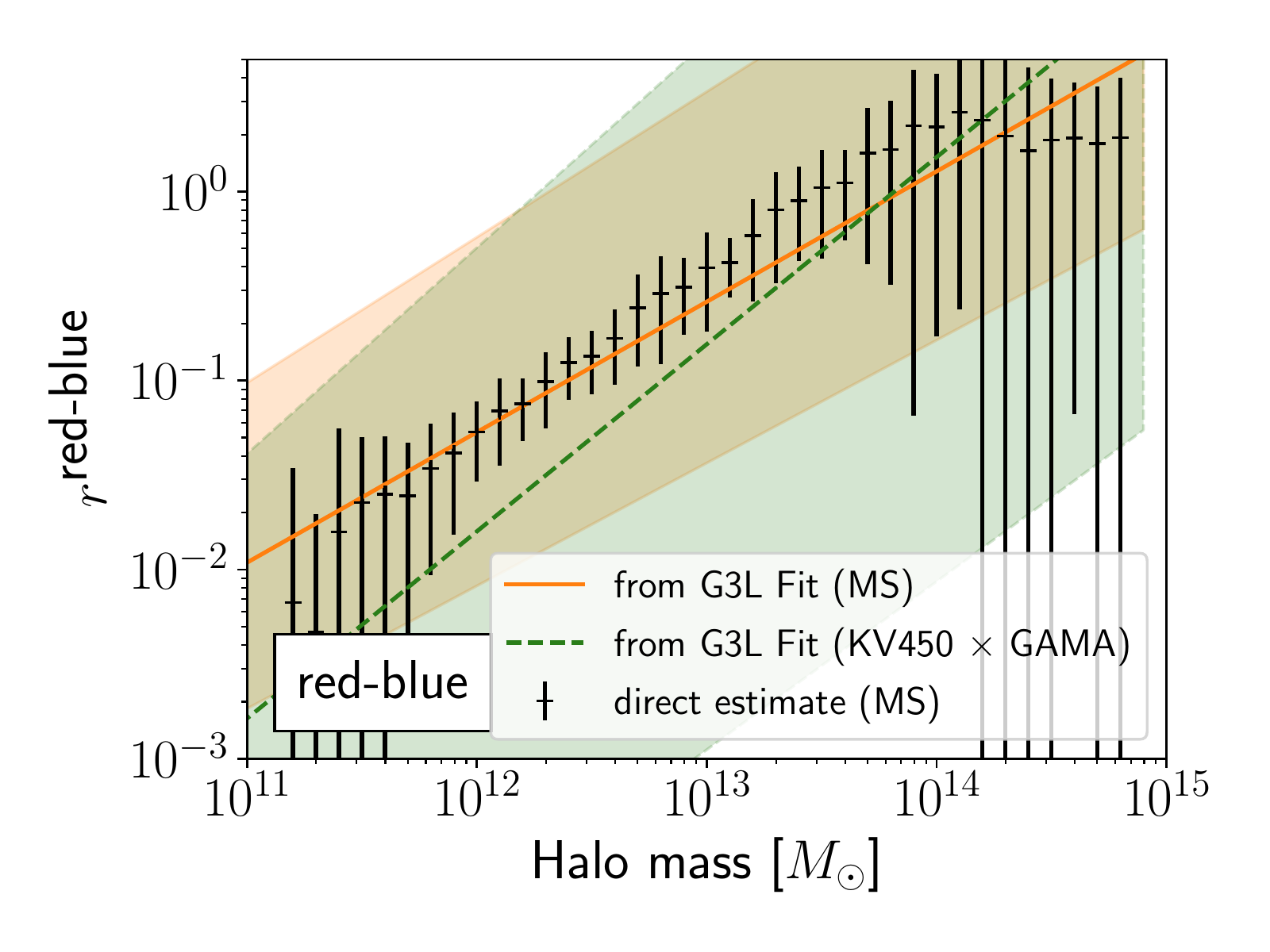}
	\caption{Correlation parameter $r^{ab}$ for red and blue galaxies in the simulation and observation as a function of halo mass. Black crosses show the direct estimate for the simulation, where the error bars are the standard deviation of the mean over the $64$ line-of-sights. The solid brown line shows the $r^{ab}$ inferred by the halo model fit to the G3L signal of the MS, and the green dashed line is the result of the fit to the KV450 $\times$ GAMA G3L signal. The shaded areas show the $68\%$ credibility bands of the fits.}
	\label{fig:r_redblue}
\end{figure}

For our science verification, we compare in Fig.~\ref{fig:HODs_MR_colour} the HODs inferred by the best-fitting G3L halo model to the directly estimated HODs of the simulated galaxies. Model prediction and direct estimate agree within the $68\%$ credibility band (shaded areas) for red and blue galaxies. Likewise, the correlation of numbers of red and blue satellites, $A^\text{red-blue}$ and $\epsilon^\text{red-blue}$, is detected at $3\sigma$ significance in the verification, and $r^{ab}(m)$ agrees in Fig.~\ref{fig:r_redblue} with the true $r^{ab}$ in the galaxy SAM within the $68\%$ credibility band. Therefore, the G3L fit recovers the galaxy HODs and $r^{ab}(m)$ sufficiently accurate within the statistical errors for a survey similar to our verification data ($\sim10^3\, \mathrm{deg}^2$, a high source number density without shape noise) and surveys with higher estimator noise, such as KV450 $\times$ GAMA.

We determine the HOD parameters of real red and blue galaxies in KV450$\times$GAMA in Fig. \ref{fig:FitResultsKV450_colour}, where we show the aperture statistics and the best-fits of our halo model, together with a decomposition into the 1-, 2-, and 3-halo terms. The model fit has  $\chi^2/{\rm dof}=0.977$, corresponding to a $p$-value of $0.53$ and an agreement with the model within the $95\%$ CL. The best-fitting parameters are reported in the second pair of columns in Table \ref{tab: fitvalues_red_blue}. They show that red galaxies clearly populate more massive halos than blue galaxies:  $M_\text{th}^\text{red}=1.5^{+6.7}_{-1.2}\times10^{12}\,M_\odot$ is roughly ten times larger than $M_\text{th}^\text{blue}=1.4^{+7.2}_{-0.9}\times10^{11}\,M_\odot$  ($68\%$ credibility intervals, CI herafter). The per-halo number of blue satellites increases slower with halo mass than for red galaxies: the mass scale $M^{\prime \mathrm{blue}}=2.0^{+3.0}_{-1.2}\times10^{14}\,M_\odot$ of blue satellites is more than five times larger than $M^{\prime \mathrm{red}}=3.6^{+3.5}_{-0.7}\times10^{13}\,M_\odot$ ($68\%$ CI). For central galaxies, the sum of $\alpha^\mathrm{red}$ and $\alpha^\mathrm{blue}$ is $0.47\pma{0.31}{0.51}$ ($68\%$ CI), which is consistent with unity ($68\%$ CI). The concentration of halo satellites is consistent with that of matter ($f^a\sim1$) in the $68\%$ CI. As expected from the qualitative analysis (Sect. \ref{sec:results:qual}) $\sigma$ cannot be constrained better than the prior.

The HODs corresponding to the best-fitting parameter values are shown in Fig.~\ref{fig:HODs_KV450_colour}. The HOD of red (blue) galaxies is non-zero for halo masses above $10^{12}\, \Msun$ ($5\times 10^{11}\, \Msun$), whereas, at lower halo masses, the constraints become essentially upper limits for $\langle N|m\rangle$. The inferred HODs for GAMA galaxies match those obtained from the fit to the simulated galaxies in Fig.~\ref{fig:HODs_KV450_colour}. This agreement reflects the similar G3L aperture statistics of mock data and observations -- the SAM predictions for $\NNM{a}{b}$ agree with the measurements in KV450 $\times$ GAMA within the errors (Fig.\ref{fig:FitResults_colour}). However, the uncertainties on the HODs and their parameters are considerably larger for the observation since our simulated data has less noise in the shear signal.

As for $r^{ab}$ of red and blue satellites in KV450$\times$GAMA, we report a $2\sigma$ to $3\sigma$ detection of a positive correlation and an amplitude increase towards more massive halos: both $A^\text{red-blue}$ and $\epsilon^\text{red-blue}$ are positive. This trend is similar to that of the simulated galaxies. However, the increase with halo mass is steeper for the observed galaxies (Fig.~\ref{fig:r_redblue}$, \epsilon^{ab}=0.99^{+0.22}_{-0.12}$ versus $\epsilon^{ab}=0.69^{+0.08}_{-0.04}$ at $68\%$ CI), while the correlation amplitude at $10^{12}\,\rm M_\odot$ is lower ($A^{ab}=1.62^{+0.62}_{-0.51}\times10^{-2}$ versus $A^{ab}=5.31^{+0.87}_{-0.92}\times10^{-2}$ at $68\%$ CI). Consequently, numbers of red and blue satellites are correlated both in the SAMs and for true galaxies, especially beyond the mass scale of galaxy groups $m\gtrsim10^{13}\,\rm M_\odot$ where $r^{\rm red-blue}\gtrsim0.16^{+0.06}_{-0.05}$ for $\epsilon^{\rm red-blue}=1$ ($68\%$ CI).

\begin{figure}
    \centering
    \includegraphics[width=\linewidth, trim={2.5cm 0.7cm 3cm 0cm}, clip]{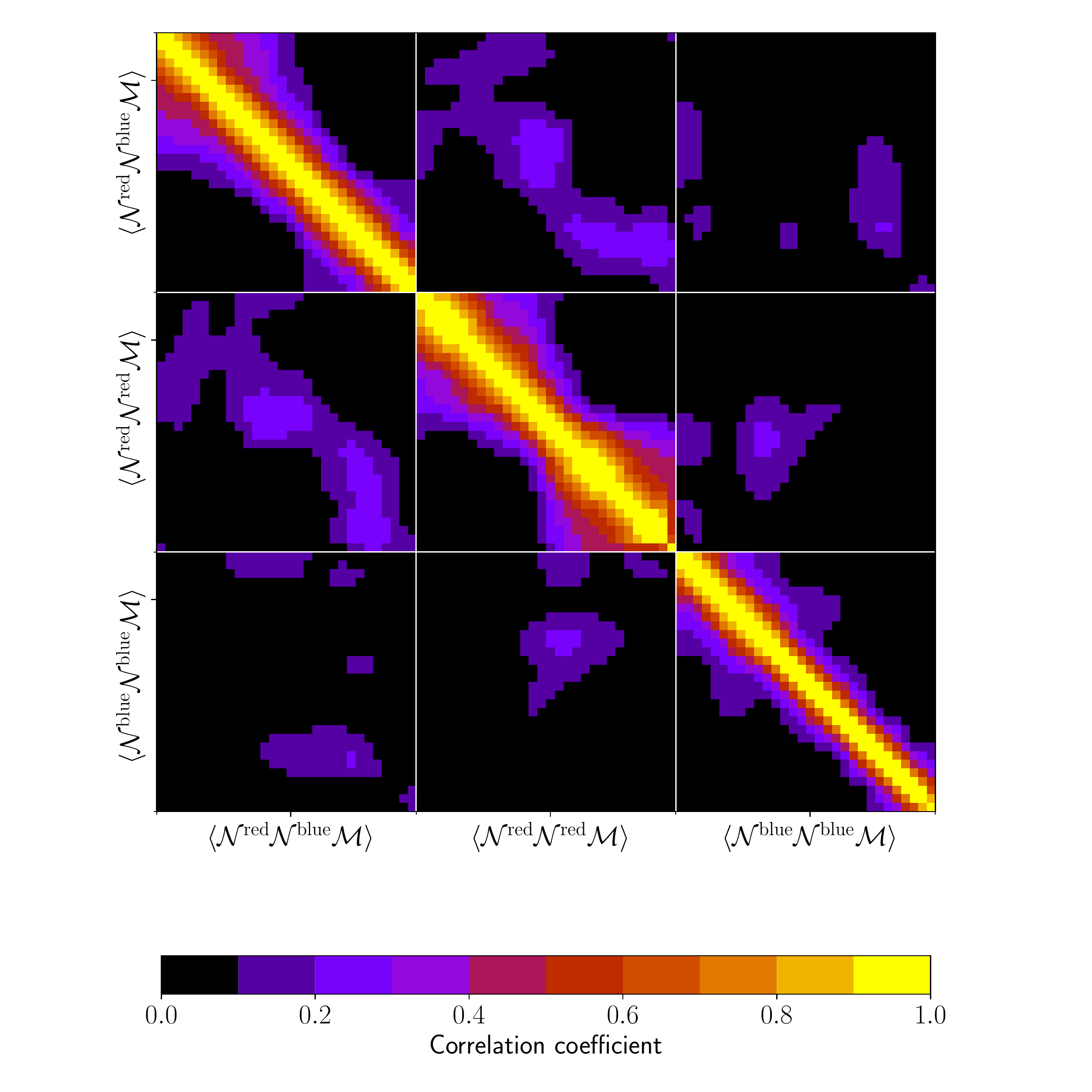}
    \caption{Correlation matrix for aperture statistics measurement in KV450$\times$GAMA for red and blue galaxy samples. The data vector is ordered as given in Eq.~\eqref{eq: order datavector}, starting with the smallest aperture radius.}
    \label{fig:corr_matrix}
\end{figure}

The correlation matrix of the G3L estimate for the KV450 $\times$ GAMA red and blue galaxies is shown in Fig.~\ref{fig:corr_matrix}.  We see that the signal for similar aperture radii is strongly correlated. The signals for red-red and blue-blue lens pairs are almost independent, while the signal for mixed galaxy pairs has correlation coefficients of up to $0.3$ with the signal for unmixed red-red pairs.  

\subsection{Stellar mass-selected lens samples}
\label{sec:results:mass}

We repeat the science verification and G3L analysis for the stellar mass-selected galaxies. We consider five stellar-mass bins m1 to m5, so there are ten distinct combinations of two samples $a$ and $b$. For each combination, we again use the statistics $\NNM{a}{a}$, $\NNM{a}{b}$, and $\NNM{b}{b}$ to infer HOD parameters. Since each galaxy sample is used in four combinations, we obtain four versions of the same HOD for each sample. For a realistic model and successful fits, these four HODs should be consistent with each other.

Fitting the halo model to two samples out of five individually neglects the correlations between the ten combinations. Consequently, better constraints on the model parameters could be obtained by fitting the model to the signal of all five stellar-mass selected samples simultaneously. However, this would increase the data vector from $90$ entries to $450$ entries. As the covariance estimated is obtained from only $180$ (quasi)-independent data realisations, it cannot be used with such a large data vector, and a simultaneous fit of all five samples is unfeasible.

\begin{figure}
    \centering
    \includegraphics[width=\linewidth]{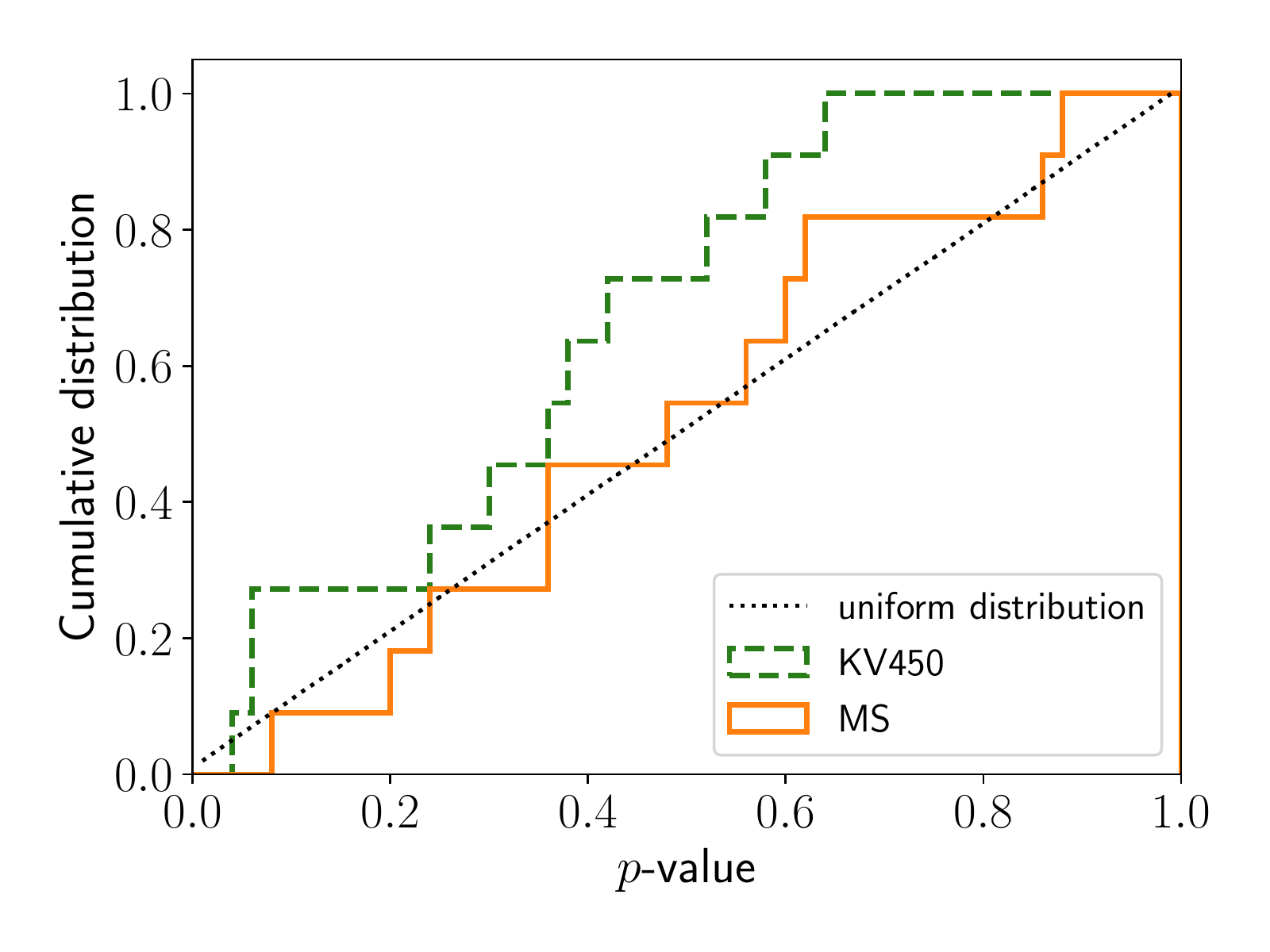}
    \caption{Cumulative distribution of $p$-values of G3L halo model fits for MS (orange, solid) and KV450 $\times$ GAMA (green, dashed). For a perfect description of G3L signal and data noise, the distributions would be consistent with a uniform distribution (black, dotted).}
    \label{fig:p_hist}
\end{figure}

The HOD parameters, $\chi^2$, and plots of the best-fitting models for the science verification with the simulated data are given in Appendix~\ref{app:results}. The $p$-values exceed $0.05$ for all fits, indicating no significant deviation between the fits and the measurements within the $95\%$ CL. To evaluate the overall agreement of all fits to the mock data, we consider the cumulative distribution of $p$-values, shown in Fig.~\ref{fig:p_hist} (solid line). This distribution should correspond to a uniform distribution between 0 and 1 (dotted line) if the measurements are unbiased realizations of the model. A Kolmogorov-Smirnov (KS) test on the distribution of $p$-values for the simulation yields a KS distance of $0.118$, which for the $11$ samples and ${\rm dof}=76$ in the distribution indicates no significant deviation from a uniform distribution at $95\%$ CL. Additionally, the four parameter sets for each stellar-mass bin agree within the $68\%$ CI. The distribution of $p$-values and the consistency of the HOD parameters for the four model fits supports the view that the model coherently and accurately reproduces the G3L signal of the verification data within the statistical errors.

\begin{figure*}
	\begin{subfigure}{\linewidth}
		\includegraphics[width=\linewidth, trim={1cm 0 3cm 1.5cm}, clip]{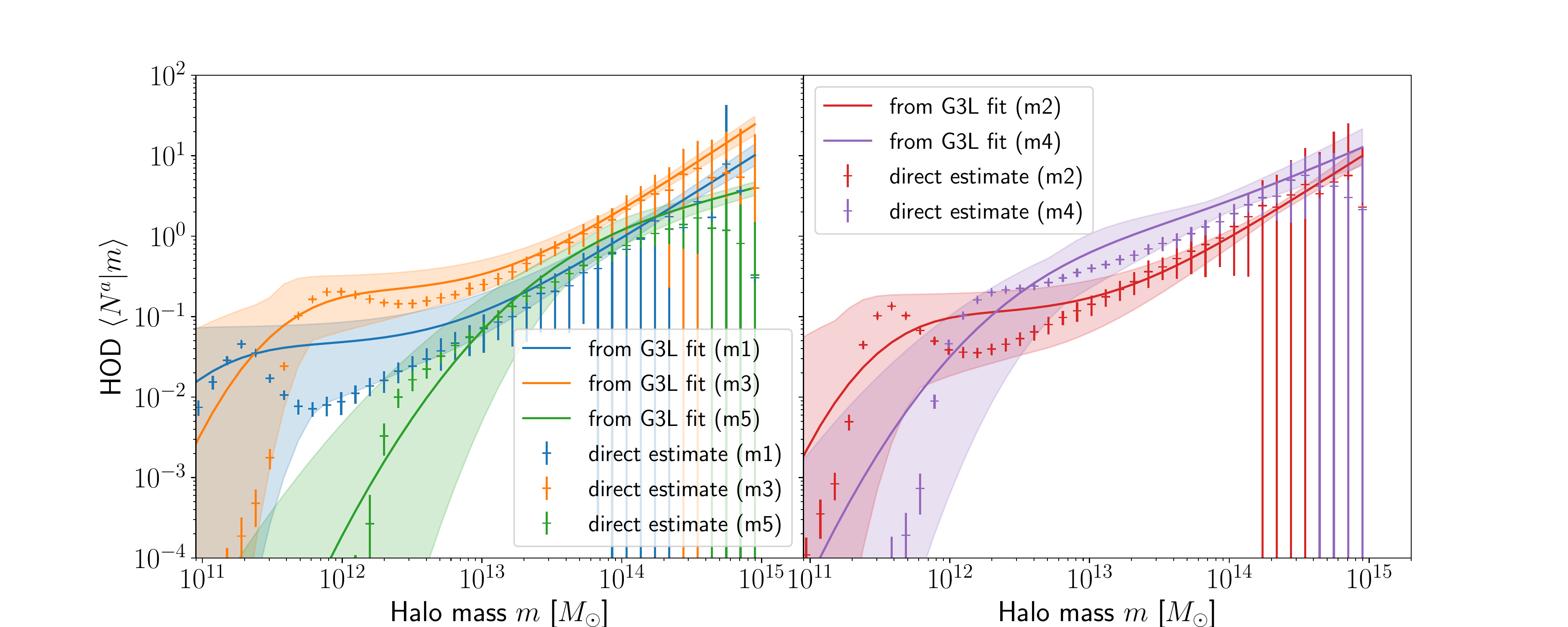}
		\caption{MS}
		\label{fig:HOD_sm_MS}
	\end{subfigure}
	\begin{subfigure}{\linewidth}
		\includegraphics[width=\linewidth, trim={1cm 0 3cm 1.5cm}, clip]{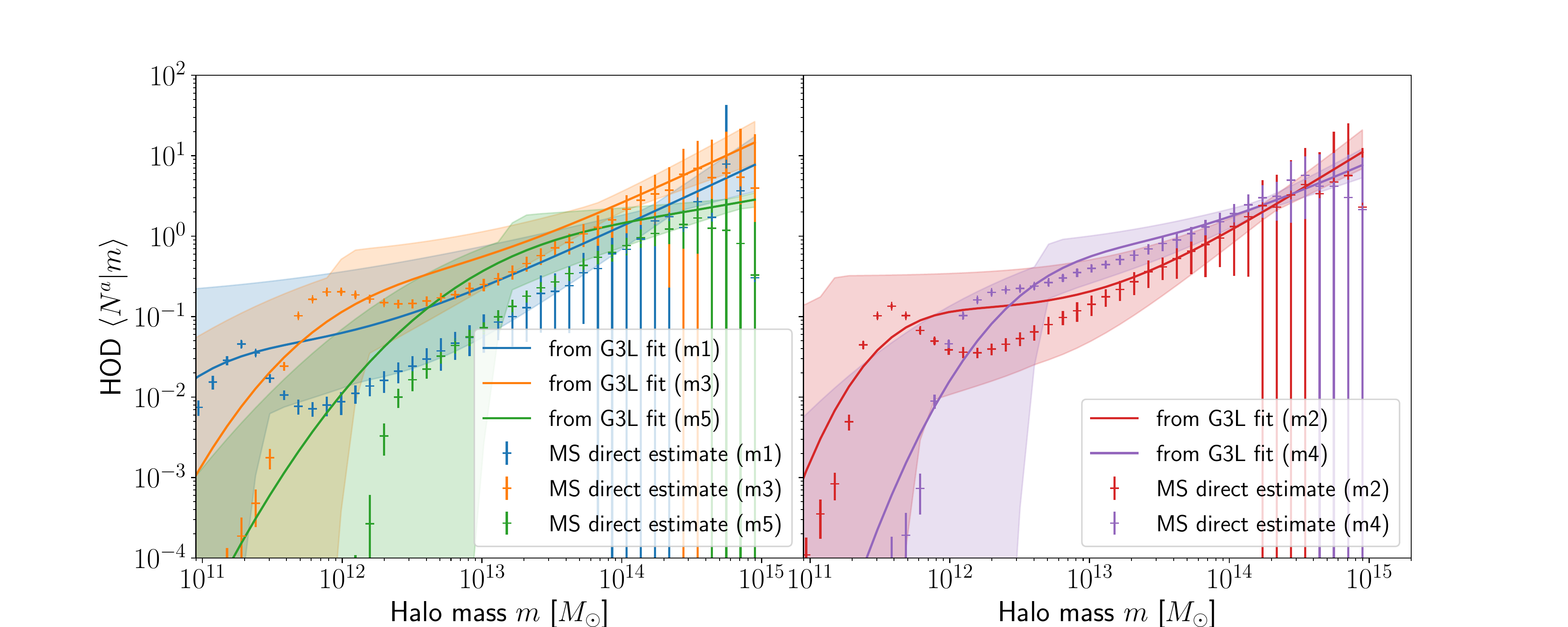}
		\caption{KV450 $\times$ GAMA}
		\label{fig:HOD_sm_KV450}
	\end{subfigure}
	\caption{Mean per-halo galaxy numbers in the simulation (top) and observation (bottom) for lens galaxies from each stellar mass bin as function of halo mass. Crosses indicate the directly estimated per-halo numbers of simulated galaxies, while lines show the predictions from the G3L fits. The shaded areas indicate the 68\% confidence areas. The left panels show the mean per-halo numbers for galaxies from stellar mass bins m1, m3, and m5 obtained from the fits to the G3L signal for m1-m5, m3-m5, and m4-m5. The right panels show the same for galaxies from stellar mass bins m2 and m5, obtained from the fits to the signal for m2-m5 and m4-m5.  The corresponding HOD parameters are listed in Table~\ref{tab:params_MS_sm}. }
	\label{fig:HOD_sm}
\end{figure*}

To verify the reconstruction of the HODs for stellar-mass samples, we compare the inferred HODs (lines and shaded areas) to the true HODs (data points) in Figure~\ref{fig:HOD_sm_MS}, which only shows the reconstructions for the combinations m1-m5, m2-m5, m3-m5, and m4-m5; the other reconstructions for the same sample but in a different combination agree with those within the $68\%$ CI. The inferred HODs agree with the true HODs within the $68\%$ credibility band. However, for the lower stellar mass galaxies from samples m1, m2, and m3, there is a `dip' for the true $\langle N|m\rangle$ with a local minimum near $6\times 10^{11}\,\Msun$, $1\times 10^{12}\,\Msun$, and $2\times 10^{12}\,\Msun$, respectively. Our halo model cannot trace this feature because, by construction, $\langle N|m\rangle$ increases monotonically with halo mass $m$; the halo model fits a smooth profile across the dip. However, this smoothing only increases the HOD reconstruction uncertainty without introducing a significant bias within the $68\%$ CI. Therefore, at the level of the precision of the verification data and for the noisier observational data, the missing HOD model feature is acceptable for KV450$\times$GAMA.

\begin{figure*}
	\includegraphics[width=\linewidth, trim={0.5cm 0.5cm 0.5cm 0.5cm}, clip]{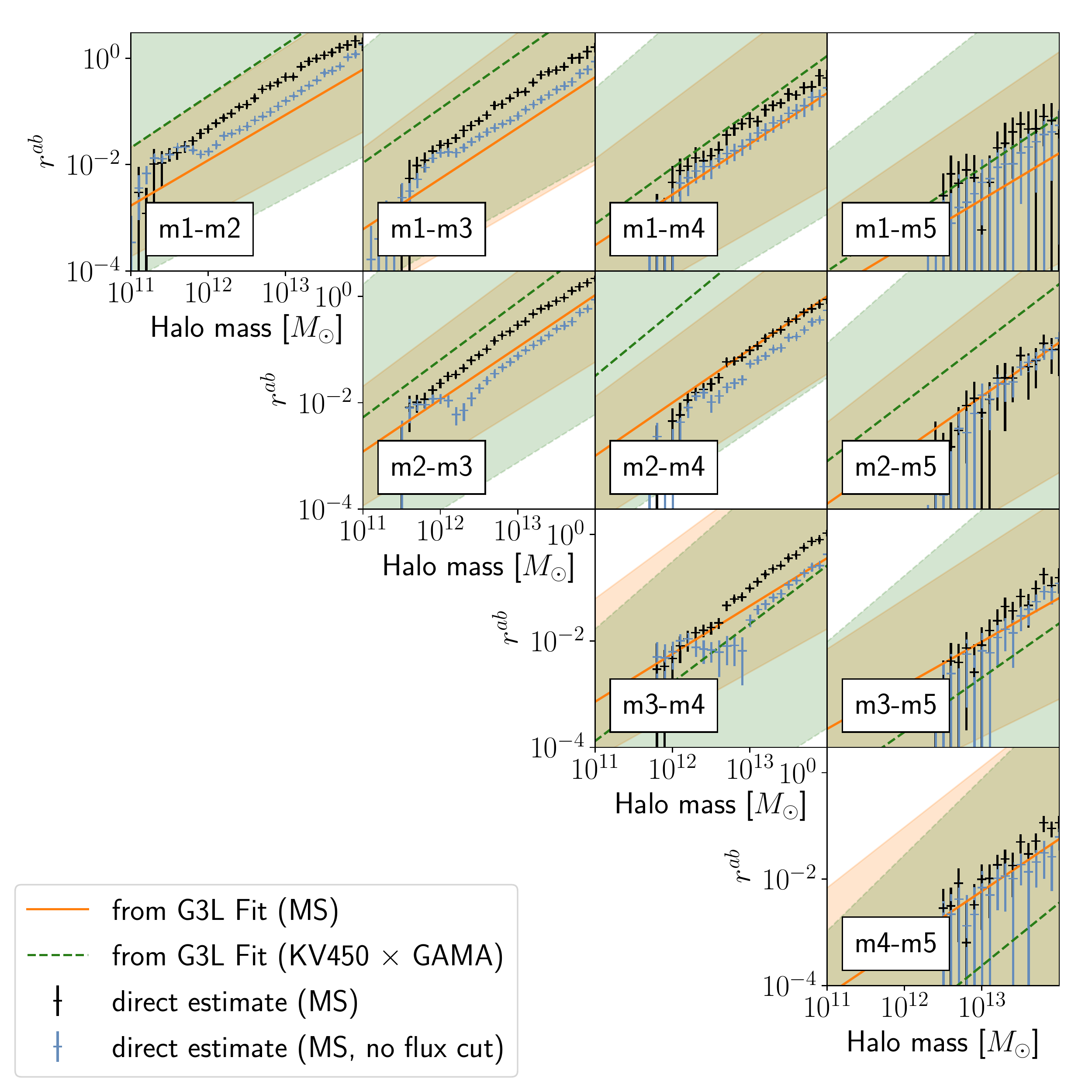}
	\caption{Correlation parameter $r^{ab}$ for stellar mass-selected galaxies in the MS and in KV450$\times$GAMA. Black crosses show the true correlation for the MS, where the error bars are the standard deviation of the mean over the 64 simulated line-of-sights. The solid brown line shows $r^{ab}$ inferred from a simulated G3L analysis of MS, and the green dashed line the inference for KV450$\times$GAMA. The shaded areas show the 68\% credibility bands of the inferences. Blue points show the correlation parameter of galaxies in the MS selected without assuming a flux limit.}
	\label{fig:r_sm}
\end{figure*}

To verify the inference of $r^{ab}$, we compare the model fits (solid lines and $68\%$ credibility regions) to the true correlation of satellite numbers (data points) in the simulation in Fig.~\ref{fig:r_sm}.  The corresponding values of $A^{ab}$ and $\epsilon^{ab}$ are listed in Table~\ref{tab:params_MS_sm} in the Appendix.  For all combinations of stellar-mass bins, $r^{ab}$ is positive and scales approximately linearly with halo mass ($\epsilon^{ab}\approx1$). The amplitude $A^{ab}$, though, drops if one of the samples is m4 or m5 in the combination. For example, $A^{\mathrm{m1m2}}$ is $1.20\pma{0.36}{0.45}\times10^{-2}$, whereas $A^{\mathrm{m1m5}}$ is $0.029\pma{0.020}{0.024}\times10^{-2}$ ($68\%$ CI). The $r^{ab}$ from the halo model fits agree with the true $r^{ab}$ for all sample combinations at the $68\%$ CI (cf. data points to lines and dark green areas).

After the science verification, we fit the model to the observed G3L signal of KV450$\times$GAMA. The best-fitting model parameters, $\chi^2$-values and plots are reported in Table \ref{tab:params_KV450_sm}. For the observations, all $p$-values exceed $0.05$, indicating no significant deviation between fit and measurement at a $95\%$ CL. Again, we perform a KS-test on the cumulative distribution of $p$-values to evaluate the overall agreement of the model and measurements (dashed lines in Fig.~\ref{fig:p_hist}). The KS distance of this distribution to a uniform distribution is now $0.31$, larger than for the science verification but still consistent with a uniform distribution at the $95\%$ CL.

The HODs inferred for KV450$\times$GAMA are shown in Fig.~\ref{fig:HOD_sm_KV450} by solid lines and $68\%$ credibility regions, again using the combinations m1--m5, m2--m5, m3--m5, and m4--m5; other combinations for the same sample yield consistent results within the $68\%$ CI. The HODs vary with the stellar mass of the lenses. In contrast to massive halos, low mass halos are mainly populated by galaxies of low stellar mass. For example, in halos with masses below $2\times 10^{11}\Msun$, m1 galaxies are the most numerous, whereas m4 galaxies dominate halos with masses above $4\times 10^{14}\Msun$. The most massive m5 galaxies, however, are never the most numerous sample between $10^{11}$ to $10^{15}\, \Msun$. Instead, they are outnumbered by satellites of lower stellar mass. Compared to the SAM HOD (data points), the HODs of KV450$\times$GAMA are consistent with the model.

Each sample HOD is an average over the sample redshift distribution, shown in Fig.~\ref{fig:nz}. Therefore, the inferred HODs are affected by the survey flux-limit. Most affected are the faint m1-galaxies. Although they are observed up to $z\sim 0.34$, for $z\gtrsim 0.15$ only the most massive m1 galaxies are seen ($M_*>2\times 10^9\, h^{-2} \Msun$). Due to this flux selection, the inferred average HOD is skewed towards the HOD of the more massive galaxies in this stellar-mass bin. Likewise, the HODS of the other samples also give more weight to the most massive galaxies in the samples, but with less bias than for the faint m1 sample.

\begin{figure}
    \centering
    \includegraphics[width=\linewidth, trim={0.7cm 0.7cm 0.5cm 0cm}, clip]{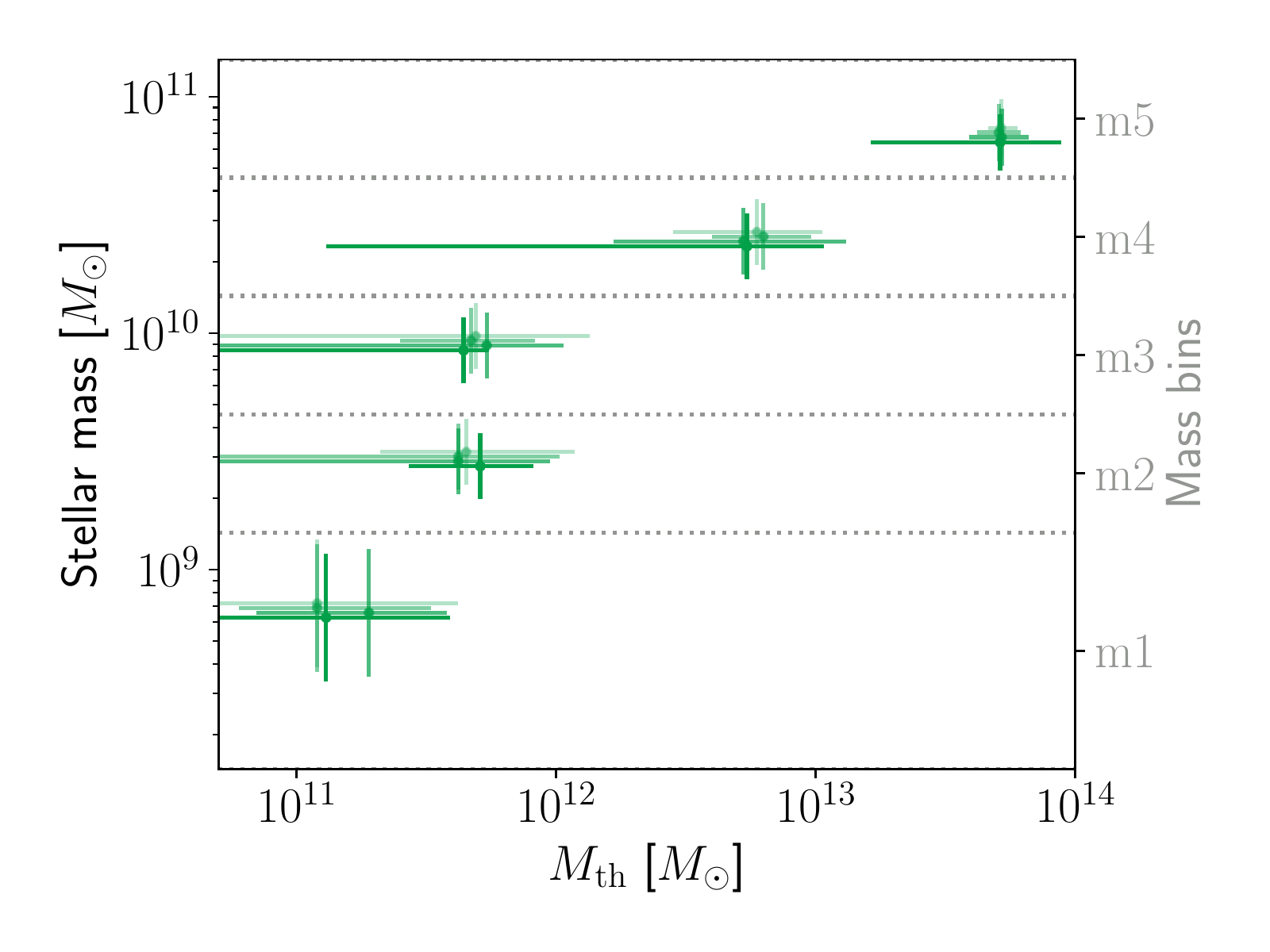}
    \caption{Threshold mass $M_\mathrm{th}^a$ measured for the GAMA galaxies as a function of the average stellar mass of each stellar mass bin (Green crosses). We show all four estimates for $M_\mathrm{th}^a$ for each sample $a$, slightly displaced along the $y$-axis for visibility. Horizontal errors correspond to $68\%$ CI of $M_\mathrm{th}^a$; vertical errors show the standard deviation of the stellar masses of galaxies within a sample.}
    \label{fig:Mth_stellar_masses}
\end{figure}

The threshold masses $M_\mathrm{th}^a$ increase with stellar mass, indicating that galaxies with higher stellar mass prefer to inhabit more massive halos. We show this trend in Fig.~\ref{fig:Mth_stellar_masses}, which plots $M_\mathrm{th}^a$ against the average stellar mass of the samples. For each stellar mass bin, there are four estimates of $M_\mathrm{th}^a$, from the four possible sample combinations, slightly displaced along the $y$-axis. As noted above, these four estimates are consistent with each other for all samples. 

The central galaxies parameter $\alpha^a$ also increases with stellar mass: The central galaxy of a halo with $m\gtrsim {\rm max}\{M_\mathrm{th}^a,M_\mathrm{th}^b\}$  is more likely to be $a$ than $b$, if $a$ is a sample with higher stellar mass than $b$. As the four $\alpha^a$s obtained for each stellar mass bin are consistent, we can average them to a $\bar{\alpha}^a$ for each sample $a$, yielding $\bar{\alpha}^\mathrm{m1}=0.11^{+0.07}_{-0.04}$, $\bar{\alpha}^\mathrm{m2}=0.14^{+0.08}_{-0.06}$, $\bar{\alpha}^\mathrm{m3}=0.22^{+0.10}_{-0.08}$, $\bar{\alpha}^\mathrm{m4}=0.47^{+0.11}_{-0.13}$, $\bar{\alpha}^\mathrm{m5}=0.54^{+0.19}_{-0.19}$. The sum of these $\alpha^a$ is $1.48_{-0.25}^{+0.26}$, which is consistent with unity within a $2\sigma$ CI. To derive the uncertainties, we assumed that each individual estimate of $\alpha^a$ is an independent measurement. This assumption is not true in our case (the data vectors for the ten fits contain multiple times the same measurements), so the derived uncertainties are probably larger in reality. A better estimate of the uncertainty on $\sum_a \bar{\alpha}^a$ requires a simultaneous fit of all $10$ combinations of the five mass bins. This is, unfortunately, not feasible in our case because our jackknife-derived covariance becomes singular for such a large data vector.

The satellite parameters $\sigma^a$, $f^a$, and $\beta^a$ do not change significantly with stellar mass ($68\%$ CI), which is not surprising for $\sigma^a$ because we cannot constrain $\sigma^a$ better than the prior range. The concentration parameter is consistent with $f^a\approx1$ for all samples ($68\%$ CI), so there is generally no detectable deviation of the satellite distribution from the matter distribution inside halos. The slope of the satellite numbers is $\beta^a\approx1$ for all mass samples.

Lastly, we show the correlation $r^{ab}$ of GAMA satellite numbers in Fig.~\ref{fig:r_sm} (green, dotted lines). There is a similar trend to the simulated galaxies (black data points), with a positive $r^{ab}$, approximately scaling linearly with halo mass ($\epsilon^{ab}\approx1$). For the sample combinations m1--m2, m1--m3, m2--m3, and m2--m4, we find \mbox{$A^{ab}>0$} ($95\%$ CI): the satellite numbers of galaxies below stellar mass of $10^{10.5}\,h^{-2}\,\rm M_\odot$ are positively correlated. Since $r^{ab}$ increases with halo mass, the correlations between low stellar masses become relevant beyond galaxy-group size halos, where $r^{m1m2}=1.9^{+0.4}_{-0.6}$, $r^{m1m3}=1.1^{+0.2}_{-0.2}$, and $r^{m2m3}=0.6^{+0.2}_{-0.2}$ for $m=10^{13}\,\Msun$ and $\epsilon=1$ ($68\%$ CI). 

We remind here that $r^{ab}>1$ are possible because the $r^{ab}$ are a Pearson correlation coefficient, $r^{ab}_\mathrm{pear}$ hereafter, only for a Poissonian variance of satellite numbers. In particular, we have the relation
\begin{equation}
    r^{ab}_\mathrm{pear} = r^{ab}\,\frac{\sqrt{\expval{\Ns{a}|m}\,\expval{\Ns{b}|m}}}{\sigma(\Ns{a}|m)\,\sigma(\Ns{b}|m)}\;,
\end{equation}
so that values of $r^{ab}>1$ indicate a super-Poissonian variance. For the strongly non-Poissonian m1 and m2 SAM galaxies at $m=10^{14}\Msun$, $r^{\mathrm{m1\, m2}}=1.2$, which corresponds to $r^{\mathrm{m1\, m2}}_\mathrm{pear}=0.18$ (see Fig.~\ref{fig:Poisson} in Appendix \ref{sec:superPoissonian} for the deviation from a Poisson statistic). For the samples m4 and m5, which are closer to Poissonian, the $r^{\mathrm{m4\, m5}}=0.07$ at the same halo mass corresponds to $r^{\mathrm{m4\, m5}}_\mathrm{pear}=0.04$. For red and blue SAM galaxies, the measured $r^{\mathrm{red\, blue}}=3.2$ at $m=10^{14}\Msun$ corresponds to $r^{\mathrm{red\, blue}}_\mathrm{pear}=0.44$.

The positive correlation coefficients are partly caused by the flux limit of the survey. To see this effect, consider as an example the correlation between m1 and m2 galaxies. At low $z \lesssim 0.1$ all galaxies with stellar masses in the m1 range are observed, so the correlation parameter at these redshifts measures the true correlation between all m1 and m2 galaxies. However, at higher redshifts, only m1 galaxies at the high end of the stellar mass bin are observed due to the flux limit. These galaxies have stellar masses closer to m2-galaxies. They are therefore more similar to m2 galaxies and stronger correlated than the overall less massive m1 galaxies. This systematically increases the inferred value of $r^{ab}$, which is an average over the correlations at different redshifts, compared to the correlation without flux limit.

To test whether the flux limit is the only cause of the correlation, we consider samples of galaxies without a flux limit in the MS. For this purpose, we select all galaxies up to  $z=0.5$, irrespective of their magnitude, and divide them by the same stellar-mass cuts as the flux-limited samples. We then directly estimate the parameter $r^{ab}$ for these samples, shown in grey in Fig.~\ref{fig:r_sm}. The $r^{ab}$ of the unlimited samples are generally lower than for the flux-limited samples, indicating that the flux limit indeed causes a higher correlation. Nevertheless, the $r^{ab}$ are still clearly positive for the sample combinations m1--m2, m1--m3, m2--m3, m2--m4, and m3--m4 without flux limit. Consequently, the measured positive correlation is not purely an effect of the survey incompleteness.

\section{Discussion}
\label{sec:discussion}
We presented a new method to measure galaxy HODs and the correlation of per-halo satellite galaxy numbers using G3L, a weak gravitational lensing effect measuring the projected galaxy-galaxy-matter bispectrum. To this end, we constructed a new halo model for the galaxy-galaxy-matter bispectrum and the G3L aperture statistics. We validated the science analysis for different selections of lens galaxies and demonstrated that it accurately recovers the true HODs within $68\%$ errors for a survey with an area of $\simeq 10^3\mathrm{deg}^2$ and lens galaxies between $0\le z\le0.5$ and brighter than $r=19.8\, \mathrm{mag}$. Therefore, our G3L analysis is accurate enough to infer HODs from KV450$\times$GAMA, which has a smaller footprint ($180\,\mathrm{deg}^2$) and similar selections for the lens galaxies.

We inferred HODs for GAMA for galaxies in two colour samples (red and blue) and five stellar mass bins between $10^{8.5}h^{-2} \Msun$ and $10^{11.5}h^{-2} \Msun$ (m1 to m5). The best-constrained HOD parameter for the KV450 $\times$ GAMA is the threshold halo mass $M_\mathrm{th}^a$, which is $M_\mathrm{th}\approx10^{11}\,M_\odot$ for our blue galaxies and $M_\mathrm{th}\approx10^{12}\,M_\odot$ for our red galaxies. These values match the expectation that red galaxies are typically group or cluster galaxies, whereas blue galaxies tend to be field galaxies. At lower halo masses, we find tight upper limits on $\expval{N|m}$, indicating that such halos rarely host galaxies satisfying our selection criteria. The transition region between this regime and halos typically containing galaxies is only poorly constrained. These poor constraints are reflected by the large uncertainty on the transition parameter $\sigma$, even for the mock data in our science verification. Therefore, G3L alone cannot constrain $\sigma$ better than our prior, and future studies may fix $\sigma$ to a fiducial value or dispense with this parameter altogether by modelling $\expval{\Nc{a}|m}$ by a step function.

The 1-halo term of the G3L signal of red-red lens pairs stretches to larger scales than for blue-blue lens pairs (Fig. \ref{fig:FitResults_colour}). This finding can be explained by the tendency of red galaxies to populate more massive halos than blue galaxies. Since massive halos are larger than less massive ones, pairs of red galaxies inside the same halo can have wider separations than pairs of blue galaxies. Accordingly, the 1-halo term extends to larger scales for red-red lens pairs than for blue-blue ones. However, mixed red-blue pairs exist in intermediate halos, large enough to host red galaxies but small enough to contain a significant fraction of blue galaxies. Consequently, the cross-over between the domination of the 1-halo and the 3-halo term occurs at larger scales than for blue-blue lens pairs.

In contrast to previous HOD studies, we measured the correlation of numbers of satellites inside halos between galaxy samples. We report a $>3\sigma$ detection of a positive correlation for red and blue GAMA galaxies ($r\gtrsim0.16^{+0.06}_{-0.05}$ for $m\gtrsim10^{13}\,M_\odot$, rising towards galaxy cluster scales). Similar positive correlations are present between the samples m1 (stellar masses below $10^{9.5} \Msun$) and m2 or m3 (stellar masses between $10^{9.5} \Msun$ and $10^{10.5} \Msun$), as well as between sample m2 (stellar masses between $10^{9.5} \Msun$ and $10^{10}\Msun$) and samples m3 or m4 (stellar masses between $10^{10}\Msun$ and $10^{11}\Msun$). In particular, galaxies with similar stellar masses (e.g., from neighbouring samples m1 and m2) show positive correlations. Towards smaller halos, correlations become irrelevant due to the almost linear decrease of $r^{ab}$, visible both in the SAMs and in our observational data (Figs. \ref{fig:r_redblue} and \ref{fig:r_sm}). This trend fits our toy model consideration in Sect.~\ref{sec:halomodel:ingredients}, where a decreasing correlation is the consequence of Poisson shot noise inside low-occupancy halos. This finding implies that the assumption of uncorrelated satellite distributions is probably still appropriate if only galaxies in low-mass halos, $m\lesssim10^{13}\, \Msun$, are considered.  

The obvious presence of correlations, especially for halos of the group- and cluster-mass scale, questions the assumption of no correlation in halo models for the cross-correlation statistics between galaxy samples in galaxy clustering studies \citep[e.g.,][]{Scranton2001, Scranton2002, Zehavi2005, Simon2008}. The correlations also raise questions about their origin and impact on galaxy models. To address the first, we note that finding strong correlations for similar galaxy samples is not unexpected. If two galaxy samples are drawn randomly from a pool of galaxies, they are statistically identical and their satellite numbers inside halos differ only by Poisson shot noise. Consequently, the satellite numbers are tightly correlated, which could explain the correlations between satellite galaxies in neighbouring stellar mass bins (Fig.~\ref{fig:r_sm}). Stellar mass varies continuously and does not uniquely define a category of galaxies. Therefore, the division of galaxies by their stellar mass is ultimately arbitrary and satellite galaxies from neighbouring stellar mass bins have similar statistical properties. In addition, errors in the estimators for stellar mass blur the stellar mass bin on the edges, similar to randomly drawing galaxies from the same pool. This effect, in combination with depth variations between the samples due to the survey flux limit (Sect. \ref{sec:results:mass}), systematically increases $r^{ab}$. Moreover, strong correlations also emerge if two samples have a (close to) fixed ratio of satellite numbers inside a halo. This effect could occur for red and blue galaxies for cluster-sized halos ($m\gtrsim 10^{14}\, \Msun$): Galaxy models commonly assume that star-forming blue galaxies falling into a galaxy cluster are quenched and turn into red galaxies with a certain probability. If this probability is roughly constant for halos of similar mass, satellite numbers of red and blue galaxies would be strongly correlated. Whatever the cause for the correlations for the GAMA galaxies, the consistency with the mock galaxies shows that galaxy SAMs already account for it at $z<0.5$. Therefore, our findings do not hint at a need to improve these models at low redshift. However, as the galaxy population in clusters evolves with time, it will be interesting to study the correlations at higher redshift in the future.

Using G3L to infer the correlations between the HODs of different galaxy populations could be interesting for other sample selections, for example, Luminous Red Galaxies (LRGs) or Emission Line Galaxies (ELGs). It would also be interesting to investigate the evolution of the cross-correlation parameter $r^{ab}$ with cosmic time, by selecting similar galaxy samples at different redshifts. Such a selection is difficult with the GAMA sample as the detection limit of $r=19.8\,\mathrm{mag}$ restricts our analysis to $z\lesssim0.5$. An analysis of the redshift evolution of the HOD parameters would consequently require a deeper lens galaxy sample, so lenses could be divided into different redshift bins. It would also require a large number of source galaxies with higher redshifts than the lens samples so that the source number density is sufficient for a significant G3L signal.

Our G3L halo model makes several unrealistic assumptions, although they are sufficient for analysing KV450$\times$GAMA. Concerning the matter distribution, we ignore halo exclusion \citep{Smith2007}, the dependency of halo properties on environment and assembly history (assembly bias, \citealp{Gao2017}), or galactic sub-halos, known to be relevant for galaxy-galaxy lensing at sub-arcmin scales \cite[e.g.,][]{Velander2014}. Concerning the lens galaxy distribution, there is a known feature of the HOD of the simulated galaxies our model does not include: a local minimum in the HODs of galaxies with stellar masses below $ 10^{10.5}\, h^{-2} \Msun$ (see Fig.~\ref{fig:HOD_sm_KV450}). This minimum occurs at the transition region between the domination of the HODs by central and satellite galaxies. There, contrary to our model for $\expval{\Nc{a}|m}$ the mean number of centrals decreases again with halo mass for galaxies from samples m1, m2, and m3 (stellar masses below $10^{10.5}\, h^{-2} \Msun$). A similar observation was reported by \citet{Berlind2003} and \citet{Zheng2005} for galaxies with young stellar populations: Due to quenching, massive halos are unlikely to have central galaxies with young stars suppressing $\expval{\Nc{a}|m}$ above a certain halo mass. Another issue is the assumption of a Poissonian variance for satellite numbers. Contrary to this assumption, the SAM galaxies by \citet{Henriques2015} have a sub-Poissonian variance for low-mass halos and super-Poissonian variance for high-mass halos, as we show in Fig.~\ref{fig:Poisson}. We investigated the impact of this effect by comparing the best-fitting G3L model to the prediction by a modified model using the actual variance $\expval{N_{\rm sat}^2|m}$ of the SAM galaxies. The difference between the two G3L signals is smaller than the statistical errors in the verification data so we can neglect non-Poissonian variances here (see Appendix~\ref{sec:superPoissonian}). However, a G3L analysis in future surveys may require a more sophisticated model with additional parameters for the satellite number variance, such as the models in \citet{Dvornik2018}. Relaxing the Poisson assumption is a new test for galaxy models with G3L. While some simulations assume Poisson satellites \citep{Kravtsov2004, Zheng2005}, recent studies suggest otherwise \citep{Dvornik2018, Gruen2018}. Despite the above model approximations, our inferred HODs agree with the true HODs of the validation data within the $68\%$ CI. Therefore, these approximations are acceptable for surveys at the same level (or worse) of statistical estimator noise as our validation data.

Aside from increasing the survey area or the source density, better HOD constraints could be achieved by incorporating the redshift-weighting scheme of lens pairs by \citet{Linke2020a}. In this scheme, lens galaxy pairs that appear close in projection but are physically distant are down-weighted compared to physically close lens pairs in estimating the G3L correlation function. This weighting increases the signal-to-noise by up to $35 \%$ for lenses with spectroscopic redshifts. However, incorporating the scheme into the halo model requires abandoning the Limber approximation and revising Eq.\eqref{limber}. Without the Limber approximation, the projected bispectrum includes three line-of-sight integrals over Bessel functions, whose evaluation is computationally challenging. There are several efforts to obtain projected matter power- and bispectra without relying on the Limber approximation \citep{Assassi2017, Campagne2017, Deshpande2020}, which might be adopted for a weighted G3L in the future.

In conclusion, G3L with our halo model is a viable new method to infer the HODs of galaxies and the correlation between the halo distributions of galaxies from different populations. This information is useful for several other analyses. First, HODs offer a fast method to obtain realistic mock galaxy catalogues from halo catalogues (e.g., \citealp{Carretero2015, Avila2018}), which is faster than using SAMs or hydrodynamical simulations. Realistic mock galaxies are vital to validate the redshift calibration (e.g., \citealp{VandenBusch2020, Hildebrandt2021}) or inference pipelines (e.g., \citealp{Ferrero2021, deRose2021}). Including the cross-correlation of galaxy samples constrained with G3L will increase the realism of the mock galaxy distribution and, therefore, the robustness of the analyses based on these simulated data. Second, HODs can be used to provide a physical interpretation of the galaxy bias, which tells us how galaxies trace the cosmic large-scale structure \citep{Cacciato2012, Simon2018, Dvornik2018}. Cosmological studies often use simple linear or quadratic models for galaxy biasing \citep{Joachimi2021, Krause2021}. They, therefore, need to exclude small scales from their analysis, where the simple models are not sufficiently accurate. HODs may provide a more accurate model on small scales so that cosmological parameters could be inferred from a larger scale range. However, the HODs obtained in this work depend on the assumed fiducial cosmology. Accordingly, using the G3L halo model for cosmological inference requires simultaneously constraining the HOD and cosmological parameters. This inference might be unfeasible in practice due to the high-dimensional parameter space. Third, HODs are important tools to constrain the stellar-to-halo mass ratio, traditionally obtained from second-order statistics, such as galaxy-galaxy-lensing \citep[][]{Velander2014, VanUitert2018, Dvornik2020}. Combining these measurements with the HOD constraints from G3L could lead to tighter constraints on the relationship between baryonic and dark matter.

\begin{acknowledgements}
We thank Elisa Chisari for her helpful comments and for acting as internal referee for the KiDS Collaboration.
This work has been supported by the Deutsche Forschungsgemeinschaft through the
project SCHN 342/15-1. LL received financial support for this research from the International Max Planck Research School (IMPRS) for Astronomy and Astrophysics at the Universities of Bonn and Cologne. AHW is supported by a European Research Council Consolidator Grant (No. 770935).
Based on data products from observations made with ESO Telescopes at the La Silla Paranal Observatory under programme IDs 177.A-3016, 177.A-3017, 177.A-3018, 179.A-2004, 298.A-5015. We also use products from the GAMA survey. GAMA is a joint European-Australasian project based around a spectroscopic campaign using the Anglo-Australian Telescope. The GAMA input catalogue is based on data taken from the Sloan Digital Sky Survey and the UKIRT Infrared Deep Sky Survey. Complementary imaging of the GAMA regions is being obtained by several independent survey programmes including GALEX MIS, VST KiDS, VISTA VIKING, WISE, Herschel-ATLAS, GMRT and ASKAP providing UV to radio coverage. GAMA is funded by the STFC (UK), the ARC (Australia), the AAO, and the participating institutions. The GAMA website is http://www.gama-survey.org/.\\
\emph{Author contributions.} All authors contributed to the development and writing of this paper. The authorship list is given in two groups: The lead authors (LL, PSi, PS), followed by an alphabetical list of contributors to either the scientific analysis or the data products.
\end{acknowledgements}

\bibliographystyle{aa}
\bibliography{biblio}

\begin{thebibliography}{93}
\expandafter\ifx\csname natexlab\endcsname\relax\def\natexlab#1{#1}\fi

\bibitem[{{Anderson}(2003)}]{Anderson2003}
{Anderson}, T.~W. 2003, An introduction to multivariate statistical analysis
  (Wiley-Interscience)

\bibitem[{{Assassi} {et~al.}(2017){Assassi}, {Simonovi{\'c}}, \&
  {Zaldarriaga}}]{Assassi2017}
{Assassi}, V., {Simonovi{\'c}}, M., \& {Zaldarriaga}, M. 2017, \jcap, 2017, 054

\bibitem[{{Avila} {et~al.}(2018){Avila}, {Crocce}, {Ross},
  {Garc{\'\i}a-Bellido}, {Percival}, {Banik}, {Camacho}, {Kokron}, {Chan},
  {Andrade-Oliveira}, {Gomes}, {Gomes}, {Lima}, {Rosenfeld}, {Salvador},
  {Friedrich}, {Abdalla}, {Annis}, {Benoit-L{\'e}vy}, {Bertin}, {Brooks},
  {Carrasco Kind}, {Carretero}, {Castander}, {Cunha}, {da Costa}, {Davis}, {De
  Vicente}, {Doel}, {Fosalba}, {Frieman}, {Gerdes}, {Gruen}, {Gruendl},
  {Gutierrez}, {Hartley}, {Hollowood}, {Honscheid}, {James}, {Kuehn},
  {Kuropatkin}, {Miquel}, {Plazas}, {Sanchez}, {Scarpine}, {Schindler},
  {Schubnell}, {Sevilla-Noarbe}, {Smith}, {Sobreira}, {Suchyta}, {Swanson},
  {Tarle}, {Thomas}, {Walker}, \& {Dark Energy Survey
  Collaboration}}]{Avila2018}
{Avila}, S., {Crocce}, M., {Ross}, A.~J., {et~al.} 2018, \mnras, 479, 94

\bibitem[{{Bartelmann} \& {Schneider}(2001)}]{Bartelmann2001}
{Bartelmann}, M. \& {Schneider}, P. 2001, \physrep, 340, 291

\bibitem[{{Berlind} \& {Weinberg}(2002)}]{Berlind2002}
{Berlind}, A.~A. \& {Weinberg}, D.~H. 2002, \apj, 575, 587

\bibitem[{{Berlind} {et~al.}(2003){Berlind}, {Weinberg}, {Benson}, {Baugh},
  {Cole}, {Dav{\'e}}, {Frenk}, {Jenkins}, {Katz}, \& {Lacey}}]{Berlind2003}
{Berlind}, A.~A., {Weinberg}, D.~H., {Benson}, A.~J., {et~al.} 2003, \apj, 593,
  1

\bibitem[{{Bernardeau} {et~al.}(2002){Bernardeau}, {Colombi}, {Gazta{\~n}aga},
  \& {Scoccimarro}}]{Bernardeau2002}
{Bernardeau}, F., {Colombi}, S., {Gazta{\~n}aga}, E., \& {Scoccimarro}, R.
  2002, Physics Reports, 367, 1

\bibitem[{{Bruzual} \& {Charlot}(2003)}]{Bruzual2003}
{Bruzual}, G. \& {Charlot}, S. 2003, \mnras, 344, 1000

\bibitem[{{Bullock} {et~al.}(2001){Bullock}, {Kolatt}, {Sigad}, {Somerville},
  {Kravtsov}, {Klypin}, {Primack}, \& {Dekel}}]{Bullock2001}
{Bullock}, J.~S., {Kolatt}, T.~S., {Sigad}, Y., {et~al.} 2001, \mnras, 321, 559

\bibitem[{{Cacciato} {et~al.}(2012){Cacciato}, {Lahav}, {van den Bosch},
  {Hoekstra}, \& {Dekel}}]{Cacciato2012}
{Cacciato}, M., {Lahav}, O., {van den Bosch}, F.~C., {Hoekstra}, H., \&
  {Dekel}, A. 2012, \mnras, 426, 566

\bibitem[{{Calzetti} {et~al.}(2000){Calzetti}, {Armus}, {Bohlin}, {Kinney},
  {Koornneef}, \& {Storchi-Bergmann}}]{Calzetti2000}
{Calzetti}, D., {Armus}, L., {Bohlin}, R.~C., {et~al.} 2000, \apj, 533, 682

\bibitem[{{Campagne} {et~al.}(2017){Campagne}, {Neveu}, \&
  {Plaszczynski}}]{Campagne2017}
{Campagne}, J.~E., {Neveu}, J., \& {Plaszczynski}, S. 2017, \aap, 602, A72

\bibitem[{{Carretero} {et~al.}(2015){Carretero}, {Castander}, {Gazta{\~n}aga},
  {Crocce}, \& {Fosalba}}]{Carretero2015}
{Carretero}, J., {Castander}, F.~J., {Gazta{\~n}aga}, E., {Crocce}, M., \&
  {Fosalba}, P. 2015, \mnras, 447, 646

\bibitem[{{Chabrier}(2003)}]{Chabrier2003}
{Chabrier}, G. 2003, \pasp, 115, 763

\bibitem[{{Clampitt} {et~al.}(2016){Clampitt}, {Miyatake}, {Jain}, \&
  {Takada}}]{Clampitt2016}
{Clampitt}, J., {Miyatake}, H., {Jain}, B., \& {Takada}, M. 2016, \mnras, 457,
  2391

\bibitem[{{Clampitt} {et~al.}(2017){Clampitt}, {S{\'a}nchez}, {Kwan}, {Krause},
  {MacCrann}, {Park}, {Troxel}, {Jain}, {Rozo}, {Rykoff}, {Wechsler}, {Blazek},
  {Bonnett}, {Crocce}, {Fang}, {Gaztanaga}, {Gruen}, {Jarvis}, {Miquel},
  {Prat}, {Ross}, {Sheldon}, {Zuntz}, {Abbott}, {Abdalla}, {Armstrong},
  {Becker}, {Benoit-L{\'e}vy}, {Bernstein}, {Bertin}, {Brooks}, {Burke},
  {Carnero Rosell}, {Carrasco Kind}, {Cunha}, {D'Andrea}, {da Costa}, {Desai},
  {Diehl}, {Dietrich}, {Doel}, {Estrada}, {Evrard}, {Fausti Neto}, {Flaugher},
  {Fosalba}, {Frieman}, {Gruendl}, {Honscheid}, {James}, {Kuehn}, {Kuropatkin},
  {Lahav}, {Lima}, {March}, {Marshall}, {Martini}, {Melchior}, {Mohr},
  {Nichol}, {Nord}, {Plazas}, {Romer}, {Sanchez}, {Scarpine}, {Schubnell},
  {Sevilla-Noarbe}, {Smith}, {Soares-Santos}, {Sobreira}, {Suchyta}, {Swanson},
  {Tarle}, {Thomas}, {Vikram}, \& {Walker}}]{Clampitt2017}
{Clampitt}, J., {S{\'a}nchez}, C., {Kwan}, J., {et~al.} 2017, \mnras, 465, 4204

\bibitem[{{Cooray} \& {Sheth}(2002)}]{Cooray2002}
{Cooray}, A. \& {Sheth}, R. 2002, \physrep, 372, 1

\bibitem[{{Crittenden} {et~al.}(2002){Crittenden}, {Natarajan}, {Pen}, \&
  {Theuns}}]{Crittenden2002}
{Crittenden}, R.~G., {Natarajan}, P., {Pen}, U.-L., \& {Theuns}, T. 2002, \apj,
  568, 20

\bibitem[{{de Jong} {et~al.}(2015){de Jong}, {Verdoes Kleijn}, {Boxhoorn},
  {Buddelmeijer}, {Capaccioli}, {Getman}, {Grado}, {Helmich}, {Huang},
  {Irisarri}, {Kuijken}, {La Barbera}, {McFarland}, {Napolitano}, {Radovich},
  {Sikkema}, {Valentijn}, {Begeman}, {Brescia}, {Cavuoti}, {Choi}, {Cordes},
  {Covone}, {Dall'Ora}, {Hildebrandt}, {Longo}, {Nakajima}, {Paolillo},
  {Puddu}, {Rifatto}, {Tortora}, {van Uitert}, {Buddendiek},
  {Harnois-D{\'e}raps}, {Erben}, {Eriksen}, {Heymans}, {Hoekstra}, {Joachimi},
  {Kitching}, {Klaes}, {Koopmans}, {K{\"o}hlinger}, {Roy}, {Sif{\'o}n},
  {Schneider}, {Sutherland}, {Viola}, \& {Vriend}}]{deJong2015}
{de Jong}, J. T.~A., {Verdoes Kleijn}, G.~A., {Boxhoorn}, D.~R., {et~al.} 2015,
  \aap, 582, A62

\bibitem[{{DeRose} {et~al.}(2021){DeRose}, {Wechsler}, {Becker}, {Rykoff},
  {Pandey}, {MacCrann}, {Amon}, {Myles}, {Krause}, {Gruen}, {Jain}, {Troxel},
  {Prat}, {Alarcon}, {S{\'a}nchez}, {Blazek}, {Crocce}, {Giannini}, {Gatti},
  {Bernstein}, {Zuntz}, {Dodelson}, {Fang}, {Friedrich}, {Secco},
  {Elvin-Poole}, {Everett}, {Choi}, {Harrison}, {Cordero}, {Rodriguez-Monroy},
  {McCullough}, {Cawthon}, {Chen}, {Alves}, {Camacho}, {Campos}, {Diehl},
  {Drlica-Wagner}, {Eifler}, {Fosalba}, {Huang}, {Porredon}, {Raveri},
  {Rosenfeld}, {Ross}, {Sanchez}, {Sheldon}, {Yanny}, {Yin}, {Aguena}, {Allam},
  {Andrade-Oliveira}, {Annis}, {Avila}, {Bacon}, {Bechtol}, {Bhargava},
  {Brooks}, {Buckley-Geer}, {Burke}, {Carnero Rosell}, {Carrasco Kind},
  {Chang}, {Costanzi}, {da Costa}, {Pereira}, {De Vicente}, {Desai},
  {Dietrich}, {Doel}, {Eckert}, {Evrard}, {Ferrero}, {Fert{\'e}}, {Flaugher},
  {Frieman}, {Garc{\'\i}a-Bellido}, {Gaztanaga}, {Giannantonio}, {Gruendl},
  {Gschwend}, {Gutierrez}, {Hartley}, {Hinton}, {Hollowood}, {Honscheid},
  {Huff}, {Huterer}, {James}, {Kuehn}, {Kuropatkin}, {Lahav}, {Lima}, {Maia},
  {Marshall}, {Melchior}, {Menanteau}, {Miquel}, {Mohr}, {Morgan}, {Palmese},
  {Paz-Chinch{\'o}n}, {Pieres}, {Plazas Malag{\'o}n}, {Sanchez}, {Scarpine},
  {Serrano}, {Sevilla-Noarbe}, {Smith}, {Soares-Santos}, {Suchyta}, {Tarle},
  {Thomas}, {To}, {Varga}, \& {Zhang}}]{deRose2021}
{DeRose}, J., {Wechsler}, R.~H., {Becker}, M.~R., {et~al.} 2021, arXiv
  e-prints, arXiv:2105.13547

\bibitem[{{Deshpande} \& {Kitching}(2020)}]{Deshpande2020}
{Deshpande}, A.~C. \& {Kitching}, T.~D. 2020, \prd, 101, 103531

\bibitem[{{Driver} {et~al.}(2009){Driver}, {Norberg}, {Baldry}, {Bamford},
  {Hopkins}, {Liske}, {Loveday}, {Peacock}, {Hill}, {Kelvin}, {Robotham},
  {Cross}, {Parkinson}, {Prescott}, {Conselice}, {Dunne}, {Brough}, {Jones},
  {Sharp}, {van Kampen}, {Oliver}, {Roseboom}, {Bland-Hawthorn}, {Croom},
  {Ellis}, {Cameron}, {Cole}, {Frenk}, {Couch}, {Graham}, {Proctor}, {De
  Propris}, {Doyle}, {Edmondson}, {Nichol}, {Thomas}, {Eales}, {Jarvis},
  {Kuijken}, {Lahav}, {Madore}, {Seibert}, {Meyer}, {Staveley-Smith},
  {Phillipps}, {Popescu}, {Sansom}, {Sutherland}, {Tuffs}, \&
  {Warren}}]{Driver2009}
{Driver}, S.~P., {Norberg}, P., {Baldry}, I.~K., {et~al.} 2009, Astronomy and
  Geophysics, 50, 5.12

\bibitem[{{Dvornik} {et~al.}(2018){Dvornik}, {Hoekstra}, {Kuijken},
  {Schneider}, {Amon}, {Nakajima}, {Viola}, {Choi}, {Erben}, {Farrow},
  {Heymans}, {Hildebrand t}, {Sif{\'o}n}, \& {Wang}}]{Dvornik2018}
{Dvornik}, A., {Hoekstra}, H., {Kuijken}, K., {et~al.} 2018, \mnras, 479, 1240

\bibitem[{{Dvornik} {et~al.}(2020){Dvornik}, {Hoekstra}, {Kuijken}, {Wright},
  {Asgari}, {Bilicki}, {Erben}, {Giblin}, {Graham}, {Heymans}, {Hildebrandt},
  {Hopkins}, {Kannawadi}, {Lin}, {Taylor}, \& {Tr{\"o}ster}}]{Dvornik2020}
{Dvornik}, A., {Hoekstra}, H., {Kuijken}, K., {et~al.} 2020, \aap, 642, A83

\bibitem[{{Edge} {et~al.}(2013){Edge}, {Sutherland}, {Kuijken}, {Driver},
  {McMahon}, {Eales}, \& {Emerson}}]{Edge2013}
{Edge}, A., {Sutherland}, W., {Kuijken}, K., {et~al.} 2013, The Messenger, 154,
  32

\bibitem[{{Eisenstein} \& {Hu}(1998)}]{Eisenstein1998}
{Eisenstein}, D.~J. \& {Hu}, W. 1998, \apj, 496, 605

\bibitem[{{Erben} {et~al.}(2005){Erben}, {Schirmer}, {Dietrich}, {Cordes},
  {Haberzettl}, {Hetterscheidt}, {Hildebrandt}, {Schmithuesen}, {Schneider},
  {Simon}, {Deul}, {Hook}, {Kaiser}, {Radovich}, {Benoist}, {Nonino}, {Olsen},
  {Prandoni}, {Wichmann}, {Zaggia}, {Bomans}, {Dettmar}, \&
  {Miralles}}]{Erben2005}
{Erben}, T., {Schirmer}, M., {Dietrich}, J.~P., {et~al.} 2005, Astronomische
  Nachrichten, 326, 432

\bibitem[{{Farrow} {et~al.}(2015){Farrow}, {Cole}, {Norberg}, {Metcalfe},
  {Baldry}, {Bland-Hawthorn}, {Brown}, {Hopkins}, {Lacey}, {Liske}, {Loveday},
  {Palamara}, {Robotham}, \& {Sridhar}}]{Farrow2015}
{Farrow}, D.~J., {Cole}, S., {Norberg}, P., {et~al.} 2015, \mnras, 454, 2120

\bibitem[{{Ferrero} {et~al.}(2021){Ferrero}, {Crocce}, {Tutusaus}, {Porredon},
  {Blot}, {Fosalba}, {Carnero Rosell}, {Avila}, {Izard}, {Elvin-Poole}, {Chan},
  {Camacho}, {Rosenfeld}, {Sanchez}, {Tallada-Cresp{\'\i}}, {Carretero},
  {Sevilla-Noarbe}, {Gaztanaga}, {Andrade-Oliveira}, {De Vicente},
  {Mena-Fern{\'a}ndez}, {Ross}, {Sanchez Cid}, {Fert{\'e}}, {Brandao-Souza},
  {Fang}, {Krause}, {Gomes}, {Aguena}, {Allam}, {Annis}, {Bertin}, {Brooks},
  {Carrasco Kind}, {Castander}, {Cawthon}, {Choi}, {Conselice}, {Costanzi}, {da
  Costa}, {Pereira}, {Diehl}, {Doel}, {Drlica-Wagner}, {Everett}, {Evrard},
  {Flaugher}, {Frieman}, {Garc{\'\i}a-Bellido}, {Gerdes}, {Gruen}, {Gruendl},
  {Gschwend}, {Gutierrez}, {Hinton}, {Hollowood}, {Honscheid}, {Hoyle},
  {Huterer}, {James}, {Kuehn}, {Lima}, {Maia}, {Marshall}, {Menanteau},
  {Miquel}, {Morgan}, {Muir}, {Ogando}, {Palmese}, {Paz-Chinch{\'o}n},
  {Percival}, {Plazas Malag{\'o}n}, {Rodriguez-Monroy}, {Scarpine},
  {Schubnell}, {Serrano}, {Smith}, {Soares-Santos}, {Suchyta}, {Swanson},
  {Tarle}, {Thomas}, {To}, {Tucker}, {Varga}, \& {DES
  Collaboration}}]{Ferrero2021}
{Ferrero}, I., {Crocce}, M., {Tutusaus}, I., {et~al.} 2021, \aap, 656, A106

\bibitem[{{Gao} \& {White}(2007)}]{Gao2017}
{Gao}, L. \& {White}, S. D.~M. 2007, \mnras, 377, L5

\bibitem[{Gough(2009)}]{Gough2009}
Gough, B. 2009, \href{https://www.gnu.org/software/gsl/doc/html/index.html}{GNU
  Scientific Library Reference Manual - Third Edition} (Network Theory Ltd.)

\bibitem[{{Gruen} {et~al.}(2018){Gruen}, {Friedrich}, {Krause}, {DeRose},
  {Cawthon}, {Davis}, {Elvin-Poole}, {Rykoff}, {Wechsler}, {Alarcon},
  {Bernstein}, {Blazek}, {Chang}, {Clampitt}, {Crocce}, {De Vicente}, {Gatti},
  {Gill}, {Hartley}, {Hilbert}, {Hoyle}, {Jain}, {Jarvis}, {Lahav}, {MacCrann},
  {McClintock}, {Prat}, {Rollins}, {Ross}, {Rozo}, {Samuroff}, {S{\'a}nchez},
  {Sheldon}, {Troxel}, {Zuntz}, {Abbott}, {Abdalla}, {Allam}, {Annis},
  {Bechtol}, {Benoit-L{\'e}vy}, {Bertin}, {Bridle}, {Brooks}, {Buckley-Geer},
  {Carnero Rosell}, {Carrasco Kind}, {Carretero}, {Cunha}, {D'Andrea}, {da
  Costa}, {Desai}, {Diehl}, {Dietrich}, {Doel}, {Drlica-Wagner}, {Fernandez},
  {Flaugher}, {Fosalba}, {Frieman}, {Garc{\'\i}a-Bellido}, {Gaztanaga},
  {Giannantonio}, {Gruendl}, {Gschwend}, {Gutierrez}, {Honscheid}, {James},
  {Jeltema}, {Kuehn}, {Kuropatkin}, {Lima}, {March}, {Marshall}, {Martini},
  {Melchior}, {Menanteau}, {Miquel}, {Mohr}, {Plazas}, {Roodman}, {Sanchez},
  {Scarpine}, {Schubnell}, {Sevilla-Noarbe}, {Smith}, {Smith}, {Soares-Santos},
  {Sobreira}, {Swanson}, {Tarle}, {Thomas}, {Vikram}, {Walker}, {Weller},
  {Zhang}, \& {DES Collaboration}}]{Gruen2018}
{Gruen}, D., {Friedrich}, O., {Krause}, E., {et~al.} 2018, \prd, 98, 023507

\bibitem[{{Hartlap} {et~al.}(2007){Hartlap}, {Simon}, \&
  {Schneider}}]{Hartlap2007}
{Hartlap}, J., {Simon}, P., \& {Schneider}, P. 2007, \aap, 464, 399

\bibitem[{{Henriques} {et~al.}(2015){Henriques}, {White}, {Thomas}, {Angulo},
  {Guo}, {Lemson}, {Springel}, \& {Overzier}}]{Henriques2015}
{Henriques}, B.~M.~B., {White}, S.~D.~M., {Thomas}, P.~A., {et~al.} 2015,
  \mnras, 451, 2663

\bibitem[{{Hilbert} {et~al.}(2009){Hilbert}, {Hartlap}, {White}, \&
  {Schneider}}]{Hilbert2009}
{Hilbert}, S., {Hartlap}, J., {White}, S.~D.~M., \& {Schneider}, P. 2009, \aap,
  499, 31

\bibitem[{{Hildebrandt} {et~al.}(2020){Hildebrandt}, {K{\"o}hlinger}, {van den
  Busch}, {Joachimi}, {Heymans}, {Kannawadi}, {Wright}, {Asgari}, {Blake},
  {Hoekstra}, {Joudaki}, {Kuijken}, {Miller}, {Morrison}, {Tr{\"o}ster},
  {Amon}, {Archidiacono}, {Brieden}, {Choi}, {de Jong}, {Erben}, {Giblin},
  {Mead}, {Peacock}, {Radovich}, {Schneider}, {Sif{\'o}n}, \&
  {Tewes}}]{Hildebrandt2020}
{Hildebrandt}, H., {K{\"o}hlinger}, F., {van den Busch}, J.~L., {et~al.} 2020,
  \aap, 633, A69

\bibitem[{{Hildebrandt} {et~al.}(2021){Hildebrandt}, {van den Busch}, {Wright},
  {Blake}, {Joachimi}, {Kuijken}, {Tr{\"o}ster}, {Asgari}, {Bilicki}, {de
  Jong}, {Dvornik}, {Erben}, {Getman}, {Giblin}, {Heymans}, {Kannawadi}, {Lin},
  \& {Shan}}]{Hildebrandt2021}
{Hildebrandt}, H., {van den Busch}, J.~L., {Wright}, A.~H., {et~al.} 2021,
  \aap, 647, A124

\bibitem[{{Ishikawa} {et~al.}(2021){Ishikawa}, {Okumura}, {Oguri}, \&
  {Lin}}]{Ishikawa2021}
{Ishikawa}, S., {Okumura}, T., {Oguri}, M., \& {Lin}, S.-C. 2021, \apj, 922, 23

\bibitem[{{Jarvis} {et~al.}(2004){Jarvis}, {Bernstein}, \& {Jain}}]{Jarvis2004}
{Jarvis}, M., {Bernstein}, G., \& {Jain}, B. 2004, \mnras, 352, 338

\bibitem[{{Joachimi} {et~al.}(2021){Joachimi}, {Lin}, {Asgari}, {Tr{\"o}ster},
  {Heymans}, {Hildebrandt}, {K{\"o}hlinger}, {S{\'a}nchez}, {Wright},
  {Bilicki}, {Blake}, {van den Busch}, {Crocce}, {Dvornik}, {Erben}, {Getman},
  {Giblin}, {Hoekstra}, {Kannawadi}, {Kuijken}, {Napolitano}, {Schneider},
  {Scoccimarro}, {Sellentin}, {Shan}, {von Wietersheim-Kramsta}, \&
  {Zuntz}}]{Joachimi2021}
{Joachimi}, B., {Lin}, C.~A., {Asgari}, M., {et~al.} 2021, \aap, 646, A129

\bibitem[{{Kaiser}(1992)}]{Kaiser1992}
{Kaiser}, N. 1992, \apj, 388, 272

\bibitem[{{Kannawadi} {et~al.}(2019){Kannawadi}, {Hoekstra}, {Miller}, {Viola},
  {Fenech Conti}, {Herbonnet}, {Erben}, {Heymans}, {Hildebrandt}, \&
  {Kuijken}}]{Kannawadi2019}
{Kannawadi}, A., {Hoekstra}, H., {Miller}, L., {et~al.} 2019, \aap, 624, A92

\bibitem[{{Krause} {et~al.}(2021){Krause}, {Fang}, {Pandey}, {Secco}, {Alves},
  {Huang}, {Blazek}, {Prat}, {Zuntz}, {Eifler}, {MacCrann}, {DeRose}, {Crocce},
  {Porredon}, {Jain}, {Troxel}, {Dodelson}, {Huterer}, {Liddle}, {Leonard},
  {Amon}, {Chen}, {Elvin-Poole}, {Fert{\'e}}, {Muir}, {Park}, {Samuroff},
  {Brandao-Souza}, {Weaverdyck}, {Zacharegkas}, {Rosenfeld}, {Campos},
  {Chintalapati}, {Choi}, {Di Valentino}, {Doux}, {Herner}, {Lemos},
  {Mena-Fern{\'a}ndez}, {Omori}, {Paterno}, {Rodriguez-Monroy}, {Rogozenski},
  {Rollins}, {Troja}, {Tutusaus}, {Wechsler}, {Abbott}, {Aguena}, {Allam},
  {Andrade-Oliveira}, {Annis}, {Bacon}, {Baxter}, {Bechtol}, {Bernstein},
  {Brooks}, {Buckley-Geer}, {Burke}, {Carnero Rosell}, {Carrasco Kind},
  {Carretero}, {Castander}, {Cawthon}, {Chang}, {Costanzi}, {da Costa},
  {Pereira}, {De Vicente}, {Desai}, {Diehl}, {Doel}, {Everett}, {Evrard},
  {Ferrero}, {Flaugher}, {Fosalba}, {Frieman}, {Garc{\'\i}a-Bellido},
  {Gaztanaga}, {Gerdes}, {Giannantonio}, {Gruen}, {Gruendl}, {Gschwend},
  {Gutierrez}, {Hartley}, {Hinton}, {Hollowood}, {Honscheid}, {Hoyle}, {Huff},
  {James}, {Kuehn}, {Kuropatkin}, {Lahav}, {Lima}, {Maia}, {Marshall},
  {Martini}, {Melchior}, {Menanteau}, {Miquel}, {Mohr}, {Morgan}, {Myles},
  {Palmese}, {Paz-Chinch{\'o}n}, {Petravick}, {Pieres}, {Plazas Malag{\'o}n},
  {Sanchez}, {Scarpine}, {Schubnell}, {Serrano}, {Sevilla-Noarbe}, {Smith},
  {Soares-Santos}, {Suchyta}, {Tarle}, {Thomas}, {To}, {Varga}, \&
  {Weller}}]{Krause2021}
{Krause}, E., {Fang}, X., {Pandey}, S., {et~al.} 2021, arXiv e-prints,
  arXiv:2105.13548

\bibitem[{{Kravtsov} {et~al.}(2004){Kravtsov}, {Berlind}, {Wechsler}, {Klypin},
  {Gottl{\"o}ber}, {Allgood}, \& {Primack}}]{Kravtsov2004}
{Kravtsov}, A.~V., {Berlind}, A.~A., {Wechsler}, R.~H., {et~al.} 2004, \apj,
  609, 35

\bibitem[{{Kuijken} {et~al.}(2015){Kuijken}, {Heymans}, {Hildebrandt},
  {Nakajima}, {Erben}, {de Jong}, {Viola}, {Choi}, {Hoekstra}, {Miller}, {van
  Uitert}, {Amon}, {Blake}, {Brouwer}, {Buddendiek}, {Conti}, {Eriksen},
  {Grado}, {Harnois-D{\'e}raps}, {Helmich}, {Herbonnet}, {Irisarri},
  {Kitching}, {Klaes}, {La Barbera}, {Napolitano}, {Radovich}, {Schneider},
  {Sif{\'o}n}, {Sikkema}, {Simon}, {Tudorica}, {Valentijn}, {Verdoes Kleijn},
  \& {van Waerbeke}}]{Kuijken2015}
{Kuijken}, K., {Heymans}, C., {Hildebrandt}, H., {et~al.} 2015, \mnras, 454,
  3500

\bibitem[{{Linke} {et~al.}(2020b){Linke}, {Simon}, {Schneider}, {Erben},
  {Farrow}, {Heymans}, {Hildebrandt}, {Hopkins}, {Kannawadi}, {Napolitano},
  {Sif{\'o}n}, \& {Wright}}]{Linke2020b}
{Linke}, L., {Simon}, P., {Schneider}, P., {et~al.} 2020b, \aap, 640, A59

\bibitem[{{Linke} {et~al.}(2020a){Linke}, {Simon}, {Schneider}, \&
  {Hilbert}}]{Linke2020a}
{Linke}, L., {Simon}, P., {Schneider}, P., \& {Hilbert}, S. 2020a, \aap, 634,
  A13

\bibitem[{{Liske} {et~al.}(2015){Liske}, {Baldry}, {Driver}, {Tuffs},
  {Alpaslan}, {Andrae}, {Brough}, {Cluver}, {Grootes}, {Gunawardhana},
  {Kelvin}, {Loveday}, {Robotham}, {Taylor}, {Bamford}, {Bland-Hawthorn},
  {Brown}, {Drinkwater}, {Hopkins}, {Meyer}, {Norberg}, {Peacock}, {Agius},
  {Andrews}, {Bauer}, {Ching}, {Colless}, {Conselice}, {Croom}, {Davies}, {De
  Propris}, {Dunne}, {Eardley}, {Ellis}, {Foster}, {Frenk}, {H{\"a}u{\ss}ler},
  {Holwerda}, {Howlett}, {Ibarra}, {Jarvis}, {Jones}, {Kafle}, {Lacey},
  {Lange}, {Lara-L{\'o}pez}, {L{\'o}pez-S{\'a}nchez}, {Maddox}, {Madore},
  {McNaught-Roberts}, {Moffett}, {Nichol}, {Owers}, {Palamara}, {Penny},
  {Phillipps}, {Pimbblet}, {Popescu}, {Prescott}, {Proctor}, {Sadler},
  {Sansom}, {Seibert}, {Sharp}, {Sutherland}, {V{\'a}zquez-Mata}, {van Kampen},
  {Wilkins}, {Williams}, \& {Wright}}]{Liske2015}
{Liske}, J., {Baldry}, I.~K., {Driver}, S.~P., {et~al.} 2015, \mnras, 452, 2087

\bibitem[{Liu(2004)}]{Liu2004}
Liu, J.~S. 2004, Monte Carlo Strategies in Scientific Computing, 1st edn. (New
  York, NY: Springer), 31--36

\bibitem[{{Mandelbaum} {et~al.}(2006){Mandelbaum}, {Hirata}, {Broderick},
  {Seljak}, \& {Brinkmann}}]{Mandelbaum2006}
{Mandelbaum}, R., {Hirata}, C.~M., {Broderick}, T., {Seljak}, U., \&
  {Brinkmann}, J. 2006, \mnras, 370, 1008

\bibitem[{{Maraston}(2005)}]{Maraston2005}
{Maraston}, C. 2005, \mnras, 362, 799

\bibitem[{{Martin}(2019)}]{Martin2019}
{Martin}, S.~M. 2019, PhD thesis, University of Bonn, Germany

\bibitem[{{Mead} {et~al.}(2015){Mead}, {Peacock}, {Heymans}, {Joudaki}, \&
  {Heavens}}]{Mead2015}
{Mead}, A.~J., {Peacock}, J.~A., {Heymans}, C., {Joudaki}, S., \& {Heavens},
  A.~F. 2015, \mnras, 454, 1958

\bibitem[{{Miller} {et~al.}(2013){Miller}, {Heymans}, {Kitching}, {van
  Waerbeke}, {Erben}, {Hildebrandt}, {Hoekstra}, {Mellier}, {Rowe}, \&
  {Coupon}}]{Miller2013}
{Miller}, L., {Heymans}, C., {Kitching}, T.~D., {et~al.} 2013, \mnras, 429,
  2858

\bibitem[{{Mo} \& {White}(1996)}]{Mo1996}
{Mo}, H.~J. \& {White}, S.~D.~M. 1996, \mnras, 282, 347

\bibitem[{{Nakamura} \& {Suto}(1997)}]{Nakamura1997}
{Nakamura}, T.~T. \& {Suto}, Y. 1997, Progress of Theoretical Physics, 97, 49

\bibitem[{{Navarro} {et~al.}(1996){Navarro}, {Frenk}, \& {White}}]{Navarro1996}
{Navarro}, J.~F., {Frenk}, C.~S., \& {White}, S. D.~M. 1996, \apj, 462, 563

\bibitem[{Nelder \& Mead(1965)}]{Nelder1965}
Nelder, J.~A. \& Mead, R. 1965, The Computer Journal, 7, 308

\bibitem[{{Planck Collaboration: Aghanim} {et~al.}(2020){Planck Collaboration:
  Aghanim}, {Akrami}, {Arroja}, {Ashdown}, {Aumont}, {Baccigalupi},
  {Ballardini}, {Banday}, {Barreiro}, {Bartolo}, {Basak}, {Battye}, {Benabed},
  {Bernard}, {Bersanelli}, {Bielewicz}, {Bock}, {Bond}, {Borrill}, {Bouchet},
  {Boulanger}, {Bucher}, {Burigana}, {Butler}, {Calabrese}, {Cardoso},
  {Carron}, {Casaponsa}, {Challinor}, {Chiang}, {Colombo}, {Combet},
  {Contreras}, {Crill}, {Cuttaia}, {de Bernardis}, {de Zotti}, {Delabrouille},
  {Delouis}, {D{\'e}sert}, {Di Valentino}, {Dickinson}, {Diego}, {Donzelli},
  {Dor{\'e}}, {Douspis}, {Ducout}, {Dupac}, {Efstathiou}, {Elsner},
  {En{\ss}lin}, {Eriksen}, {Falgarone}, {Fantaye}, {Fergusson},
  {Fernandez-Cobos}, {Finelli}, {Forastieri}, {Frailis}, {Franceschi},
  {Frolov}, {Galeotta}, {Galli}, {Ganga}, {G{\'e}nova-Santos}, {Gerbino},
  {Ghosh}, {Gonz{\'a}lez-Nuevo}, {G{\'o}rski}, {Gratton}, {Gruppuso},
  {Gudmundsson}, {Hamann}, {Handley}, {Hansen}, {Helou}, {Herranz},
  {Hildebrandt}, {Hivon}, {Huang}, {Jaffe}, {Jones}, {Karakci}, {Keih{\"a}nen},
  {Keskitalo}, {Kiiveri}, {Kim}, {Kisner}, {Knox}, {Krachmalnicoff}, {Kunz},
  {Kurki-Suonio}, {Lagache}, {Lamarre}, {Langer}, {Lasenby}, {Lattanzi},
  {Lawrence}, {Le Jeune}, {Leahy}, {Lesgourgues}, {Levrier}, {Lewis},
  {Liguori}, {Lilje}, {Lilley}, {Lindholm}, {L{\'o}pez-Caniego}, {Lubin}, {Ma},
  {Mac{\'\i}as-P{\'e}rez}, {Maggio}, {Maino}, {Mandolesi}, {Mangilli},
  {Marcos-Caballero}, {Maris}, {Martin}, {Martinelli},
  {Mart{\'\i}nez-Gonz{\'a}lez}, {Matarrese}, {Mauri}, {McEwen}, {Meerburg},
  {Meinhold}, {Melchiorri}, {Mennella}, {Migliaccio}, {Millea}, {Mitra},
  {Miville-Desch{\^e}nes}, {Molinari}, {Moneti}, {Montier}, {Morgante}, {Moss},
  {Mottet}, {M{\"u}nchmeyer}, {Natoli}, {N{\o}rgaard-Nielsen}, {Oxborrow},
  {Pagano}, {Paoletti}, {Partridge}, {Patanchon}, {Pearson}, {Peel}, {Peiris},
  {Perrotta}, {Pettorino}, {Piacentini}, {Polastri}, {Polenta}, {Puget},
  {Rachen}, {Reinecke}, {Remazeilles}, {Renault}, {Renzi}, {Rocha}, {Rosset},
  {Roudier}, {Rubi{\~n}o-Mart{\'\i}n}, {Ruiz-Granados}, {Salvati}, {Sandri},
  {Savelainen}, {Scott}, {Shellard}, {Shiraishi}, {Sirignano}, {Sirri},
  {Spencer}, {Sunyaev}, {Suur-Uski}, {Tauber}, {Tavagnacco}, {Tenti},
  {Terenzi}, {Toffolatti}, {Tomasi}, {Trombetti}, {Valiviita}, {Van Tent},
  {Vibert}, {Vielva}, {Villa}, {Vittorio}, {Wandelt}, {Wehus}, {White},
  {White}, {Zacchei}, \& {Zonca}}]{Planck2018}
{Planck Collaboration: Aghanim}, N., {Akrami}, Y., {Arroja}, F., {et~al.} 2020,
  \aap, 641, A1

\bibitem[{{R{\"o}diger}(2009)}]{Roedinger2009}
{R{\"o}diger}, J. 2009, PhD thesis, University of Bonn, Germany

\bibitem[{{Ross} \& {Brunner}(2009)}]{Ross2009}
{Ross}, A.~J. \& {Brunner}, R.~J. 2009, \mnras, 399, 878

\bibitem[{{Saghiha} {et~al.}(2017){Saghiha}, {Simon}, {Schneider}, \&
  {Hilbert}}]{Saghiha2017}
{Saghiha}, H., {Simon}, P., {Schneider}, P., \& {Hilbert}, S. 2017, \aap, 601,
  A98

\bibitem[{{Schirmer}(2013)}]{Schirmer2013}
{Schirmer}, M. 2013, \apjs, 209, 21

\bibitem[{{Schneider} {et~al.}(2005){Schneider}, {Kilbinger}, \&
  {Lombardi}}]{SchneiderKilbinger2005}
{Schneider}, P., {Kilbinger}, M., \& {Lombardi}, M. 2005, \aap, 431, 9

\bibitem[{{Schneider} \& {Watts}(2005)}]{Schneider2005}
{Schneider}, P. \& {Watts}, P. 2005, \aap, 432, 783

\bibitem[{{Scoccimarro} {et~al.}(2001){Scoccimarro}, {Sheth}, {Hui}, \&
  {Jain}}]{Scoccimarro2001}
{Scoccimarro}, R., {Sheth}, R.~K., {Hui}, L., \& {Jain}, B. 2001, \apj, 546, 20

\bibitem[{Scranton(2001)}]{Scranton2001}
Scranton, R. 2001, \mnras, 332, 697

\bibitem[{Scranton(2002)}]{Scranton2002}
Scranton, R. 2002, \mnras, 339, 410

\bibitem[{{Sheth} \& {Tormen}(1999)}]{Sheth1999}
{Sheth}, R.~K. \& {Tormen}, G. 1999, \mnras, 308, 119

\bibitem[{{Simon} {et~al.}(2013){Simon}, {Erben}, {Schneider}, {Heymans},
  {Hildebrandt}, {Hoekstra}, {Kitching}, {Mellier}, {Miller}, {Van Waerbeke},
  {Bonnett}, {Coupon}, {Fu}, {Hudson}, {Kuijken}, {Rowe}, {Schrabback},
  {Semboloni}, \& {Velander}}]{Simon2013}
{Simon}, P., {Erben}, T., {Schneider}, P., {et~al.} 2013, \mnras, 430, 2476

\bibitem[{{Simon} {et~al.}(2009){Simon}, {Hetterscheidt}, {Wolf},
  {Meisenheimer}, {Hildebrandt}, {Schneider}, {Schirmer}, \&
  {Erben}}]{Simon2009}
{Simon}, P., {Hetterscheidt}, M., {Wolf}, C., {et~al.} 2009, \mnras, 398, 807

\bibitem[{{Simon} \& {Hilbert}(2018)}]{Simon2018}
{Simon}, P. \& {Hilbert}, S. 2018, \aap, 613, A15

\bibitem[{{Simon} {et~al.}(2008){Simon}, {Watts}, {Schneider}, {Hoekstra},
  {Gladders}, {Yee}, {Hsieh}, \& {Lin}}]{Simon2008}
{Simon}, P., {Watts}, P., {Schneider}, P., {et~al.} 2008, \aap, 479, 655

\bibitem[{{Smith} {et~al.}(2007){Smith}, {Scoccimarro}, \& {Sheth}}]{Smith2007}
{Smith}, R.~E., {Scoccimarro}, R., \& {Sheth}, R.~K. 2007, \prd, 75, 063512

\bibitem[{{Springel} {et~al.}(2005){Springel}, {White}, {Jenkins}, {Frenk},
  {Yoshida}, {Gao}, {Navarro}, {Thacker}, {Croton}, {Helly}, {Peacock}, {Cole},
  {Thomas}, {Couchman}, {Evrard}, {Colberg}, \& {Pearce}}]{Springel2005}
{Springel}, V., {White}, S. D.~M., {Jenkins}, A., {et~al.} 2005, \nat, 435, 629

\bibitem[{{Szapudi} \& {Szalay}(1998)}]{Szapudi1998}
{Szapudi}, I. \& {Szalay}, A.~S. 1998, \apjl, 494, L41

\bibitem[{{Taylor} {et~al.}(2011){Taylor}, {Hopkins}, {Baldry}, {Brown},
  {Driver}, {Kelvin}, {Hill}, {Robotham}, {Bland-Hawthorn}, {Jones}, {Sharp},
  {Thomas}, {Liske}, {Loveday}, {Norberg}, {Peacock}, {Bamford}, {Brough},
  {Colless}, {Cameron}, {Conselice}, {Croom}, {Frenk}, {Gunawardhana},
  {Kuijken}, {Nichol}, {Parkinson}, {Phillipps}, {Pimbblet}, {Popescu},
  {Prescott}, {Sutherland}, {Tuffs}, {van Kampen}, \&
  {Wijesinghe}}]{Taylor2011}
{Taylor}, E.~N., {Hopkins}, A.~M., {Baldry}, I.~K., {et~al.} 2011, \mnras, 418,
  1587

\bibitem[{{Tegmark} {et~al.}(1997){Tegmark}, {Taylor}, \&
  {Heavens}}]{Tegmark1997}
{Tegmark}, M., {Taylor}, A.~N., \& {Heavens}, A.~F. 1997, \apj, 480, 22

\bibitem[{{van den Busch} {et~al.}(2020){van den Busch}, {Hildebrandt},
  {Wright}, {Morrison}, {Blake}, {Joachimi}, {Erben}, {Heymans}, {Kuijken}, \&
  {Taylor}}]{VandenBusch2020}
{van den Busch}, J.~L., {Hildebrandt}, H., {Wright}, A.~H., {et~al.} 2020,
  \aap, 642, A200

\bibitem[{{van Uitert} {et~al.}(2018){van Uitert}, {Joachimi}, {Joudaki},
  {Amon}, {Heymans}, {K{\"o}hlinger}, {Asgari}, {Blake}, {Choi}, {Erben},
  {Farrow}, {Harnois-D{\'e}raps}, {Hildebrandt}, {Hoekstra}, {Kitching},
  {Klaes}, {Kuijken}, {Merten}, {Miller}, {Nakajima}, {Schneider}, {Valentijn},
  \& {Viola}}]{VanUitert2018}
{van Uitert}, E., {Joachimi}, B., {Joudaki}, S., {et~al.} 2018, \mnras, 476,
  4662

\bibitem[{{Velander} {et~al.}(2014){Velander}, {van Uitert}, {Hoekstra},
  {Coupon}, {Erben}, {Heymans}, {Hildebrandt}, {Kitching}, {Mellier}, {Miller},
  {Van Waerbeke}, {Bonnett}, {Fu}, {Giodini}, {Hudson}, {Kuijken}, {Rowe},
  {Schrabback}, \& {Semboloni}}]{Velander2014}
{Velander}, M., {van Uitert}, E., {Hoekstra}, H., {et~al.} 2014, \mnras, 437,
  2111

\bibitem[{{Venemans} {et~al.}(2015){Venemans}, {Verdoes Kleijn}, {Mwebaze},
  {Valentijn}, {Ba{\~n}ados}, {Decarli}, {de Jong}, {Findlay}, {Kuijken}, {La
  Barbera}, {McFarland}, {McMahon}, {Napolitano}, {Sikkema}, \&
  {Sutherland}}]{Venemans2015}
{Venemans}, B.~P., {Verdoes Kleijn}, G.~A., {Mwebaze}, J., {et~al.} 2015,
  \mnras, 453, 2259

\bibitem[{{Vogelsberger} {et~al.}(2020){Vogelsberger}, {Marinacci}, {Torrey},
  \& {Puchwein}}]{Vogelsberger2020}
{Vogelsberger}, M., {Marinacci}, F., {Torrey}, P., \& {Puchwein}, E. 2020,
  Nature Reviews Physics, 2, 42

\bibitem[{{Wang} {et~al.}(2007){Wang}, {Yang}, {Mo}, \& {van den
  Bosch}}]{Wang2007}
{Wang}, Y., {Yang}, X., {Mo}, H.~J., \& {van den Bosch}, F.~C. 2007, \apj, 664,
  608

\bibitem[{{Watts} \& {Schneider}(2005)}]{Watts2005}
{Watts}, P. \& {Schneider}, P. 2005, in IAU Symposium, Vol. 225, Gravitational
  Lensing Impact on Cosmology, ed. Y.~{Mellier} \& G.~{Meylan}, 243--248

\bibitem[{Weisstein(2022)}]{MathWorld}
Weisstein, E.~W. 2022, Delta Function. {From MathWorld---A Wolfram Web
  Resource}, \url{http://mathworld.wolfram.com/DeltaFunction.html} Last visited
  on 02/3/2022

\bibitem[{{White} \& {Rees}(1978)}]{White1978}
{White}, S.~D.~M. \& {Rees}, M.~J. 1978, \mnras, 183, 341

\bibitem[{{Wright} {et~al.}(2019){Wright}, {Hildebrandt}, {Kuijken}, {Erben},
  {Blake}, {Buddelmeijer}, {Choi}, {Cross}, {de Jong}, {Edge},
  {Gonzalez-Fernandez}, {Gonz{\'a}lez Solares}, {Grado}, {Heymans}, {Irwin},
  {Kupcu Yoldas}, {Lewis}, {Mann}, {Napolitano}, {Radovich}, {Schneider},
  {Sif{\'o}n}, {Sutherland}, {Sutorius}, \& {Verdoes Kleijn}}]{Wright2019}
{Wright}, A.~H., {Hildebrandt}, H., {Kuijken}, K., {et~al.} 2019, \aap, 632,
  A34

\bibitem[{{Wright} {et~al.}(2016){Wright}, {Robotham}, {Bourne}, {Driver},
  {Dunne}, {Maddox}, {Alpaslan}, {Andrews}, {Bauer}, {Bland-Hawthorn},
  {Brough}, {Brown}, {Clarke}, {Cluver}, {Davies}, {Grootes}, {Holwerda},
  {Hopkins}, {Jarrett}, {Kafle}, {Lange}, {Liske}, {Loveday}, {Moffett},
  {Norberg}, {Popescu}, {Smith}, {Taylor}, {Tuffs}, {Wang}, \&
  {Wilkins}}]{Wright2016}
{Wright}, A.~H., {Robotham}, A.~S.~G., {Bourne}, N., {et~al.} 2016, \mnras,
  460, 765

\bibitem[{{Zehavi} {et~al.}(2011){Zehavi}, {Zheng}, {Weinberg}, {Blanton},
  {Bahcall}, {Berlind}, {Brinkmann}, {Frieman}, {Gunn}, {Lupton}, {Nichol},
  {Percival}, {Schneider}, {Skibba}, {Strauss}, {Tegmark}, \&
  {York}}]{Zehavi2011}
{Zehavi}, I., {Zheng}, Z., {Weinberg}, D.~H., {et~al.} 2011, \apj, 736, 59

\bibitem[{{Zehavi} {et~al.}(2005){Zehavi}, {Zheng}, {Weinberg}, {Frieman},
  {Berlind}, {Blanton}, {Scoccimarro}, {Sheth}, {Strauss}, {Kayo}, {Suto},
  {Fukugita}, {Nakamura}, {Bahcall}, {Brinkmann}, {Gunn}, {Hennessy},
  {Ivezi{\'c}}, {Knapp}, {Loveday}, {Meiksin}, {Schlegel}, {Schneider},
  {Szapudi}, {Tegmark}, {Vogeley}, {York}, \& {SDSS
  Collaboration}}]{Zehavi2005}
{Zehavi}, I., {Zheng}, Z., {Weinberg}, D.~H., {et~al.} 2005, \apj, 630, 1

\bibitem[{{Zheng} {et~al.}(2005){Zheng}, {Berlind}, {Weinberg}, {Benson},
  {Baugh}, {Cole}, {Dav{\'e}}, {Frenk}, {Katz}, \& {Lacey}}]{Zheng2005}
{Zheng}, Z., {Berlind}, A.~A., {Weinberg}, D.~H., {et~al.} 2005, \apj, 633, 791

\bibitem[{{Zheng} {et~al.}(2007){Zheng}, {Coil}, \& {Zehavi}}]{Zheng2007}
{Zheng}, Z., {Coil}, A.~L., \& {Zehavi}, I. 2007, \apj, 667, 760

\end{thebibliography}

\begin{appendix} 
	
\onecolumn	
\allowdisplaybreaks
\section{Calculation of galaxy-galaxy-matter bispectrum}
\label{app:calculationBispectrum_halo model}
		
Here we derive the galaxy-galaxy-matter bispectrum for mixed ($a\neq b$) and unmixed ($a=b$) galaxy pairs in a halo model. While the bispectrum for unmixed pairs has already been derived partially in \citet{Watts2005}, \citet{Roedinger2009}, and \citet{Martin2019}, a model for mixed pairs is, to our knowledge, still pending and therefore shown in detail here. 

Our model assumes, that all statistics of matter and galaxies inside halos depend only on halo mass, halo positions are clustered according to a deterministic linear bias, matter and galaxy number density profiles are spherically symmetric, satellite galaxies have no sub-halos and that galaxy positions inside halos are statistically independent from each other. These approximations greatly simplify the model formalism, while still being suffienciently accurate to describe the G3L aperture statistics in the science verification data. For brevity, we drop all explicit redshift dependencies and all quantities are understood to be evaluated at the same redshift.
	
The halo model averages over the probability density functions (PDFs) of the halos masses, positions, galaxy numbers and galaxy positions. The average of an arbitrary quantity $f(\dots)$ depending on the halo variables is, for $H$ halos distributed in the volume $V$ and containing two \emph{distinct} galaxy samples $a\neq b$,
\begin{align}
	\label{eq:average halo model distinct samples}
	\expval{f} 
	&=  \int_0^\infty \dd{m_1}\; \dots\int_0^\infty \dd{m_{H}}\;\underbrace{P_\textrm{m}(m_1, \dots, m_{H})}_{{\text{PDF of halo masses }m_1, \dots, m_H}}\, \int _V \dd[3]{x_1}\; \dots \int_V \dd[3]{x_{H}}\; \underbrace{P_\textrm{c}(\vec{x}_1, \dots, \vec{x}_{H}\,|\, m_1, \dots , m_{H})}_{{\text{PDF of halo centres }\vec{x}_1, \dots, \vec{x}_{H}}} \\
	&\notag\quad \times \prod_{h=1}^{H}\Bigg[ \sum_{\Ncc{a}{h}=0}^\infty  \sum_{\Nss{a}{h}=0}^\infty \sum_{\Ncc{b}{h}=0}^\infty \sum_{\Nss{b}{h}=0}^\infty \underbrace{P^{ab}_\mathrm{N}(\Ncc{a}{h}, \Nss{a}{h}, \Ncc{b}{h}, \Nss{b}{h}\,|\,m_h)}_{\substack{\text{Probability that halo $h$ has $\Ncc{a}{h}$ centrals and $\Nss{a}{h}$ satellites of sample $a$,}\\ \text{$\Ncc{b}{h}$ centrals and $\Nss{b}{h}$ satellites of sample $b$}}}\\
	&\notag\quad\quad\quad \times \prod_{v=1}^{\Nss{a}{h}}\left(\int \dd[3]{\Delta {x}^{a}_{hv}}\right)\, \prod_{w=1}^{\Nss{b}{h}}\left(\int \dd[3]{\Delta {x}^{b}_{hw}}\right)\, \underbrace{P^{ab}_\mathrm{g}(\Delta \vec{x}^{a}_{h1}, \dots, \Delta \vec{x}^{a}_{h\Nss{a}{h}}, \Delta \vec{x}^{b}_{h1}, \dots, \Delta \vec{x}^{b}_{h\Nss{b}{h}}\,|\, m_h)}_{\text{PDF of satellite offsets $\Delta \vec{x}$ relative to halo centre}}\Bigg]\, f(\dots)\;.
\end{align}
For two \emph{identical} galaxy samples $a=b$, this expression simplifies to
\begin{align}
	\label{eq:average halo model equal samples}
	\expval{f} 
	&=  \int_0^\infty\dd{m_1}\; \dots\int_0^\infty\dd{m_{H}}\;{P_\textrm{m}(m_1, \dots, m_{H})}\, \int _V \dd[3]{x_1}\; \dots \int_V \dd[3]{x_{H}}\; {P_\textrm{c}(\vec{x}_1, \dots, \vec{x}_{H}\,|\, m_1, \dots , m_{H})} \\
	&\notag\quad \times \prod_{h=1}^{H}\Bigg[  \sum_{\Ncc{a}{h}=0}^\infty  \sum_{\Nss{a}{h}=0}^\infty {P^a_\mathrm{N}(\Ncc{a}{h}, \Nss{a}{h}\,|\,m_h)} \prod_{v=1}^{\Nss{a}{h}}\left(\int \dd[3]{\Delta {x}^{a}_{hv}}\right)\, {P^a_\mathrm{g}(\Delta \vec{x}^{a}_{h1}, \dots, \Delta \vec{x}^{a}_{h\Nss{a}{h}}\,|\, m_h)}\Bigg]\, f(\dots)\;,
\end{align}
where $P^a_\mathrm{N}(\Ncc{a}{h}, \Nss{a}{h}\,|\,m_h)$ is the joint probability for halo $h$  to have $\Ncc{a}{h}$ central and $\Nss{a}{h}$ satellite galaxies from sample $a$, and $P^a_\mathrm{g}(\Delta \vec{x}^{a}_{h1}, \dots, \Delta \vec{x}^{a}_{h\Nss{a}{h}}\,|\, m_h)$ is the PDF for the positions $\Delta \vec{x}^{a}_{h1}, \dots, \Delta \vec{x}^{a}_{h\Nss{a}{h}}$ of these satellites relative to the halo centre. The joint PDF $P_\mathrm{m}$ of halo masses, assuming that halo masses are independent of each other, is a product of the {HMF},
\begin{equation}
	P_\mathrm{m}(m_1 \dots m_H) = \bar{n}_\mathrm{h}^{-H} \, n(m_1) \dots n(m_H)\;,
\end{equation}
where $\bar{n}_\mathrm{h}=H/V$ is the mean halo number density \footnote{We note that an improper PDF $P_\mathrm{m}$ is irrelevant here, as long as its moments are well defined, which is the case for the HMF.}. The PDF $P^{ab}_\mathrm{g}$ is a product of the normalized spatial distributions $u\rmg^a(\Delta \vec{x}|m)$ of satellites in each halo,
\begin{align}
	P^{ab}_\mathrm{g}(\Delta \vec{x}^{a}_{h1}, \dots, \Delta \vec{x}^{a}_{h \Nss{a}{h}}, \Delta \vec{x}^{b}_{h1}, \dots, \Delta \vec{x}^{b}_{ h\Nss{b}{h}}\,|\, m_h)
	&= 	P^a_\mathrm{g}(\Delta \vec{x}^{a}_{h1}, \dots, \Delta \vec{x}^{a}_{h \Nss{a}{h}}\,|\, m_h)\, P^b_\mathrm{g}(\Delta \vec{x}^{b}_{h1}, \dots, \Delta \vec{x}^{b}_{h\Nss{b}{h}}\,|\, m_h)\\
	&= \prod_{v=1}^{\Nss{a}{h}} \left[ u\rmg^{a}(\Delta \vec{x}^{a}_{hv}\,|\,m_h)  \right]\, \prod_{w=1}^{\Nss{b}{h}} \left[  u\rmg^{b}(\Delta \vec{x}^{b}_{hw}\,|\,m_h)  \right]\;,
\end{align}
under the assumption of independent satellite positions inside a halo. The joint probability $P_\mathrm{N}$ of per-halo galaxy numbers has the moments
\begin{equation}
	\label{eq:HOD PN}
	\expval{(\Nc{a})^p\,(\Ns{a})^q\,(\Nc{b})^r\,(\Nc{b})^s\,|\, m} = \sum_{\Ncc{a}{h}=0}^\infty  \sum_{\Nss{a}{h}=0}^\infty \sum_{\Ncc{b}{h}=0}^\infty \sum_{\Nss{b}{h}=0}^\infty (\Ncc{a}{h})^p\, (\Nss{a}{h})^q\, (\Ncc{b}{h})^r\, (\Nss{b}{h})^s \,{P_\mathrm{N}(\Ncc{a}{h}, \Nss{a}{h}, \Ncc{b}{h}, \Nss{b}{h}\,|\,m_h)}\,,
\end{equation}
where $p, q, r, s \in \mathbb{N}_0$.

After inserting Eqs.~\eqref{eq:ft halo matter density} and \eqref{eq:ft galaxy number density} into \eqref{eq:bispectrum halo model} and using Eq.~\eqref{eq:average halo model distinct samples}, the bispectrum $\Bggdab$ for mixed pairs, $a\neq b$, is given by
\begin{align}
\label{eq:bispec_allterms}
	&\notag	(2\pi)^3 \Bggdab(\vec{k}_1, \vec{k}_2, \vec{k}_3)\, \dirac(\vec{k}_1+\vec{k}_2+\vec{k}_3) + \text{unconnected terms}\\
	&=  \frac{1}{\bar{n}_\mathrm{h}^H \bar{n}^a\rmg \bar{n}^b\rmg\, \bar{\rho}} \sum_{i,j,k=1}^H \int_0^\infty\dd{m_1}\; \dots\int_0^\infty\dd{m_{H}}\; n(m_1) \dots n(m_H)\, \int _V \dd[3]{x_1}\; \dots \int_V \dd[3]{x_{H}}\; P_\textrm{c}(\vec{x}_1, \dots, \vec{x}_{H}\,|\, m_1, \dots , m_{H}) \\
	&\notag \times \prod_{h=1}^{H}\Biggl\{ \sum_{\Ncc{a}{h}=0}^\infty  \sum_{\Nss{a}{h}=0}^\infty \sum_{\Ncc{b}{h}=0}^\infty \sum_{\Nss{b}{h}=0}^\infty {P_\mathrm{N}(\Ncc{a}{h}, \Nss{a}{h}, \Ncc{b}{h}, \Nss{b}{h}\,|\,m_h)}\, \prod_{v=1}^{\Nss{a}{h}} \left[ \int \dd[3]{\Delta {x}^{a}_{hv}}\, u\rmg^{a}(\Delta \vec{x}^{a}_{hv}\,|\,m_h)  \right]\, \prod_{w=1}^{\Nss{b}{h}} \left[ \int \dd[3]{\Delta {x}^{b}_{hw}}\, u\rmg^{b}(\Delta \vec{x}^{b}_{hw}\,|\,m_h)  \right]\Biggr\}\\
	&\notag \times  m_k\, \hat{u}(\vec{k}_3\,|\,m_k)\, \mathrm{e}^{-\I \vec{k}_3\cdot \vec{x}_k}\, \left[\Ncc{a}{i}\,\mathrm{e}^{-\I\vec{k}_1\cdot \vec{x}_i} + \sum_{l=1}^{\Nss{a}{i}} \mathrm{e}^{-\I \vec{k}_1\cdot\vec{x}_i-\I \vec{k}_1\cdot\Delta \vec{x}^a_{il}}\right]\, \left[\Ncc{b}{j}\,\mathrm{e}^{-\I\vec{k}_2\cdot \vec{x}_j} + \sum_{m=1}^{\Nss{b}{j}} \mathrm{e}^{-\I \vec{k}_2\cdot\vec{x}_j-\I \vec{k}_2\cdot\Delta \vec{x}^b_{jm}}\right]\;,
\end{align}	
where \emph{unconnected terms} do not contain $\dirac(\vec{k}_1+\vec{k}_2+\vec{k}_2)$,  $\bar{n}^a\rmg$ is the mean galaxy number density and $\bar{\rho}$ is the mean matter density (as in Eqs.~\ref{eq:definition delta} and \ref{eq:definition delta_g}).
For unmixed pairs, $a=b$, the bispectrum is given by
\begin{align}
	&\notag	(2\pi)^3 \Bggd^{aa}(\vec{k}_1, \vec{k}_2, \vec{k}_3)\, \dirac(\vec{k}_1+\vec{k}_2+\vec{k}_3) + \text{unconnected terms}\\
	&=  \frac{1}{\bar{n}_\mathrm{h}^H\, (\bar{n}^a\rmg)^2 \, \bar{\rho}} \sum_{i,j,k=1}^H \int_0^\infty\dd{m_1}\; \dots\int_0^\infty\dd{m_{H}}\;  n(m_1) \dots n(m_H)\, \int _V \dd[3]{x_1}\; \dots \int_V \dd[3]{x_{H}}\; P_\textrm{c}(\vec{x}_1, \dots, \vec{x}_{H}\,|\, m_1, \dots , m_{H}) \\
	&\notag \quad \times \prod_{h=1}^{H}\Biggl\{\sum_{\Ncc{a}{h}=0}^\infty  \sum_{\Nss{a}{h}=0}^\infty {P_\mathrm{N}(\Ncc{a}{h}, \Nss{a}{h}\,|\,m_h)}\, \prod_{v=1}^{\Nss{a}{h}} \left[ \int \dd[3]{\Delta {x}^{a}_{hv}}\, u\rmg^{a}(\Delta \vec{x}^{a}_{hv}\,|\,m_h)  \right]\Biggr\}\\
	&\notag \quad \times  m_k\, \hat{u}(\vec{k}_3\,|\,m_k)\, \mathrm{e}^{-\I \vec{k}_3\cdot \vec{x}_k}\,\Bigg[\Ncc{a}{i}\,(\Ncc{a}{j} - \kronecker{i}{j})\,\mathrm{e}^{-\I\vec{k}_1\cdot \vec{x}_i-\I\vec{k}_2\cdot \vec{x}_j} + \Ncc{a}{j}\,\sum_{l=1}^{\Nss{a}{i}} \mathrm{e}^{-\I \vec{k_2}\cdot \vec{x}_j-\I \vec{k}_1\cdot\vec{x}_i-\I \vec{k}_1\cdot\Delta \vec{x}^a_{il}}\\
	&\notag \quad \quad +  \Ncc{a}{i}\,\sum_{m=1}^{\Nss{a}{j}} \mathrm{e}^{-\I \vec{k_1}\cdot \vec{x}_i-\I \vec{k}_2\cdot\vec{x}_j-\I \vec{k}_2\cdot\Delta \vec{x}^a_{jm}}  +  \sum_{l=1}^{\Nss{a}{i}} \sum_{m=1, m\neq l}^{\Nss{a}{j}} \mathrm{e}^{-\I \vec{k}_1\cdot\vec{x}_i - \I \vec{k}_1\cdot\Delta \vec{x}_{il}^a-\I \vec{k}_2\cdot\vec{x}_j-\I \vec{k}_2\cdot\Delta \vec{x}^a_{jm}}\Bigg]\;,
\end{align}
where $\kronecker{i}{j}$ is the Kronecker symbol. 

We divide the sum over $i,j$ and $k$ into the 1-halo term $\Bggdabh{1}$ for $i=j=k$, the 2-halo term $\Bggdabh{2}$ for $i=j\neq k$, $i=k \neq j$, and $j=k\neq i$, and the 3-halo term $\Bggdabh{3}$ for $i\neq j\neq k$. In the following, we calculate these terms individually. 

\subsection{1-halo term}
To calculate the 1-halo term $\Bggdabh{1}$ with $i=j=k$, we distinguish between mixed ($a\neq b$) and unmixed ($a=b$) pairs. For $a=b$, the 1-halo term is given by
\begin{align}
	&\notag	(2\pi)^3 \Bggdaah{1}(\vec{k}_1, \vec{k}_2, \vec{k}_3)\, \dirac(\vec{k}_1+\vec{k}_2+\vec{k}_3)\\
	&= \frac{1}{\bar{n}_\mathrm{h}^H\,(\bar{n}^a\rmg)^2\, \bar{\rho}} \sum_{i=1}^H \int_0^\infty\dd{m_1}\; \dots \int_0^\infty\dd{m_H}\; n(m_1)\dots n(m_H) \, \int _V \dd[3]{x_1}\; \dots \int_V \dd[3]{x_{H}}\; P_\textrm{c}(\vec{x}_1, \dots, \vec{x}_{H}\,|\, m_1, \dots , m_{H}) \\
	&\notag \quad \times \prod_{h=1}^{H}\Biggl\{ \sum_{\Ncc{a}{h}=0}^\infty  \sum_{\Nss{a}{h}=0}^\infty  {P_\mathrm{N}(\Ncc{a}{h}, \Nss{a}{h}\,|\,m_h)}\, \prod_{v=1}^{\Nss{a}{h}} \left[ \int \dd[3]{\Delta {x}^{a}_{hv}}\, u\rmg^{a}(\Delta \vec{x}^{a}_{hv}\,|\,m_h)  \right]\Biggr\}\, m_i\, \hat{u}(\vec{k}_3\,|\,m_i)\, \mathrm{e}^{-\I \vec{k}_3\cdot \vec{x}_i}\, \\
	&\notag \quad \times \Bigg[\Ncc{a}{i}\,(\Ncc{a}{i} - 1)\,\mathrm{e}^{-\I(\vec{k}_1+\vec{k}_2)\cdot \vec{x}_i} + \Ncc{a}{i}\,\sum_{l=1}^{\Nss{a}{i}} \mathrm{e}^{-\I (\vec{k}_1+\vec{k}_2)\cdot \vec{x}_i-\I \vec{k}_1\cdot\Delta \vec{x}^a_{il}} +  \Ncc{a}{i}\,\sum_{m=1}^{\Nss{a}{i}} \mathrm{e}^{-\I (\vec{k_1}+\vec{k_2})\cdot \vec{x}_i-\I \vec{k}_2\cdot\Delta \vec{x}^a_{im}}  +  \sum_{l=1}^{\Nss{a}{i}} \sum^{\Nss{a}{i}}_{\substack{m=1,\\m\neq l}} \mathrm{e}^{-\I (\vec{k}_1+\vec{k}_2)\cdot\vec{x}_i - \I \vec{k}_1\cdot\Delta\vec{x}^a_{il}-\I \vec{k}_2\cdot\Delta \vec{x}^a_{im}}\Bigg]\,.
\end{align}
Evaluating all $m$-integrals aside from the one over $m_i$ ($H-1$ integrals in total) leads to
\begin{align}
	&\notag	(2\pi)^3 \Bggdaah{1}(\vec{k}_1, \vec{k}_2, \vec{k}_3)\, \dirac(\vec{k}_1+\vec{k}_2+\vec{k}_3)\\
	&= \frac{1}{\bar{n}_\mathrm{h}^H\,(\bar{n}^a\rmg)^2\, \bar{\rho}} \sum_{i=1}^H \bar{n}_\mathrm{h}^{H-1} \int_0^\infty\dd{m_i}\; n(m_i) \, \int _V \dd[3]{x_1}\; \dots \int_V \dd[3]{x_{H}}\; P_\textrm{c}(\vec{x}_1, \dots, \vec{x}_{H}\,|\, m_1, \dots , m_{H})\, \mathrm{e}^{-\I (\vec{k}_1 + \vec{k}_2 +\vec{k}_3)\cdot \vec{x}_i}\, \\
	&\notag \quad \times \sum_{\Ncc{a}{h}=0}^\infty  \sum_{\Nss{a}{h}=0}^\infty {P_\mathrm{N}(\Ncc{a}{i}, \Nss{a}{i}\,|\,m_i)}\, \prod_{v=1}^{\Nss{a}{i}} \left[ \int \dd[3]{\Delta {x}^{a}_{iv}}\, u\rmg^{a}(\Delta \vec{x}^{a}_{iv}\,|\,m_i)  \right]\, m_i\, \hat{u}(\vec{k}_3\,|\,m_i) \\
	&\notag \quad \times \Bigg[\Ncc{a}{i}\,(\Ncc{a}{i} - 1) + \Ncc{a}{i}\,\sum_{l=1}^{\Nss{a}{i}} \mathrm{e}^{-\I \vec{k}_1\cdot\Delta \vec{x}^a_{il}} +  \Ncc{a}{i}\,\sum_{m=1}^{\Nss{a}{i}} \mathrm{e}^{-\I \vec{k}_2\cdot\Delta \vec{x}^a_{im}}  +  \sum_{l=1}^{\Nss{a}{i}} \sum_{m=1, m\neq l}^{\Nss{a}{i}} \mathrm{e}^{ - \I \vec{k}_1\cdot\Delta\vec{x}^a_{il}-\I \vec{k}_2\cdot\Delta \vec{x}^a_{im}}\Bigg]\;.
\end{align}
Using Eqs.~\eqref{eq:HOD PN}, $\int \dd[3]{x}\; u\rmg^a(\vec{x}\,|\,m) = 1$, and  $\int \dd[3]{x}\; u\rmg^a(\vec{x}\,|\,m)\, \exp(-\I\vec{k}\cdot \vec{x}) = \hat{u}\rmg^a(\vec{k}\,|\,m)$, we find
\begin{align}
	&\notag	(2\pi)^3 \Bggdaah{1}(\vec{k}_1, \vec{k}_2, \vec{k}_3)\, \dirac(\vec{k}_1+\vec{k}_2+\vec{k}_3)\\
	&= \frac{1}{\bar{n}_\mathrm{h}\,(\bar{n}^a\rmg)^2\, \bar{\rho}} \sum_{i=1}^H \int_0^\infty\dd{m_i}\; n(m_i) \, \int _V \dd[3]{x_1}\; \dots \int_V \dd[3]{x_{H}}\; P_\textrm{c}(\vec{x}_1, \dots, \vec{x}_{H}\,|\, m_1, \dots , m_{H})\, \mathrm{e}^{-\I (\vec{k}_1 + \vec{k}_2 +\vec{k}_3)\cdot \vec{x}_i}\, m_i\, \hat{u}(\vec{k}_3\,|\,m_i) \\
	&\notag \quad \times  \left\lbrace\expval{\Nc{a}\,(\Nc{a} - 1)\,|\,m_i} + \expval{\Nc{a}\, \Ns{a}\,|\,m_i} \left[ \hat{u}\rmg^a(\vec{k}_1\,|\,m_i) + \hat{u}\rmg^a(\vec{k}_2\,|\,m_i) \right] +\expval{\Ns{a}\; (\Ns{a}-1)\,|\, m_i}\, \hat{u}\rmg^a(\vec{k}_1\,|\,m_i)\, \hat{u}\rmg^a(\vec{k}_2\,|\,m_i)\right\rbrace\;,
\end{align}
explicitly using the second-order moments of $P_\mathrm{N}$ in the last line. 

We consider large volumes $V$ here, rendering the $V$-integrals  over the halo positions $\vec{x}$ with $\exp{-\I\vec{k}\cdot\vec{x}}$ in the integrand essentially Fourier transforms. This can, for example, be seen for a cubic volume $V_s$ with side length $s$,
\begin{equation}
    \int_{V_s} \dd[3]{x} \, \exp(-\I\vec{k}\cdot\vec{x}) =  \frac{8}{k_x k_y k_z} \sin(\frac{s k_x}{2})\,\sin(\frac{s k_y}{2})\,\sin(\frac{s k_z}{2})\;,
\end{equation}
where the $k_i$ are the components of $\vec{k}$. For large $V_s$ ($s\rightarrow \infty$), this expression approximates a Dirac-distribution (e.g., \citealp{MathWorld}), as
\begin{equation}
    \lim_{s\rightarrow \infty} \frac{\sin(s\,k)}{k} = \pi \dirac(k)
\end{equation}
and hence
\begin{equation}
\label{eq:large volume approximation}
    \lim_{s\rightarrow \infty}\int_{V_s} \dd[3]{x} \, \exp(-\I\vec{k}\cdot\vec{x})=(2\pi)^3 \, \dirac(\vec{x})\;.
\end{equation}

In this limit (assumed henceforth),
\begin{align}
	\label{eq:x_integral 1haloterm}
\int_V \dd[3]{x_1} \, \dots\, \dd[3]{x_H}\; {P_\textrm{c}(\vec{x}_1, \dots, \vec{x}_{H}\,|\, m_1, \dots , m_{H})}\,  \mathrm{e}^{-\I (\vec{k}_1 + \vec{k}_2 +\vec{k}_3)\cdot \vec{x}_i} = \int_V \dd[3]{x_i}\; \frac{1}{V} \,  \mathrm{e}^{-\I (\vec{k}_1 + \vec{k}_2 +\vec{k}_3)\cdot \vec{x}_i}= \frac{(2\pi)^3 \, \bar{n}_\mathrm{h}}{H}\, \dirac(\vec{k}_1+\vec{k}_2+\vec{k}_3)\;,
\end{align}
which leads to
\begin{align}
	\Bggdaah{1}(\vec{k}_1, \vec{k}_2, \vec{k}_3) &= \frac{1}{(\bar{n}^a\rmg)^2\, \bar{\rho}}\, \int_0^\infty\dd{m}\; n(m)\,  m\, \hat{u}(\vec{k}_3\,|\,m) \,  \Big\{\expval{\Nc{a}\,(\Nc{a} - 1)\,|\,m} + \expval{\Nc{a}\, \Ns{a}\,|\,m} \left[ \hat{u}\rmg^a(\vec{k}_1\,|\,m) + \hat{u}\rmg^a(\vec{k}_2\,|\,m) \right]\\
	&\notag \quad \quad  +\expval{\Ns{a}\; (\Ns{a}-1)\,|\, m}\, \hat{u}\rmg^a(\vec{k}_1\,|\,m)\, \hat{u}\rmg^a(\vec{k}_2\,|\,m)\Big\} \;\\
	&= \frac{1}{(\bar{n}^a\rmg)^2\, \bar{\rho}}\, \int_0^\infty\dd{m}\; n(m)\,  m\, \hat{u}(\vec{k}_3\,|\,m)\, G^{aa}(\vec{k}_1, \vec{k}_2\,|\, m) \;,
\end{align}		
with $\vec{k}_3=-\vec{k}_1-\vec{k}_2$ and
\begin{align}
	\label{eq: def G_gg}
	&\notag G^{ab}(\vec{k}_1, \vec{k}_2\,|\,m)\\
	&= \expval{\Nc{a}\,(\Nc{a}-\kronecker{a}{b})\,|\,m} + \expval{\Nc{a}\, \Ns{b}\,|\,m}\, \hat{u}\rmg^{b}(\vec{k}_2\,|\,m) + \expval{\Nc{b}\, \Ns{a}\,|\,m}\, \hat{u}\rmg^{a}(\vec{k}_1\,|\,m) + \expval{\Ns{a}\,(\Ns{b} - \kronecker{a}{b})}\, \hat{u}\rmg^{a}(\vec{k}_1\,|\,m)\,\hat{u}\rmg^{b}(\vec{k}_2\,|\,m)\;.
\end{align}

For mixed pairs, $a\neq b$, an analogous calculation gives
\begin{align}
	\Bggdabh{1}(\vec{k}_1, \vec{k}_2, \vec{k}_3)	&= \frac{1}{\bar{n}^a\rmg\,\bar{n}^b\rmg\, \bar{\rho}}\, \int_0^\infty\dd{m}\; n(m)\,  m\, \hat{u}(\vec{k}_3\,|\,m) \Big[\expval{\Nc{a}\,\Nc{b}\,|\,m} + \expval{\Nc{a}\, \Ns{b}\,|\,m} \, \hat{u}\rmg^b(\vec{k}_2\,|\,m)  \\
	&\notag \quad \quad  + \expval{\Nc{b}\, \Ns{a}\,|\,m} \, \hat{u}\rmg^a(\vec{k}_1\,|\,m)   + \expval{\Ns{a}\; \Ns{b}\,|\, m}\, \hat{u}\rmg^a(\vec{k}_1\,|\,m)\, \hat{u}\rmg^b(\vec{k}_2\,|\,m)\Big]\;\\
	&= \frac{1}{\bar{n}^a\rmg\,\bar{n}^b\rmg\, \bar{\rho}}\, \int_0^\infty\dd{m}\; n(m)\,  m\, \hat{u}(\vec{k}_3\,|\,m)\, G^{ab}(\vec{k}_1, \vec{k}_2\,|\, m)\;.	
\end{align}	

In the main text, we denote $\Bggdabh{1}(\vec{k}_1, \vec{k}_2, \vec{k}_3)$ as $\Bggdabh{1}(\vec{k}_1, \vec{k}_2)$ for brevity.

\subsection{2-halo term}
The 2-halo term $\Bggdabh{2}$ in Eq.~\eqref{eq:bispec_allterms} contains the contributions from $i=j\neq k$, $i=k\neq j$, and $k=j\neq i$. For unmixed pairs, $a=b$, the 2-halo term is given by
\begin{align}
	&\notag	(2\pi)^3 \Bggdaah{2}(\vec{k}_1, \vec{k}_2, \vec{k}_3)\, \dirac(\vec{k}_1+\vec{k}_2+\vec{k}_3) + \text{unconnected terms}\\
	&= \frac{1}{\bar{n}_\mathrm{h}^H\,(\bar{n}^a\rmg)^2\, \bar{\rho}} \sum_{i,j=1, j\neq i}^H \int_0^\infty\dd{m_1}\; \dots \int_0^\infty\dd{m_H}\; n(m_1)\dots n(m_H) \, \int _V \dd[3]{x_1}\; \dots \int_V \dd[3]{x_{H}}\; P_\textrm{c}(\vec{x}_1, \dots, \vec{x}_{H}\,|\, m_1, \dots , m_{H}) \\
	&\notag \quad \times \prod_{h=1}^{H}\Biggl\{ \sum_{\Ncc{a}{h}=0}^\infty \sum_{\Nss{a}{h}=0}^\infty {P_\mathrm{N}(\Ncc{a}{h}, \Nss{a}{h}\,|\,m_h)}\, \prod_{v=1}^{\Nss{a}{h}} \left[ \int \dd[3]{\Delta {x}^{a}_{hv}}\, u\rmg^{a}(\Delta \vec{x}^{a}_{hv}\,|\,m_h)  \right] \Biggr\}\\
	&\notag \quad \times \Biggl\{ m_j\, \hat{u}(\vec{k}_3\; | \; m_j)\, \mathrm{e}^{-\I \vec{k}_3\cdot \vec{x}_j} \Big[ \Ncc{a}{i}\,(\Ncc{a}{i}-1)\,\mathrm{e}^{-\I(\vec{k}_1+\vec{k}_2)\cdot\vec{x}_i} + \Ncc{a}{i}\sum_{l=1}^{\Nss{a}{i}}\mathrm{e}^{-\I\vec{k}_1\cdot(\vec{x}_i + \Delta \vec{x}^a_{il})-\I \vec{k}_2\cdot \vec{x}_i} + \Ncc{a}{i}\sum_{m=1}^{\Nss{a}{i}}\mathrm{e}^{-\I\vec{k}_2\cdot(\vec{x}_i + \Delta \vec{x}^a_{im})-\I \vec{k}_1\cdot \vec{x}_i}\\
	&\notag \quad \quad \quad  + \sum_{l=1}^{\Nss{a}{i}}\sum_{l\neq m} \mathrm{e}^{-\I \vec{k}_1\cdot (\vec{x}_i + \Delta \vec{x}^a_{il})-\I \vec{k}_2\cdot (\vec{x}_i + \Delta \vec{x}^a_{im})}\Big]\\
	&\notag \quad \quad + m_i\, \hat{u}(\vec{k}_3\; | \; m_i)\, \mathrm{e}^{-\I \vec{k}_3\cdot \vec{x}_i} \Big(\Ncc{a}{i}\, \mathrm{e}^{-\I\vec{k}_1\cdot \vec{x}_i} + \sum_{l=1}^{\Nss{a}{i}}\mathrm{e}^{-\I\vec{k}_1(\vec{x}_i + \Delta\vec{x}^a_{il})}\Big) \Big(\Ncc{a}{j}\, \mathrm{e}^{\I\vec{k}_2\cdot \vec{x}_j} + \sum_{m=1}^{\Nss{a}{j}}\mathrm{e}^{\I\vec{k}_2(\vec{x}_j + \Delta\vec{x}^a_{jm})}\Big)	\\
	&\notag \quad \quad + m_i\, \hat{u}(\vec{k}_3\; | \; m_i)\, \mathrm{e}^{-\I \vec{k}_3\cdot \vec{x}_i} \Big(\Ncc{a}{j}\, \mathrm{e}^{-\I\vec{k}_1\cdot \vec{x}_j} + \sum_{l=1}^{\Nss{a}{i}}\mathrm{e}^{-\I\vec{k}_1(\vec{x}_i + \Delta\vec{x}^a_{jl})}\Big) \Big(\Ncc{a}{i}\, \mathrm{e}^{\I\vec{k}_2\cdot \vec{x}_i} + \sum_{m=1}^{\Nss{a}{i}}\mathrm{e}^{\I\vec{k}_2(\vec{x}_i + \Delta\vec{x}^a_{im})}\Big)\Biggr\}\;.
\end{align}
We now evaluate all $m$-integrals independent of $m_i$ and $m_j$ (these are $H(H-1)$ terms in total), use Eq.~\eqref{eq:HOD PN}, and evaluate the $\vec{\Delta x}$ integrals over the halo profiles $u\rmg^a$. This leads to
\begin{align}
	&\notag	(2\pi)^3 \Bggdaah{2}(\vec{k}_1, \vec{k}_2, \vec{k}_3)\, \dirac(\vec{k}_1+\vec{k}_2+\vec{k}_3) + \text{unconnected terms}\\
	&= \frac{1}{\bar{n}_\mathrm{h}^2\,(\bar{n}^a\rmg)^2\, \bar{\rho}} \sum_{i=1, j\neq i}^H  \int_0^\infty\dd{m_i}\;\int_0^\infty\dd{m_j}\; n(m_i)\, n(m_j) \, \int _V \dd[3]{x_1}\; \dots \int_V \dd[3]{x_{H}}\; P_\textrm{c}(\vec{x}_1, \dots, \vec{x}_{H}\,|\, m_1, \dots , m_{H}) \\
&\notag \quad \times \Biggl\{ m_j\, \hat{u}(\vec{k}_3\; | \; m_j)\, \mathrm{e}^{-\I(\vec{k}_1+\vec{k}_2)\cdot \vec{x}_i-\I \vec{k}_3\cdot \vec{x}_j} \Big[ \expval{\Nc{a}\,(\Nc{a}-1)\,|\,m_i} + \expval{\Nc{a}\, \Ns{a}\,|\, m_i}\left(\hat{u}\rmg^a(\vec{k}_1\,|\,m_i) + \hat{u}\rmg^a(\vec{k}_2\,|\,m_i)\right) + \expval{\Ns{a}\, (\Ns{a}-1)\,|\,m_i}\Big]\\
&\notag \quad \quad + m_i\, \hat{u}(\vec{k}_3\; | \; m_i)\, \mathrm{e}^{-\I (\vec{k}_1+\vec{k}_3)\cdot \vec{x}_i - \I \vec{k}_2\cdot \vec{x}_j}\, \Big[\expval{\Nc{a}\,|\,m_i} + \expval{\Ns{a}\,|\,m_i}\, \hat{u}\rmg^a(\vec{k}_1\,|\,m_i)\Big]\, \Big[\expval{\Nc{a}\,|\,m_j} + \expval{\Ns{a}\,|\,m_j}\, \hat{u}\rmg^a(\vec{k}_2\,|\,m_j)\Big]  	\\
&\notag \quad \quad + m_i\, \hat{u}(\vec{k}_3\; | \; m_i)\, \mathrm{e}^{-\I (\vec{k}_2+\vec{k}_3)\cdot \vec{x}_i - \I \vec{k}_1\cdot \vec{x}_j}\, \Big[\expval{\Nc{a}\,|\,m_i} + \expval{\Ns{a}\,|\,m_i}\, \hat{u}\rmg^a(\vec{k}_2\,|\,m_i)\Big]\, \Big[\expval{\Nc{a}\,|\,m_j} + \expval{\Ns{a}\,|\,m_j}\, \hat{u}\rmg^a(\vec{k}_1\,|\,m_j)\Big] \Biggr\} \;.
\end{align}
In the limit of an infinite volume $V$, Eq.~\eqref{eq:large volume approximation}, the halo two-point correlation function $\xi_\mathrm{h}$ and its Fourier transform $P_\mathrm{h}$, gives
\begin{align}
	&\notag\int_V \dd[3]{x_1} \, \dots\, \dd[3]{x_H}\; {P_\textrm{c}(\vec{x}_1, \dots, \vec{x}_{H}\,|\, m_1, \dots , m_{H})}\, \mathrm{e}^{-\I(\vec{k}_1+\vec{k}_2)\, \vec{x}_i -\I \vec{k}_3 \cdot \vec{x}_j} \\
	&= \int_V \dd[3]{x_i} \int_V \dd[3]{x_j}\; \frac{1}{V^2} \left[1 + \xi_\mathrm{h}(|\vec{x}_i- \vec{x}_j|\,|\,m_1, m_2)\right] \,  \mathrm{e}^{-\I(\vec{k}_1+\vec{k}_2)\, \vec{x}_i -\I \vec{k}_3 \cdot \vec{x}_j}\\
	&=\frac{\bar{n}_\mathrm{h}^2}{H^2}\, \left[(2\pi)^6\,\dirac(\vec{k}_1+\vec{k}_2)\, \dirac(\vec{k}_3) + (2\pi)^3\, \dirac(\vec{k}_1+\vec{k}_2+\vec{k}_3)\, P_\mathrm{h}(|\vec{k}_1+\vec{k}_2|\,|\,m_1, m_2)\right]\;,
\end{align}
With this expression, neglecting unconnected terms proportional to $\dirac(\vec{k}_1)$ and $\dirac(\vec{k}_2)$ and the approximation $H(H-1)/H^2\simeq 1$, we find
\begin{align}
	&\notag	\Bggdaah{2}(\vec{k}_1, \vec{k}_2, \vec{k}_3)\, \\
	&\simeq \frac{1}{(\bar{n}^a\rmg)^2\, \bar{\rho}} \, \int_0^\infty\dd{m_1}\;\int_0^\infty\dd{m_2}\; n(m_1)\, n(m_2) \, \Biggl\{ m_2\, \hat{u}(\vec{k}_3\; | \; m_2)\, P_\mathrm{h}(\vec{k}_3\,|\,m_1, m_2) \Big[ \expval{\Nc{a}\,(\Nc{a}-1)\,|\,m_1} \\
	&\notag \quad \quad \quad + \expval{\Nc{a}\, \Ns{a}\,|\, m_1}\left(\hat{u}\rmg^a(\vec{k}_1\,|\,m_1) + \hat{u}\rmg^a(\vec{k}_2\,|\,m_1)\right) + \expval{\Ns{a}\, (\Ns{a}-1)\,|\,m_1}\, \hat{u}\rmg^a(\vec{k}_1\,|\,m_1)\, \hat{u}\rmg^a(\vec{k}_2\,|\,m_1)\Big]\\
	&\notag \quad \quad + m_1\, \hat{u}(\vec{k}_3\; | \; m_1)\, \Big[\expval{\Nc{a}\,|\,m_1} + \expval{\Ns{a}\,|\,m_1}\, \hat{u}\rmg^a(\vec{k}_1\,|\,m_1)\Big]\, \Big[\expval{\Nc{a}\,|\,m_2} + \expval{\Ns{a}\,|\,m_2}\, \hat{u}\rmg^a(\vec{k}_2\,|\,m_2)\Big]  	\\
	&\notag \quad \quad + m_1\, \hat{u}(\vec{k}_3\; | \; m_1)\, \Big[\expval{\Nc{a}\,|\,m_1} + \expval{\Ns{a}\,|\,m_1}\, \hat{u}\rmg^a(\vec{k}_2\,|\,m_1)\Big]\, \Big[\expval{\Nc{a}\,|\,m_2} + \expval{\Ns{a}\,|\,m_2}\, \hat{u}\rmg^a(\vec{k}_1\,|\,m_2)\Big] \Biggr\}\\
	&= \frac{1}{(\bar{n}^a\rmg)^2\, \bar{\rho}}\, \int_0^\infty\dd{m_1}\;\int_0^\infty\dd{m_2}\; n(m_1)\, n(m_2) \\
	&\notag \quad \times \Big[ m_2\, \hat{u}(\vec{k}_3\; | \; m_2)\, P_\mathrm{h}(\vec{k}_3\,|\,m_1, m_2)\, G^{aa}(\vec{k}_1, \vec{k}_2\,|\, m_1) + m_1\, \hat{u}(\vec{k}_3\; | \; m_1)\, G^a(\vec{k}_1\,|\, m_1)\, G^a(\vec{k}_2\,|\,m_2)  + m_1\, \hat{u}(\vec{k}_3\; | \; m_1)\, G^a(\vec{k}_1\,|\, m_2)\, G^a(\vec{k}_2\,|\,m_1) \Big]\;,
\end{align}
where again $\vec{k}_3=-\vec{k}_1-\vec{k}_2$ and
\begin{equation}
	\label{eq: def G_g}
	G^{a}(\vec{k}\,|\,m) = \expval{\Nc{a}\,|\,m}+ \expval{\Ns{a}\,|\,m} \hat{u}^a\rmg(\vec{k}\,|\,m)\;,
\end{equation}

For mixed pairs, $a\neq b$, a similar calculation yields
\begin{align}
	&\notag	\Bggdabh{2}(\vec{k}_1, \vec{k}_2, \vec{k}_3)\\
	&\simeq \frac{1}{(\bar{n}^a\rmg)^2\, \bar{\rho}} \, \int_0^\infty\dd{m_1}\;\int_0^\infty\dd{m_2}\; n(m_1)\, n(m_2) \\
	&\notag \quad \times \Big[ m_2\, \hat{u}(\vec{k}_3\; | \; m_2)\, P_\mathrm{h}(\vec{k}_3\,|\,m_1, m_2)\, G^{ab}(\vec{k}_1, \vec{k}_2\,|\, m_1) + m_1\, \hat{u}(\vec{k}_3\; | \; m_1)\, G^a(\vec{k}_1\,|\, m_1)\, G^b(\vec{k}_2\,|\,m_2)   + m_1\, \hat{u}(\vec{k}_3\; | \; m_1)\, G^a(\vec{k}_1\,|\, m_2)\, G^b(\vec{k}_2\,|\,m_1) \Big]\;.
\end{align}
In the main text we just denote $\Bggdabh{2}(\vec{k}_1, \vec{k}_2)$.

\subsection{3-halo term}
The 3-halo term for unmixed pairs $a=b$ is given by
\begin{align}
	&\notag	(2\pi)^3 \Bggdaah{3}(\vec{k}_1, \vec{k}_2, \vec{k}_3)\, \dirac(\vec{k}_1+\vec{k}_2+\vec{k}_3) + \text{unconnected terms}\\
	&= \frac{1}{\bar{n}_\mathrm{h}^H\,(\bar{n}^a\rmg)^2\, \bar{\rho}} \sum_{\substack{i,j,k=1, j\neq i,\\ k\neq i,j}}^H \int_0^\infty\dd{m_1}\; \dots \int_0^\infty\dd{m_H}\; n(m_1)\dots n(m_H) \, \int _V \dd[3]{x_1}\; \dots \int_V \dd[3]{x_{H}}\; P_\textrm{c}(\vec{x}_1, \dots, \vec{x}_{H}\,|\, m_1, \dots , m_{H}) \\
	&\notag \quad \times \prod_{h=1}^{H}\Biggl\{ \sum_{\Ncc{a}{h}}^\infty  \sum_{\Nss{a}{h}}^\infty {P_\mathrm{N}(\Ncc{a}{h}, \Nss{a}{h}\,|\,m_h)}\, \prod_{v=1}^{\Nss{a}{h}} \left[ \int \dd[3]{\Delta {x}^{a}_{hv}}\, u\rmg^{a}(\Delta \vec{x}^{a}_{hv}\,|\,m_h)  \right]\Biggr\}\, m_k\, \hat{u}(\vec{k}_3\,|\,m_k)\, \mathrm{e}^{-\I \vec{k}_3\cdot \vec{x}_k}\, \\
	&\notag \quad \times\Big[\Ncc{a}{i}\, \mathrm{e}^{-\I\vec{k}_1\cdot \vec{x}_i} + \sum_{l=1}^{\Nss{a}{i}}\mathrm{e}^{-\I\vec{k}_1(\vec{x}_i + \Delta\vec{x}^a_{il})}\Big] \Big[\Ncc{a}{j}\, \mathrm{e}^{-\I\vec{k}_2\cdot \vec{x}_j} + \sum_{m=1}^{\Nss{a}{j}}\mathrm{e}^{-\I\vec{k}_2(\vec{x}_j + \Delta\vec{x}^a_{jm})}\Big]\;.
\end{align}

We evaluate all $m$-integrals independent of $m_i, m_j$, and $m_k$ ($H(H-1)(H-2)$ terms in total), and evaluate the $\vec{\Delta x}$ integrals. This leads to
\begin{align}
	&\notag \Bggdaah{3}(\vec{k}_1, \vec{k}_2, \vec{k}_3) + \text{unconnected terms}\\
	&= \frac{1}{\bar{n}_\mathrm{h}^3\,(\bar{n}^a\rmg)^2\, \bar{\rho}} \sum_{\substack{i=1, j\neq i,\\ k\neq i,j}}^H \int_0^\infty\dd{m_i} \int_0^\infty\dd{m_j}\int_0^\infty\dd{m_k}\; n(m_i)\, n(m_j)\, n(m_k) \, \int _V \dd[3]{x_1}\; \dots \int_V \dd[3]{x_{H}}\; P_\textrm{c}(\vec{x}_1, \dots, \vec{x}_{H}\,|\, m_1, \dots , m_{H}) \\
	&\notag\quad \times m_k\, \hat{u}(\vec{k}_3\,|\,m_k)\, \Big[\expval{\Nc{a}\,|\,m_i} + \expval{\Ns{a}\,|\,m_i}\, \hat{u}\rmg^a(\vec{k}_1\,|\,m_i)\Big] \Big[\expval{\Nc{a}\,|\,m_j} + \expval{\Ns{a}\,|\,m_j}\, \hat{u}\rmg^a(\vec{k}_2\,|\,m_j)\Big]\, \mathrm{e}^{-\I \vec{k}_1\cdot \vec{x}_i -\I \vec{k}_2\cdot \vec{x}_j -\I \vec{k}_3\cdot \vec{x}_k}  \;.
\end{align}
Now, again in the limit of an infinite volume $V$, Eq.~\eqref{eq:large volume approximation}, we use the two- and three-point functions  $\xi_\mathrm{h}$ and $\zeta_\mathrm{h}$ of halo clustering and their Fourier transforms $P_\mathrm{h}$ and $B_ \mathrm{h}$ to find
	\begin{align}
	&\notag\int_V \dd[3]{x_1} \, \dots\, \dd[3]{x_H}\; {P_\textrm{c}(\vec{x}_1, \dots, \vec{x}_{H}\,|\, m_1, \dots , m_{H})}\, \mathrm{e}^{-\I \vec{k}_1\cdot \vec{x}_i -\I \vec{k}_2\cdot \vec{x}_j -\I \vec{k}_3\cdot \vec{x}_k}\\
	&= \int_V \dd[3]{x_i} \int_V \dd[3]{x_j} \int_V \dd[3]{x_k}\; \frac{\mathrm{e}^{-\I \vec{k}_1\cdot \vec{x}_i -\I \vec{k}_2\cdot \vec{x}_j -\I \vec{k}_3\cdot \vec{x}_k}}{V^3} \left[1 + \xi_\mathrm{h}(|\vec{x}_i- \vec{x}_j| \,|\, m_1, m_2) + \xi_\mathrm{h}(|\vec{x}_i- \vec{x}_k| \,|\, m_1, m_3) + \xi_\mathrm{h}(|\vec{x}_j- \vec{x}_k| \,|\, m_2, m_3)\right.\\
	&\notag \quad \quad \left. + \zeta_\mathrm{h}(|\vec{x}_i-\vec{x}_j|, |\vec{x}_i-\vec{x}_k| , |\vec{x}_j-\vec{x}_k| \,|\, m_1, m_2, m_3)\right] \\
	&= \frac{\bar{n}_\mathrm{h}^3}{H^3}\, \big[(2\pi)^9\,\dirac(\vec{k}_1)\, \dirac(\vec{k}_2)\, \dirac(\vec{k}_3) +  (2\pi)^6\,\dirac(\vec{k}_2)\, \dirac(\vec{k}_1+\vec{k}_3)\, P_\mathrm{h}(|\,\vec{k}_1+\vec{k}_3\,|\, \,|\, m_1, m_3) \\
	&\quad \quad  \notag + (2\pi)^6\,\dirac(\vec{k}_3)\, \dirac(\vec{k}_1+\vec{k}_2)\, P_\mathrm{h}(|\,\vec{k}_1+\vec{k}_2\,|\, \,|\, m_1, m_2) + (2\pi)^6\,\dirac(\vec{k}_1)\, \dirac(\vec{k}_2+\vec{k}_3)\, P_\mathrm{h}(|\,\vec{k}_2+\vec{k}_3\,|\, \,|\, m_2, m_3)\\
	&\quad \quad \notag +
	(2\pi)^3\,\dirac(\vec{k}_1+\vec{k}_2+\vec{k}_3)\, B_\mathrm{h}(\vec{k}_1, \vec{k}_2, \vec{k}_3 \,|\, m_1, m_2, m_3)\big]\;,
\end{align}
As before, all terms not proportional to $\dirac(\vec{k}_1+\vec{k}_2+\vec{k}_3)$ are unconnected and therefore neglected. Using this expression and the approximation $H(H-1)(H-2)/H^3 \simeq 1$, we obtain
\begin{align}
	&\notag\Bggdaah{3}(\vec{k}_1, \vec{k}_2, \vec{k}_3) \\
	&\simeq \frac{1}{(\bar{n}^a\rmg)^2\, \bar{\rho}} \, \int_0^\infty\dd{m_1}\int_0^\infty\dd{m_2}\int_0^\infty\dd{m_3}\; n(m_1)\, n(m_2)\, n(m_3)\, B_\mathrm{h}(\vec{k}_1, \vec{k}_2, \vec{k}_3\,|\, m_1, m_2, m_3)\, m_3\, \hat{u}(\vec{k}_3\; | \; m_3) \\
	&\notag \quad \times   \Big[\expval{\Nc{a}\,|\,m_1} + \expval{\Ns{a}\,|\,m_1}\, \hat{u}\rmg^a(\vec{k}_1\,|\,m_1)\Big]\, \Big[\expval{\Nc{a}\,|\,m_2} + \expval{\Ns{a}\,|\,m_2}\, \hat{u}\rmg^a(\vec{k}_2\,|\,m_2)\Big] \\
	&= \frac{1}{(\bar{n}^a\rmg)^2\, \bar{\rho}} \, \int_0^\infty\dd{m_1}\int_0^\infty\dd{m_2}\int_0^\infty\dd{m_3}\; n(m_1)\, n(m_2)\, n(m_3)\, B_\mathrm{h}(\vec{k}_1, \vec{k}_2, \vec{k}_3\,|\, m_1, m_2, m_3)\, m_3\, \hat{u}(\vec{k}_3\; | \; m_3)\, G^a(\vec{k}_1\,|\,m_1) \, G^a(\vec{k}_2\,|\,m_2)\;.
\end{align}
For mixed pairs $a\neq b$ a similar calculation yields
\begin{align}
	&\notag	\Bggdabh{3}(\vec{k}_1, \vec{k}_2, \vec{k}_3) \\
	&\simeq \frac{1}{\bar{n}^a\rmg\,\bar{n}^b\rmg\, \bar{\rho}} \, \int_0^\infty\dd{m_1}\int_0^\infty\dd{m_2}\int_0^\infty\dd{m_3}\; n(m_1)\, n(m_2)\, n(m_3)\, B_\mathrm{h}(\vec{k}_1, \vec{k}_2, \vec{k}_3\,|\, m_1, m_2, m_3)\, m_3\, \hat{u}(\vec{k}_3\; | \; m_3)\, G^a(\vec{k}_1\,|\,m_1) \, G^b(\vec{k}_2\,|\,m_2)\;.
\end{align}
Again, we just denote $\Bggdabh{3}(\vec{k}_1, \vec{k}_2)$ in the main text.

\section{Poissonianity of satellite galaxies}
\label{sec:superPoissonian}

\begin{figure}
	\includegraphics[width=0.49\linewidth]{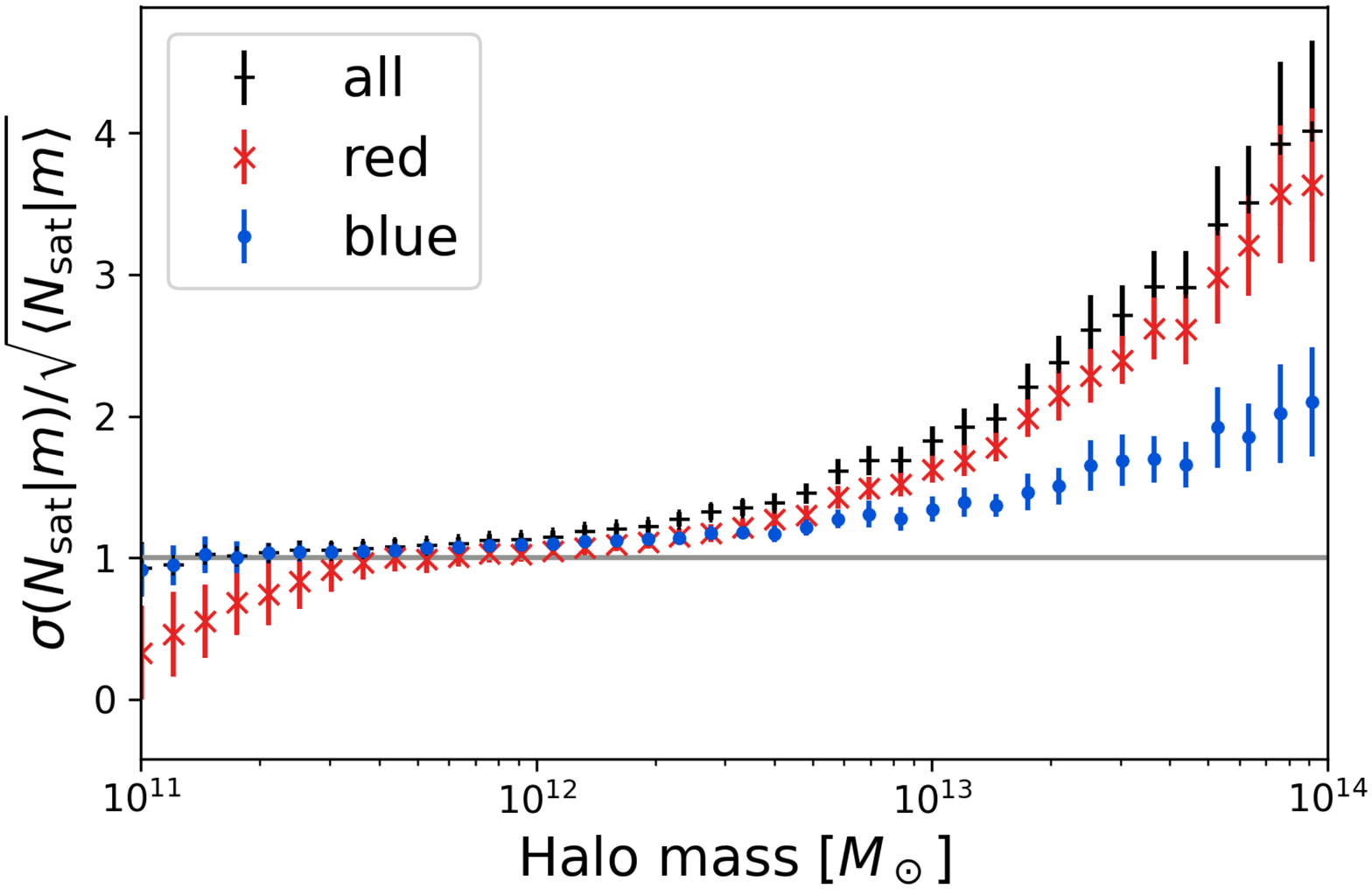}
	\includegraphics[width=0.49\linewidth]{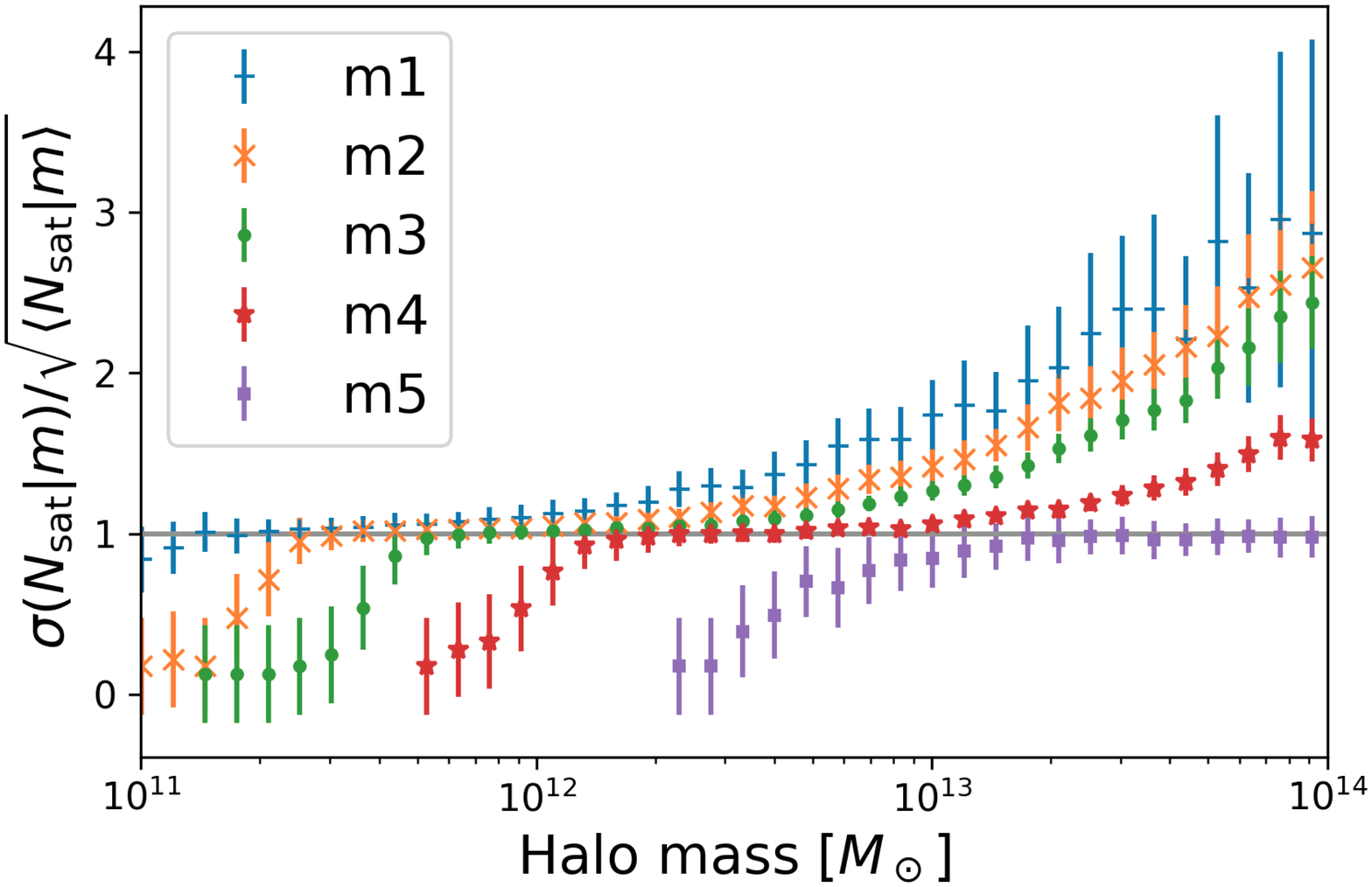}	
	\caption{Ratio of the standard deviation of the galaxy satellite number per halo and the Poisson variance. For Poissonian satellites this ratio is unity. \textit{Left:} Ratio for all simulated galaxies (black crosses), red simulated galaxies (red crosses), and blue simulated galaxies (blue dots). \textit{Right:} Ratio for stellar mass-selected samples.}
	\label{fig:Poisson}
\end{figure}

Our HOD model explicitly assumes Poisson satellites -- satellite numbers inside halos that vary according to a Poisson statistic. In this section, we explore the bias due to this assumption in the presence of non-Poisson satellites. To some degree, deviations from Poisson satellites are indeed visible in our SAM galaxies: Figure~\ref{fig:Poisson} shows the ratio of $\sigma(\Ns{a}|m)$ to the Poisson variance $\sqrt{\expval{\Ns{a}|m}}$ for different samples of the simulated galaxies. Poisson satellites, where this ratio is unity, are present in the intermediate halo-mass range, such as between $2\times 10^{11} \Msun\lesssim m\lesssim 10^{12} \Msun$ for galaxies from stellar-mass bin m2. On the low-mass end, satellites are distributed sub-Poissonian (ratio is below unity), and on the high-mass end they are super-Poissonian (ratio exceeds unity). The trend is similar across all galaxy populations, although with a shift of the intermediate range depending on the stellar mass of the sample.

We estimate the bias induced by the mismatch between the true satellite variance and the model assumption by computing the fractional change of $\NNM{a}{b}_\mathrm{mod}$ when switching from Poisson satellites to the true variance in the SAM. This means: For each best-fit to the simulated G3L signal, we calculate new model predictions $\NNM{a}{b}_\mathrm{mod}$, using the HOD parameters from the best-fit but updating the number of satellite pairs inside a halo to
\begin{equation}
    \expval{\Ns{a}(\Ns{a}-1)|m} = \left(1-\frac{\sigma^2(\Ns{a}|m)}{\expval{\Ns{a}|m}}\right) \expval{\Ns{a}|m} + \expval{\Ns{a}|m}^2\;,
\end{equation}
where $\sigma^2(\Ns{a}|m)$ is the true satellite variance in the simulation. Clearly, for $\sigma(\Ns{a}|m) = \sqrt{\expval{\Ns{a}|m}}$, i.e., Poisson satellites, this reduces to Eq.~\eqref{eq:HOD_NsatNsat_same}. 

\begin{figure*}
    \centering
    \includegraphics[width=0.49\linewidth]{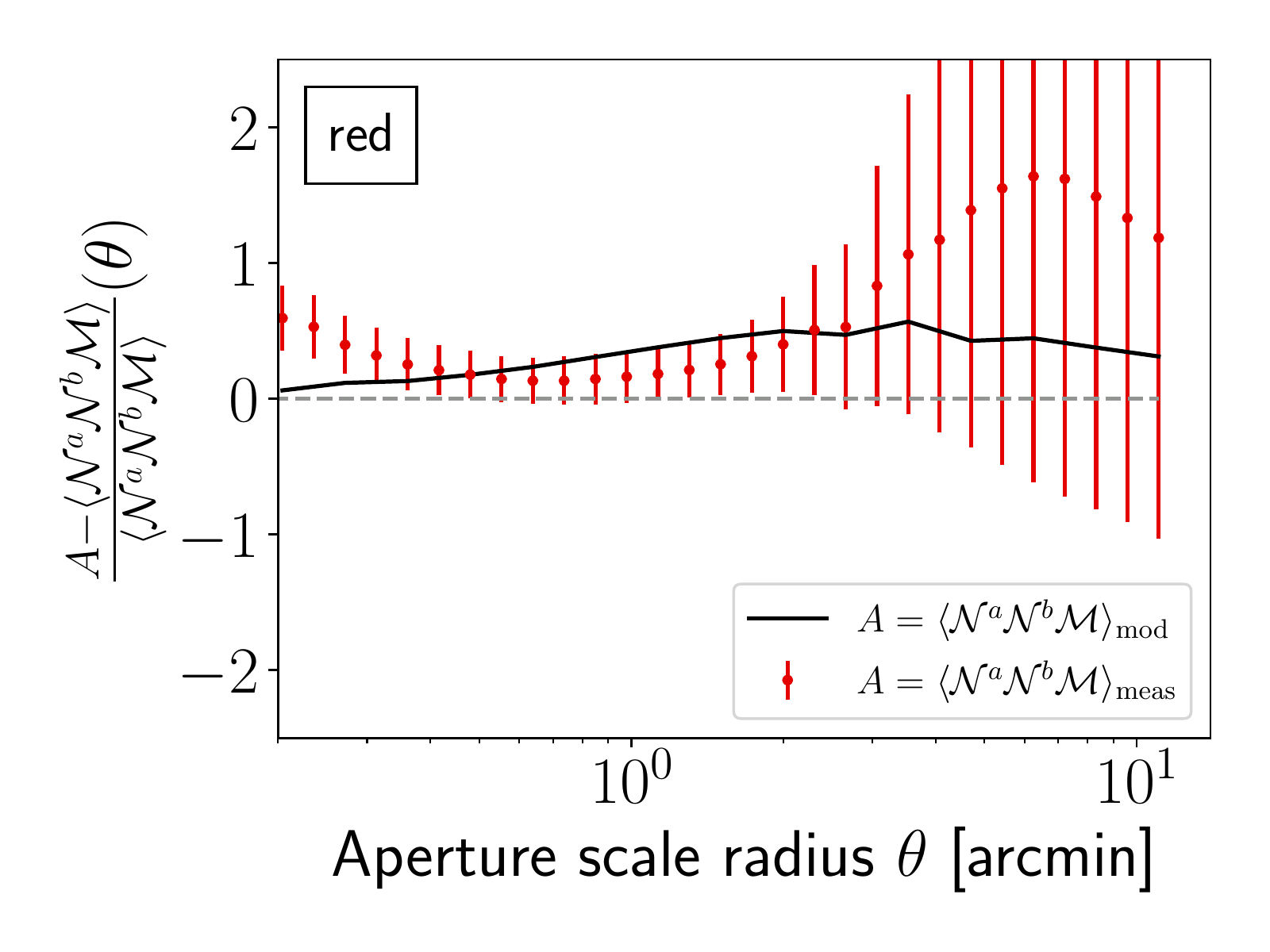}
    \includegraphics[width=0.49\linewidth]{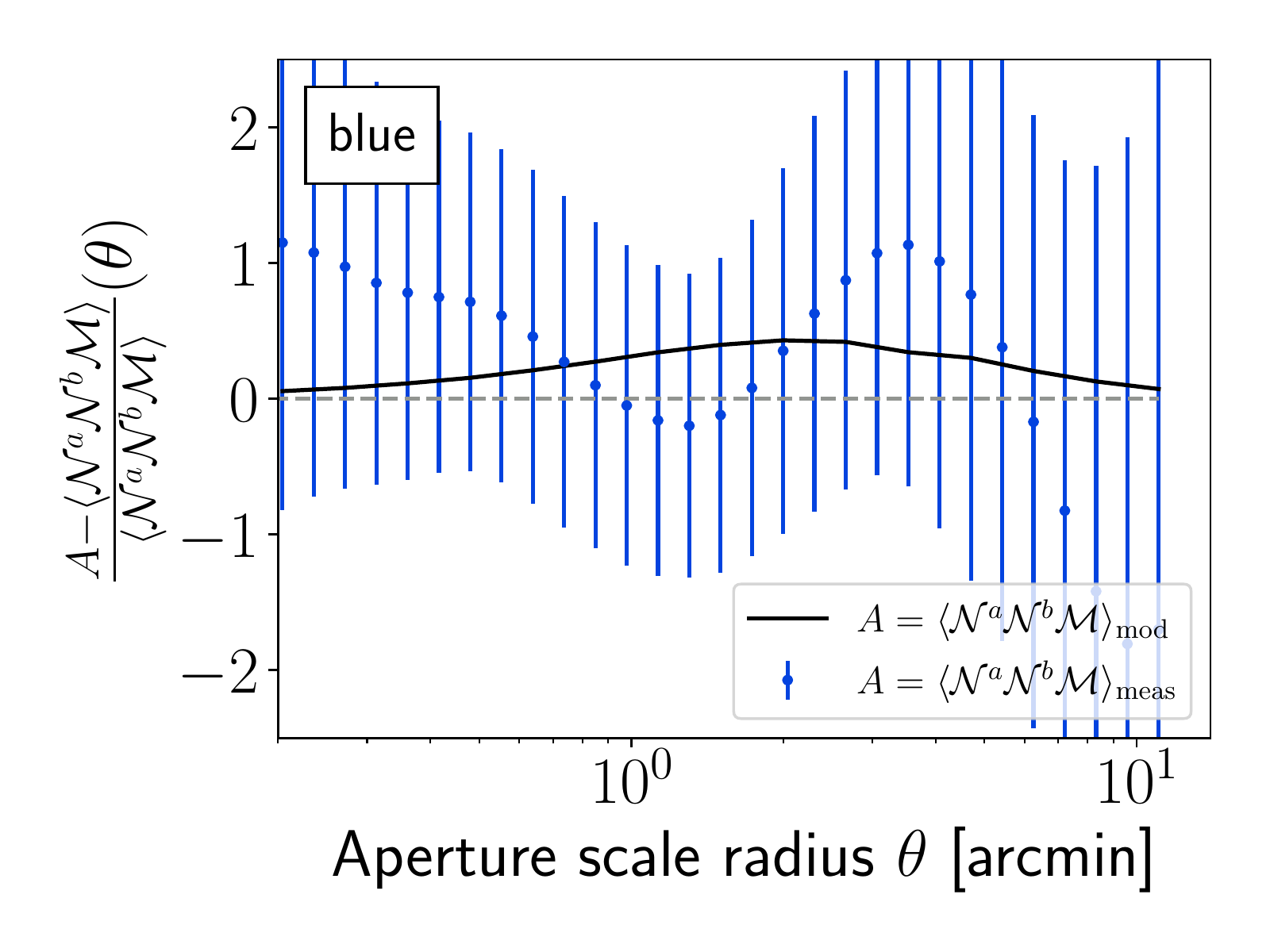}
    \caption{Fractional difference of aperture statistics measurement ${\NNM{a}{b}}_\mathrm{meas}$ (points) and modified aperture statistics model ${\NNM{a}{b}}_\mathrm{mod}$(lines) to the original model ${\NNM{a}{b}}$ for red-red lens pairs (left) and blue-blue lens pairs (right) in the MS. The original model assumes a Poissonian satellite distribution, while the modified model uses the directly measured distribution of satellite galaxies.}
    \label{fig:rescaled_ratio_colours}
\end{figure*}

Figure~\ref{fig:rescaled_ratio_colours} shows the fractional change between the updated model and the original best-fit with Poisson satellites for the red and blue galaxies. Also shown is the fractional difference of the measured aperture statistics to the best fit. For red galaxies, the updated model differs from the original strongest at \ang[astroang]{;4;} by roughly $50\%$. For blue galaxies, the difference is strongest at \ang[astroang]{;2;} and has a similar magnitude. It is to be expected that the aperture statistics at scales between \ang[astroang]{;1;} and \ang[astroang]{;10;} are most affected because the 1-halo term, containing $\expval{\Ns{a}(\Ns{a}-1)|m}$, dominates here. Smaller scales are dominated by the halo term containing $\expval{\Nc{a}\Ns{a}|m}$, whereas larger scales are dominated by the 3-halo term. Although a bias is visible by the solid line in Fig.~\ref{fig:rescaled_ratio_colours}, it is of the order of or smaller than the uncertainties on the aperture statistics measurements (error bars). In particular, for blue galaxies, the measurements cannot discriminate between the two models. Therefore, while the presence of non-Poisson satellites affects the model prediction, a Poisson model is accurate enough for the G3L analysis of measurements in this work.

We have repeated this test also for the satellite galaxies from the stellar mass-selected samples. Galaxies from samples m4 and m5 are closer to a Poisson satellite model than blue galaxies and therefore show less bias (below $10\%$). For the other samples, the bias is larger (up to $61\%$ for m1, $57\%$ for m2, $19\%$ for m3 satellites), but since the measured aperture statistics have larger uncertainties, the bias is of even less relevance than for red and blue galaxies. 

Finally, a deviation of the variance in satellite numbers from a Poisson statistic also affects the interpretation of the correlation parameter $r^{ab}$, as defined in Eq.~\eqref{eq:HOD_NsatNsat}. For Poisson satellites, $r^{ab}$ corresponds to a Pearson coefficient, i.e., $r^{ab}=1$ for perfectly correlated $\Ns{a}$ and $\Ns{b}$, $r^{ab}=-1$ for perfectly anti-correlated satellite numbers, and $r^{ab}=0$ for uncorrelated samples.  However, for a super-Poisson variance in the high-mass regime, our $r^{ab}$ is larger than the Pearson coefficient. Conversely, for low-mass halos with sub-Poisson variance, $r^{ab}$ is smaller than the Pearson coefficient. This in combination increases the slope $\epsilon^{ab}$ of our $r^{ab}(m)$ compared to a Pearson definition of correlation. Nevertheless, the sign of $r^{ab}$ remains unchanged, and if the samples $a$ and $b$ are uncorrelated, both $r^{ab}$ and the Pearson coefficient vanish.

\section{Results of model fit to stellar mass-selected galaxies}
\label{app:results}

In Fig \ref{fig:fitresults_masses_ms} we show the model fits to the G3L aperture statistics from stellar mass-selected lenses in the MS and the $\chi^2$ of the fits. All fits have 76 degrees-of-freedom (the same as the fit to the colour-selected samples in Sect.~\ref{sec:results:colour}), so a $\chi^2<97.4$ signifies agreement between fit and data at the $95\%$ CL. The HOD parameters of the best-fitting models are given in Table~\ref{tab:params_MS_sm}; the parameters $A^{ab}$ and $\epsilon^{ab}$ that determine the cross-correlation between satellite numbers together with the $\chi^2$ and $p$-values of the fits are given in Table~\ref{tab:params_MS_sm_chisq}. Figure \ref{fig:fitresults_masses_KV450} shows the model fits to the G3L aperture statistics from stellar mass-selected lenses in KV450$\times$ GAMA and the $\chi^2$ of the fits. The parameters of the best-fitting models are given in Tables~\ref{tab:params_KV450_sm} and ~\ref{tab:params_sm_kv450_chisq}.

\begin{center}
	\includegraphics[width=\textwidth, trim={5cm 0 5cm 2cm}, clip]{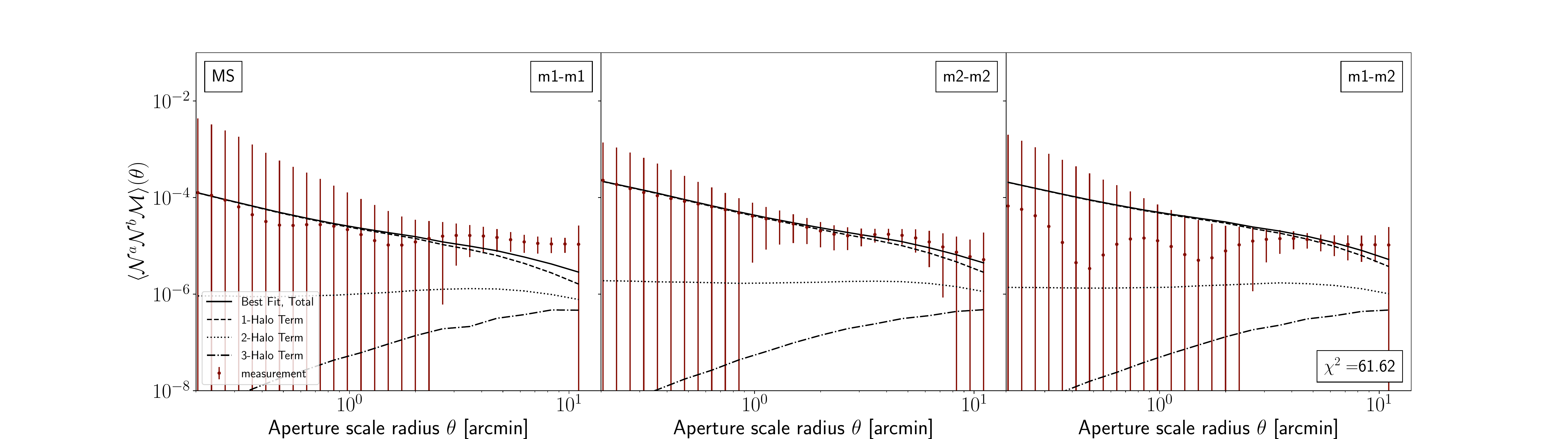}
	\includegraphics[width=\textwidth, trim={5cm 0 5cm 2cm}, clip]{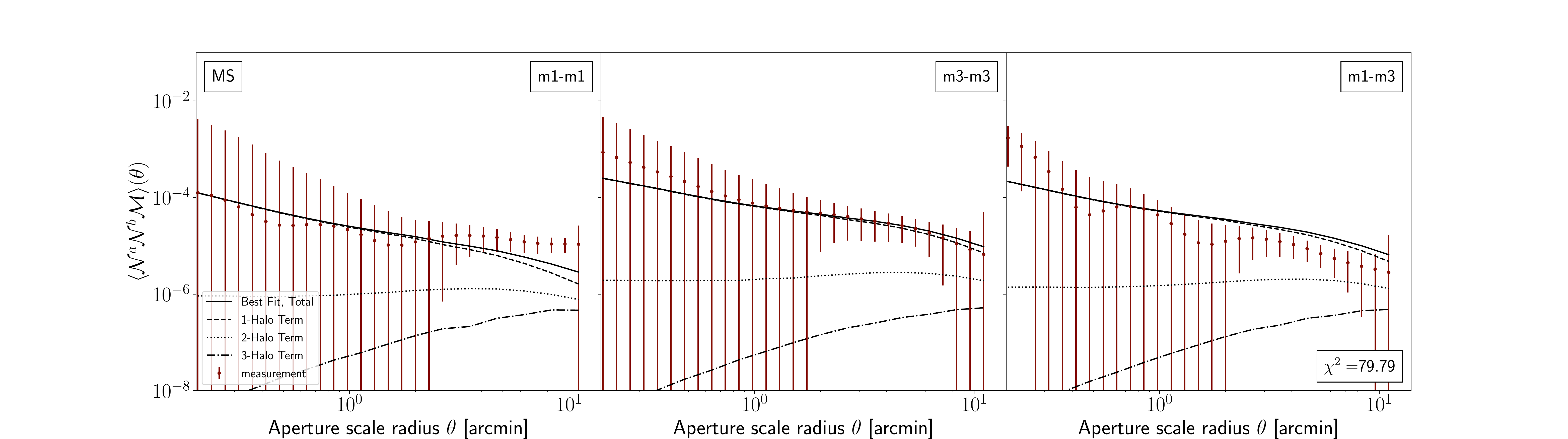}
	\includegraphics[width=\textwidth, trim={5cm 0 5cm 2cm}, clip]{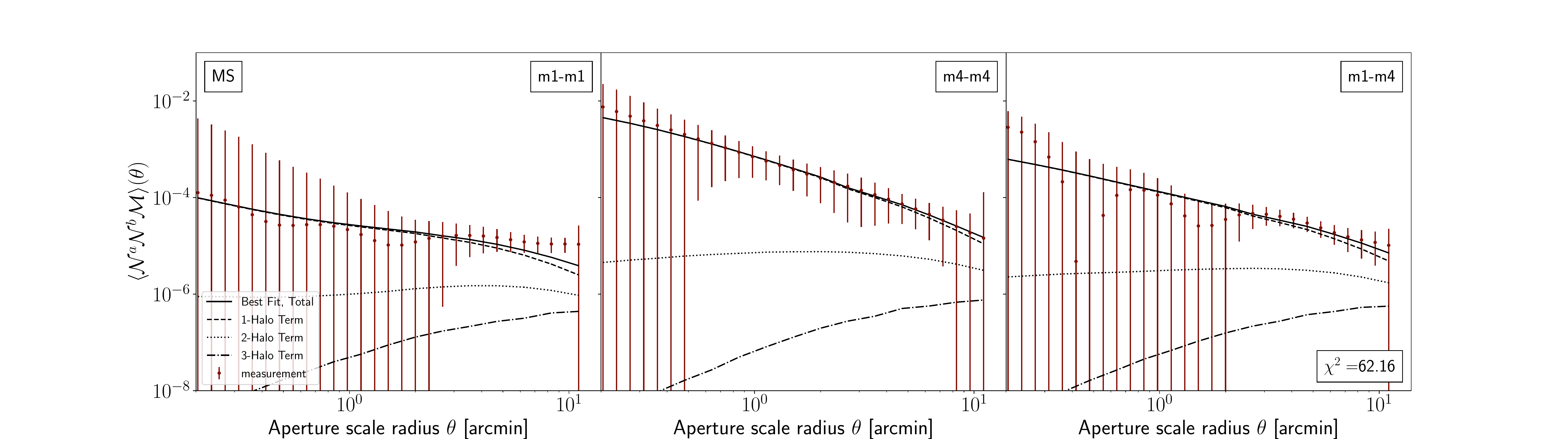}
	\includegraphics[width=\textwidth, trim={5cm 0 5cm 2cm}, clip]{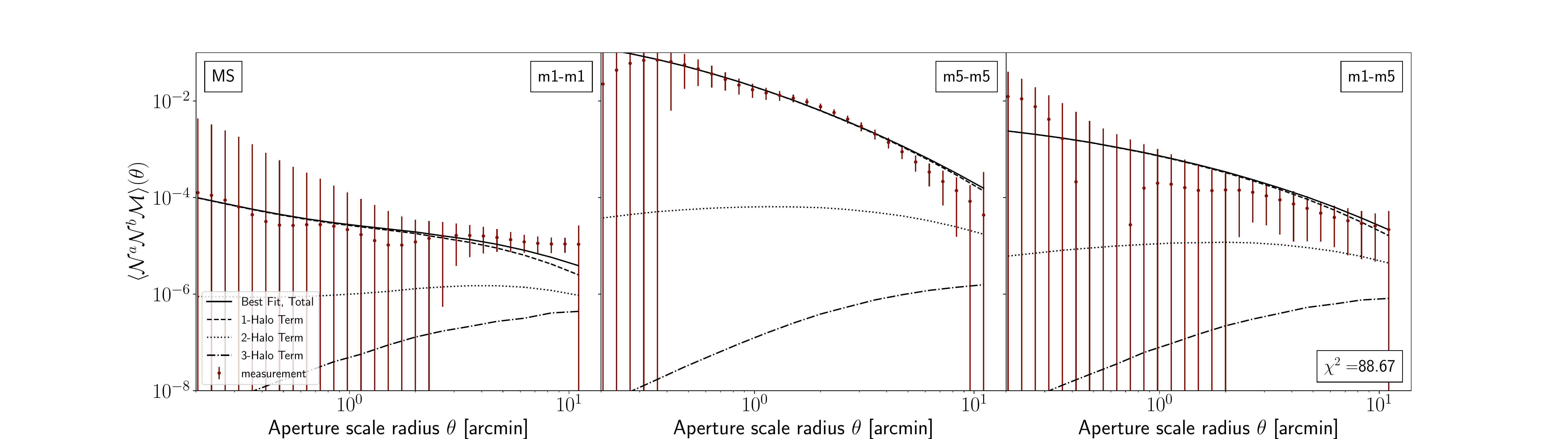}
	\includegraphics[width=\textwidth, trim={5cm 0 5cm 2cm}, clip]{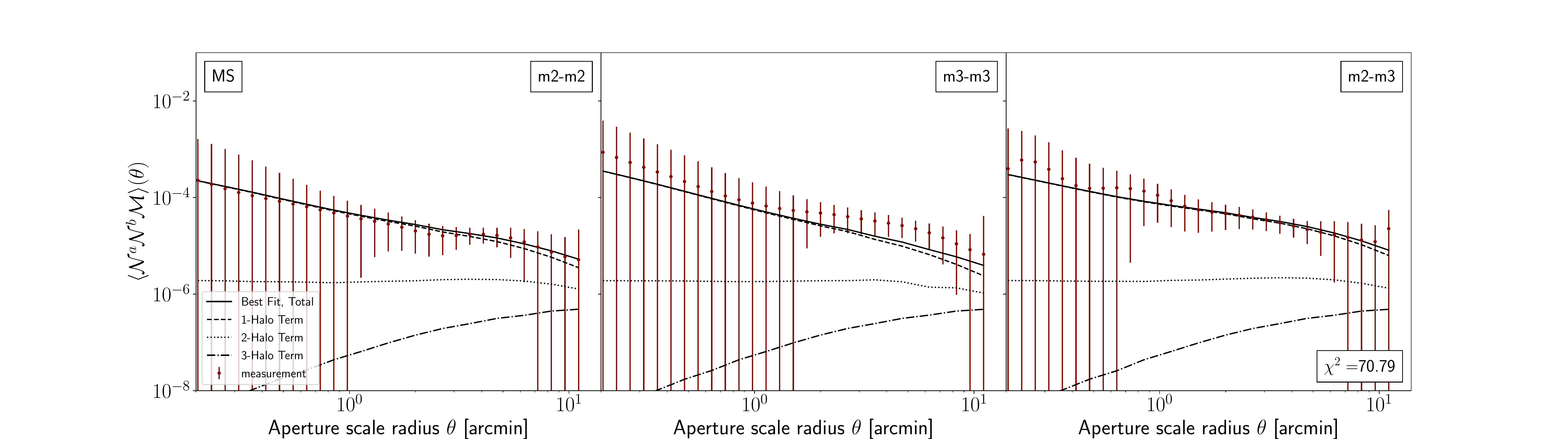}
	\includegraphics[width=\textwidth, trim={5cm 0 5cm 2cm}, clip]{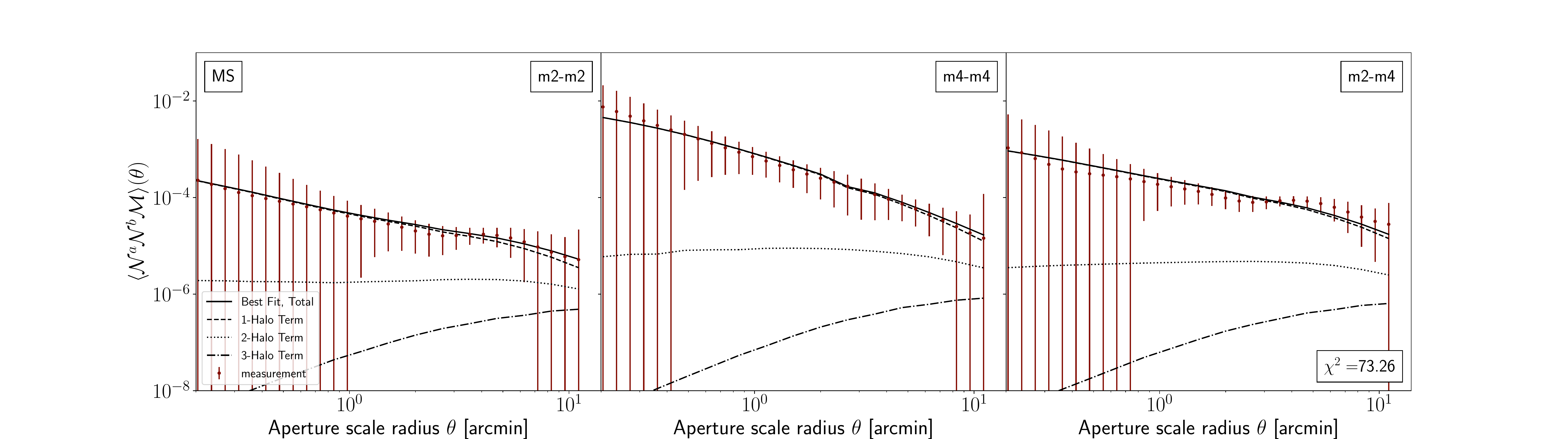}
	\includegraphics[width=\textwidth, trim={5cm 0 5cm 2cm}, clip]{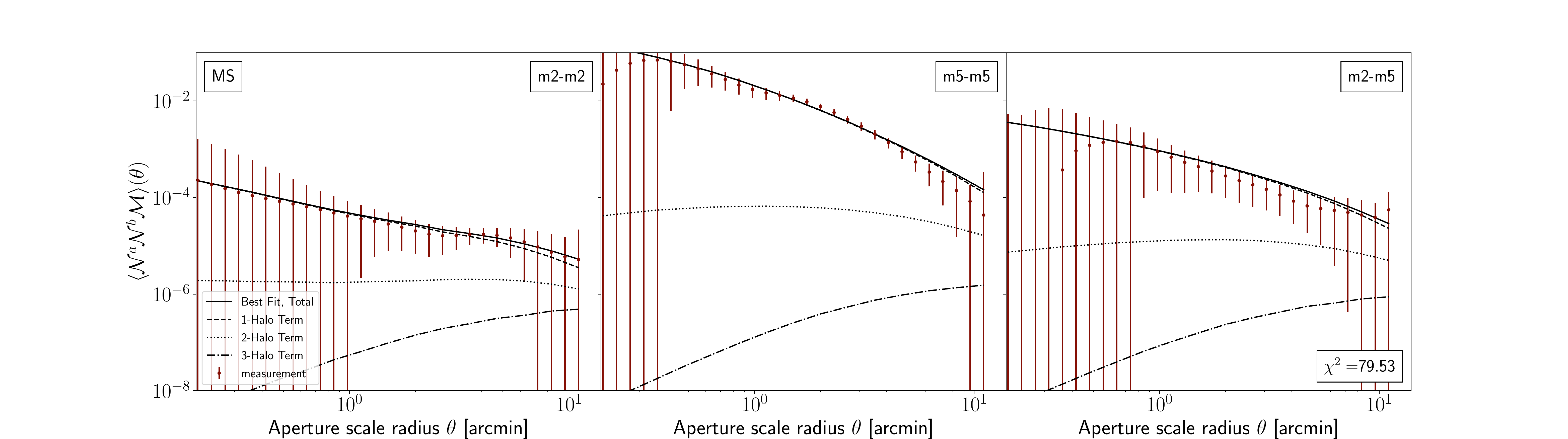}
	\includegraphics[width=\textwidth, trim={5cm 0 5cm 2cm}, clip]{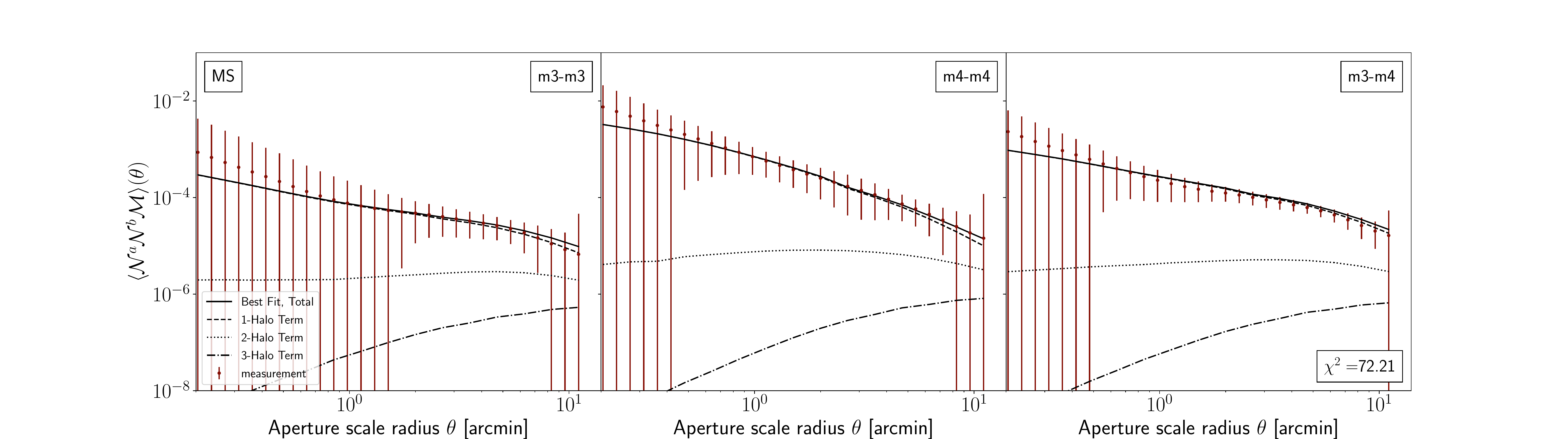}
	\includegraphics[width=\textwidth, trim={5cm 0 5cm 2cm}, clip]{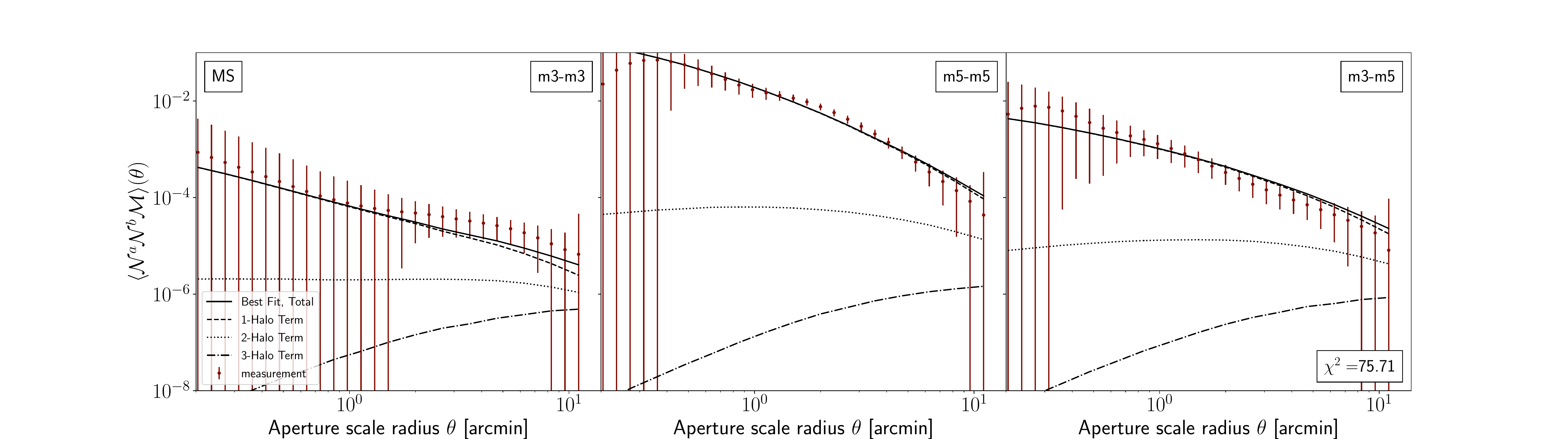}
	\includegraphics[width=\textwidth, trim={5cm 0 5cm 2cm}, clip]{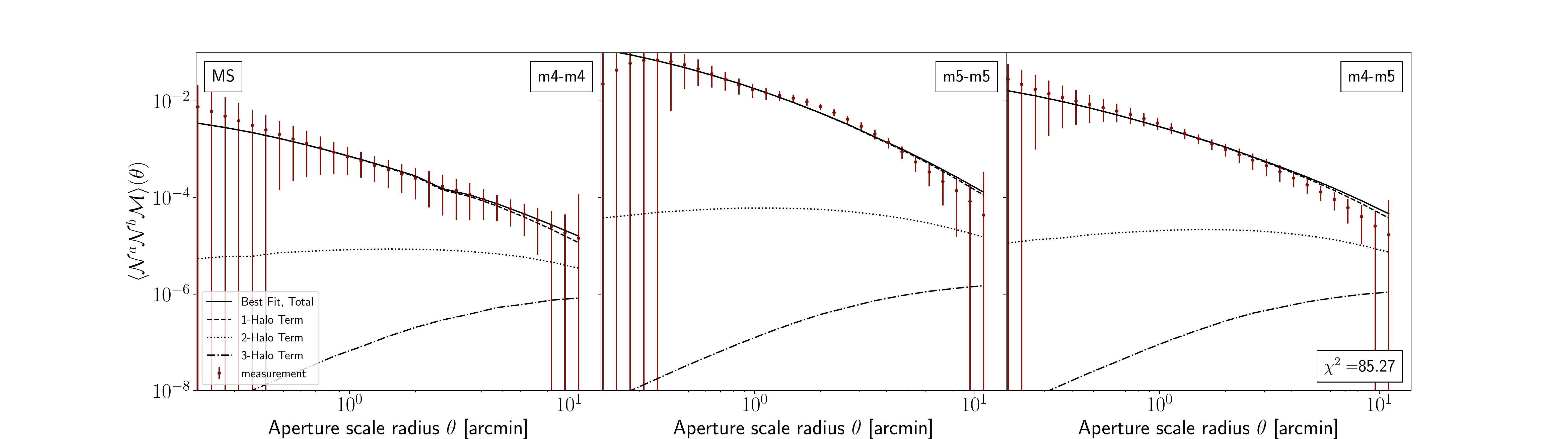}
	\captionof{figure}{{G3L} measurement in MS (points) and best fitting halo model (lines) for stellar mass-selected lens samples, as defined in Table~\ref{tab:sub-samples}. Solid lines indicate the total aperture statistics, dashed lines the 1-halo, dotted lines the 2-halo, and dash-dotted lines the 3-halo term of the fit. Each row was fitted individually, leading to the $\chi^2$ values in the last panel. The corresponding halo model parameters are given in Table~\ref{tab:params_MS_sm}.}
	\label{fig:fitresults_masses_ms}
\end{center}

\begin{table}[p]
	\caption{Best-fit values for halo model parameters for stellar-mass-selected lenses in MS for each stellar mass sample $a$.}
	\centering
	\label{tab:params_MS_sm}
	\begin{tabularx}{\textwidth}{llXXXXX}
		\hline
		                        &                                             & $b=$m1                       & $b=$m2                   & $b=$m3                      & $b=$m4                      & $b=$m5                                 \\ \hline
		\multirow{9}{*}{$a=$m1} & $\alpha^a$                                  &                              & $0.05\pma{0.04}{0.08}$   & $0.06\pma{0.05}{0.09} $     & $ 0.04\pma{0.03}{0.03}    $ & $ 0.10\pma{0.09}{0.10}  $              \\
		                        & $\sigma^a$                                  &                              & $0.57\pma{0.49}{0.48} $  & $0.53\pma{0.43}{0.41}     $ & $ 0.51\pma{0.38}{0.48}   $  & $ 0.51\pma{0.43}{0.44}   $             \\
		                        & $M_\mathrm{th}^a [10^{11}\mathrm{M}_\odot]$ &                              & $1.3\pma{1.1}{1.3}     $ & $1.9\pma{1.0}{1.4}    $     & $ 1.2\pma{0.8}{3.4}       $ & $ 1.2\pma{1.5}{2.0}       $            \\
		                        & $\beta^a$                                   &                              & $0.99\pma{0.15}{0.10}$   & $0.97\pma{0.19}{0.11} $     & $ 1.09\pma{0.10}{0.12}  $   & $ 1.09\pma{0.13}{0.17}   $             \\
		                        & $M^{\prime a} [10^{13}\Msun]$               &                              & $21.4 \pma{1.0}{1.2}$    & $21.5\pma{1.3}{1.8}       $ & $ 20.4\pma{2.7}{2.8}   $    & $ 20.5\pma{1.5}{1.1}                $  \\
		                        & $f^a$                                       &                              & $1.02\pma{0.34}{0.35}  $ & $1.01\pma{0.31}{0.46}$      & $ 1.09\pma{0.54}{0.60}$     & $ 1.11\pma{0.31}{0.33}               $ \\
		\hline
		\multirow{9}{*}{$a=$m2} & $\alpha^a$                                  & $0.11 \pma{0.10}{0.09}$      &                          & $0.10 \pma{0.12}{0.14}$     & $0.10 \pma{0.08}{0.08}$     & $0.13 \pma{0.14}{0.09}$                \\
		                        & $\sigma^a$                                  & $0.45 \pma{0.43}{0.42} $     &                          & $0.44 \pma{0.41}{0.44}$     & $0.45 \pma{0.43}{0.53}$     & $0.47 \pma{0.41}{0.48}$                \\
		                        & $M_\mathrm{th}^a [10^{11}\mathrm{M}_\odot]$ & $5.1 \pma{1.5}{2.1}    $     &                          & $4.2 \pma{2.8}{3.1} $       & $4.2 \pma{2.6}{2.8}$        & $4.5 \pma{1.2}{2.3}$                   \\
		                        & $\beta^a$                                   & $1.09 \pma{0.15}{0.09}  $    &                          & $1.11 \pma{0.09}{0.07}$     & $1.11 \pma{0.10}{0.14}$     & $1.21 \pma{0.11}{0.09}$                \\
		                        & $M^{\prime a} [10^{13}\Msun]$               & $20.0 \pma{1.3}{1.4}      $  &                          & $21.5 \pma{1.4}{1.1} $      & $21.3 \pma{1.1}{0.5}$       & $25.1 \pma{1.7}{1.1}$                  \\
		                        & $f^a$                                       & $0.77 \pma{0.13}{0.13}     $ &                          & $0.81 \pma{0.17}{0.09}$     & $0.79 \pma{0.09}{0.13}$     & $0.84 \pma{0.19}{0.17}$                \\
		 \hline
		\multirow{9}{*}{$a=$m3} & $\alpha^a$                                  & $0.22 \pma{0.13}{0.09}   $   & $0.22 \pma{0.08}{0.11} $ &                             & $0.19 \pma{0.11}{0.10}$     & $0.23 \pma{0.11}{0.12}$                \\
		                        & $\sigma^a$                                  & $0.45 \pma{0.38}{0.47}$      & $0.44 \pma{0.44}{0.49}$  &                             & $0.46 \pma{0.37}{0.51}$     & $0.48 \pma{0.33}{0.45}$                \\
		                        & $M_\mathrm{th}^a [10^{11}\mathrm{M}_\odot]$ & $4.4 \pma{2.2}{1.6}$         & $5.4 \pma{1.4}{1.5}$     &                             & $4.7\pma{1.8}{1.3}$         & $4.9 \pma{1.5}{1.6}$                   \\
		                        & $\beta^a$                                   & $1.18 \pma{0.09}{0.13}$      & $1.11 \pma{0.12}{0.16}$  &                             & $1.15 \pma{0.06}{0.11}$     & $0.97 \pma{0.19}{0.12}$                \\
		                        & $M^{\prime a} [10^{13}\Msun]$               & $10.0 \pma{0.6}{1.1}$        & $9.4 \pma{0.5}{1.2}$     &                             & $10.2  \pma{0.8}{0.7}$      & $9.6 \pma{0.8}{1.1}$                   \\
		                        & $f^a$                                       & $0.87 \pma{0.22}{0.45}$      & $0.94 \pma{0.27}{0.39}$  &                             & $0.83 \pma{0.19}{0.25}$     & $0.92 \pma{0.24}{0.23}$                \\
		 \hline
		\multirow{9}{*}{$a=$m4} & $\alpha^a$                                  & $0.52 \pma{0.12}{0.30}$      & $0.56 \pma{0.09}{0.12} $ & $0.37 \pma{0.09}{0.11}$     &                             & $0.53 \pma{0.09}{0.18}$                \\
		                        & $\sigma^a$                                  & $0.65 \pma{0.51}{0.31}$      & $0.64 \pma{0.34}{0.34}$  & $0.72 \pma{0.62}{0.29}$     &                             & $0.68 \pma{0.65}{0.41}$                \\
		                        & $M_\mathrm{th}^a [10^{11}\mathrm{M}_\odot]$ & $54.3 \pma{2.8}{2.6}$        & $52.6 \pma{2.0}{2.1}$    & $62.9 \pma{1.6}{2.2}$       &                             & $59.3  \pma{3.4}{2.8}$                 \\
		                        & $\beta^a$                                   & $0.79 \pma{0.11}{0.11}$      & $0.82 \pma{0.11}{0.11}$  & $0.65 \pma{0.28}{0.23}$     &                             & $0.77 \pma{0.19}{0.11}$                \\
		                        & $M^{\prime a} [10^{13}\Msun]$               & $6.7 \pma{1.2}{2.2}$         & $8.5 \pma{1.3}{0.9}$     & $8.4 \pma{0.6}{1.2}$        &                             & $8.6\pma{1.4}{1.5} $                   \\
		                        & $f^a$                                       & $1.17 \pma{0.23}{0.27}$      & $0.94 \pma{0.16}{0.13}$  & $0.74 \pma{0.19}{0.21}$     &                             & $0.62 \pma{0.17}{0.20}$                \\
		\hline
		\multirow{6}{*}{$a=$m5} & $\alpha^a$                                  & $0.51 \pma{0.21}{0.24}$      & $0.56 \pma{0.15}{0.23}$  & $0.64 \pma{0.19}{0.23}$     & $0.54 \pma{0.21}{0.19}$     &                                        \\
		                        & $\sigma^a$                                  & $0.65 \pma{0.25}{0.25}$      & $0.69 \pma{0.26}{0.22}$  & $0.65 \pma{0.25}{0.26}$     & $0.71 \pma{0.65}{0.24}$     &                                        \\
		                        & $M_\mathrm{th}^a [10^{11}\mathrm{M}_\odot]$ & $513 \pma{23}{19} $          & $520 \pma{18}{22}$       & $507 \pma{17}{13}$          & $517  \pma{15}{22}$         &                                        \\
		                        & $\beta^a$                                   & $0.59 \pma{0.13}{0.14}$      & $0.55 \pma{0.08}{0.11}$  & $0.42 \pma{0.11}{0.14}$     & $0.52 \pma{0.07}{0.12}$     &                                        \\
		                        & $M^{\prime a} [10^{13}\Msun]$               & $27.0 \pma{2.8}{2.7}$        & $25.1 \pma{1.2}{1.9}$    & $34.7 \pma{1.7}{3.2}$       & $30.7 \pma{3.0}{2.5}$       &                                        \\
		                        & $f^a$                                       & $1.22 \pma{0.19}{0.21}$      & $1.11 \pma{0.19}{0.12}$  & $1.20 \pma{0.25}{0.15}$     & $1.21  \pma{0.23}{0.19}$    &                                        \\ \hline
	\end{tabularx}
\end{table}

\begin{table}[]
\caption{Best-fit values of HOD parameters describing satellite number cross-correlation, and $\chi^2$ and $p$-values for G3L Halomodel fit to MS}
\label{tab:params_MS_sm_chisq}
\begin{tabular}{llllll}
\hline
                    &                      & $b$=m2                     & $b$=m3                     & $b$=m4                        & $b$=m5                        \\ \hline
\multirow{4}{*}{$a$=m1} & $A^{ab} $[$10^{-2}$] & $1.20\pma{0.36}{0.45}$ & $0.54\pma{0.11}{0.14}$ & $0.273\pma{0.052}{0.057}$ & $0.029\pma{0.024}{0.020}$ \\
                    & $\epsilon^{ab}$      & $0.85\pma{0.07}{0.09}$ & $0.95\pma{0.11}{0.13}$ & $0.94\pma{0.14}{0.12}$    & $0.87\pma{0.11}{0.12}$    \\
                    & $\chi^2$             & $61.62$                & $79.79$                & $62.16$                   & $88.67$                   \\
                    & $p$                  & $0.884$                & $0.361$                & $0.874$                   & $0.152$                   \\\hline
\multirow{4}{*}{$a$=m2} & $A^{ab}$             &                        & $1.15\pma{0.32}{0.40}$ & $1.00\pma{0.67}{0.71}$    & $0.129\pma{0.071}{0.081}$ \\
                    & $\epsilon^{ab}$      &                        & $0.98\pma{0.07}{0.08}$ & $0.99\pma{0.10}{0.08}$    & $1.101\pma{0.15}{0.12}$   \\
                    & $\chi^2$             &                        & $71.66$                & $73.26$                   & $79.53$                   \\
                    & $p$                  &                        & $0.620$                & $0.568$                   & $0.368$                   \\\hline
\multirow{4}{*}{$a$=m3} & $A^{ab}$             &                        &                        & $0.570\pma{0.084}{0.095}$ & $0.147\pma{0.083}{0.076}$ \\
                    & $\epsilon^{ab}$      &                        &                        & $1.09\pma{0.17}{0.19}$    & $1.02\pma{0.11}{0.12}$    \\
                    & $\chi^2$             &                        &                        & $72.21$                   & $75.71$                   \\
                    & $p$                  &                        &                        & $0.602$                   & $0.488$                   \\\hline
\multirow{4}{*}{$a$={m4}} & $A^{ab}$             &                        &                        &                           & $0.062\pma{0.063}{0.052}$ \\
                    & $\epsilon^{ab}$      &                        &                        &                           & $0.98\pma{0.14}{0.16}$    \\
                    & $\chi^2$             &                        &                        &                           & $85.27$                   \\
                    & $p$                  &                        &                        &                           & $0.219$                   \\ \hline
\end{tabular}
\end{table}

\begin{center}
	\includegraphics[width=\textwidth, trim={5cm 0 5cm 2cm}, clip]{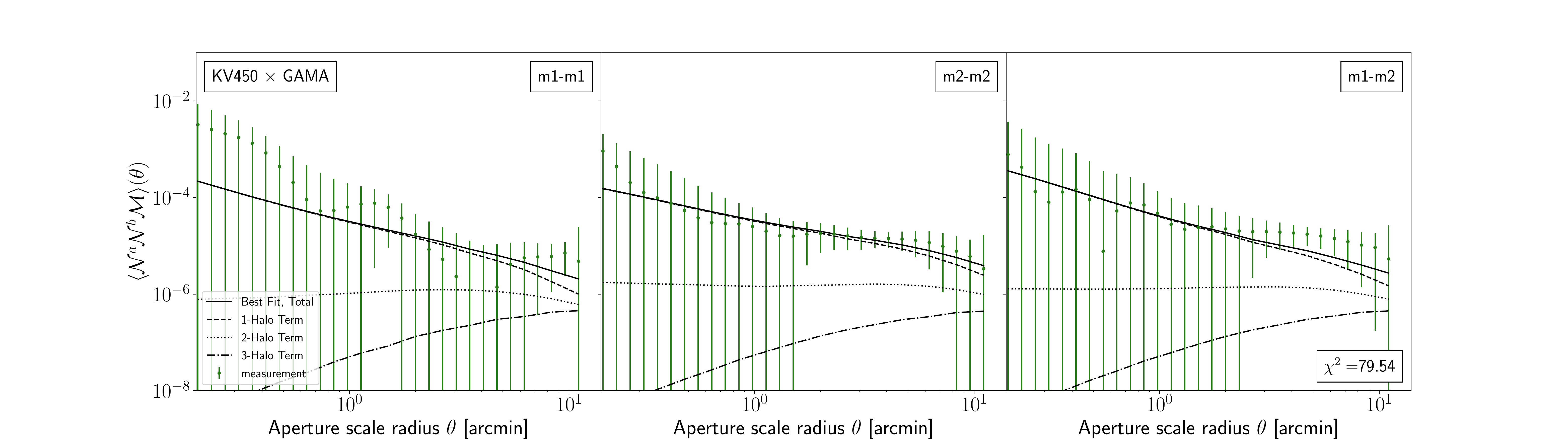}
	\includegraphics[width=\textwidth, trim={5cm 0 5cm 2cm}, clip]{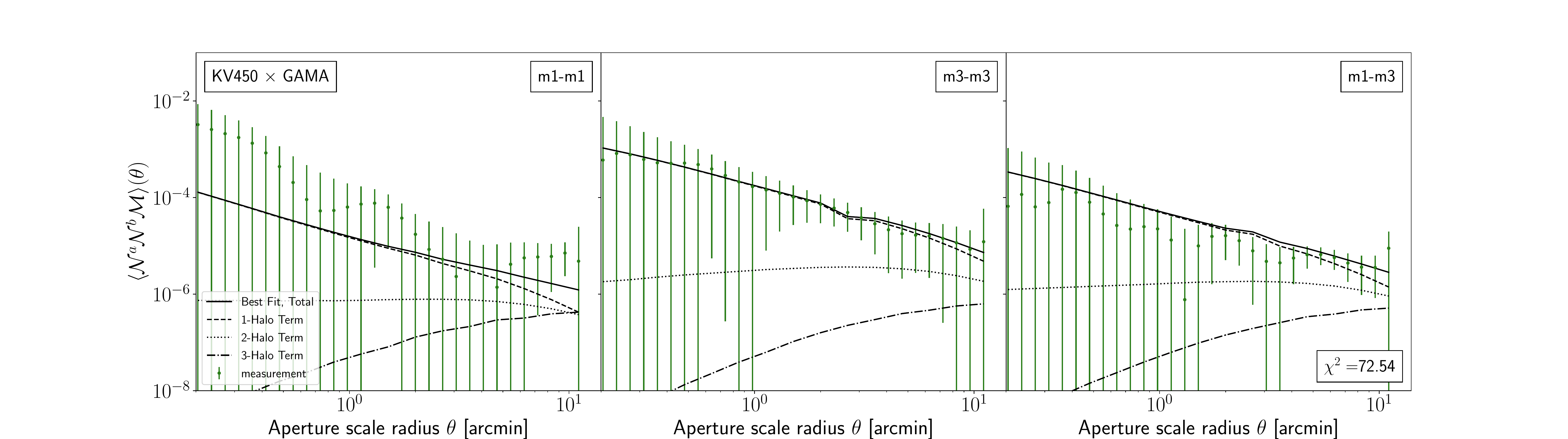}
	\includegraphics[width=\textwidth, trim={5cm 0 5cm 2cm}, clip]{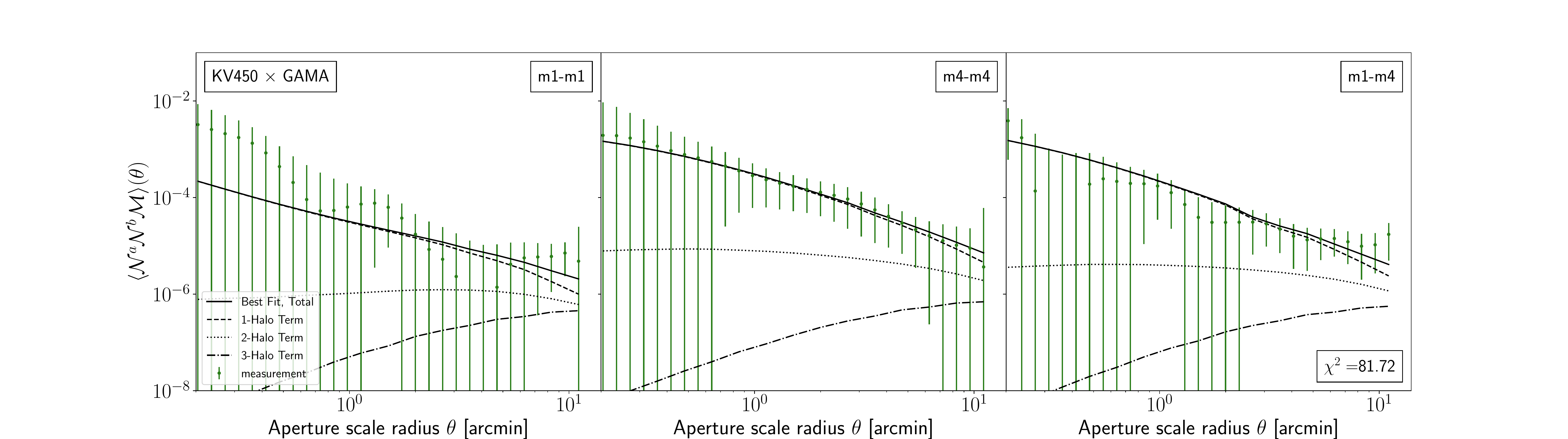}
	\includegraphics[width=\textwidth, trim={5cm 0 5cm 2cm}, clip]{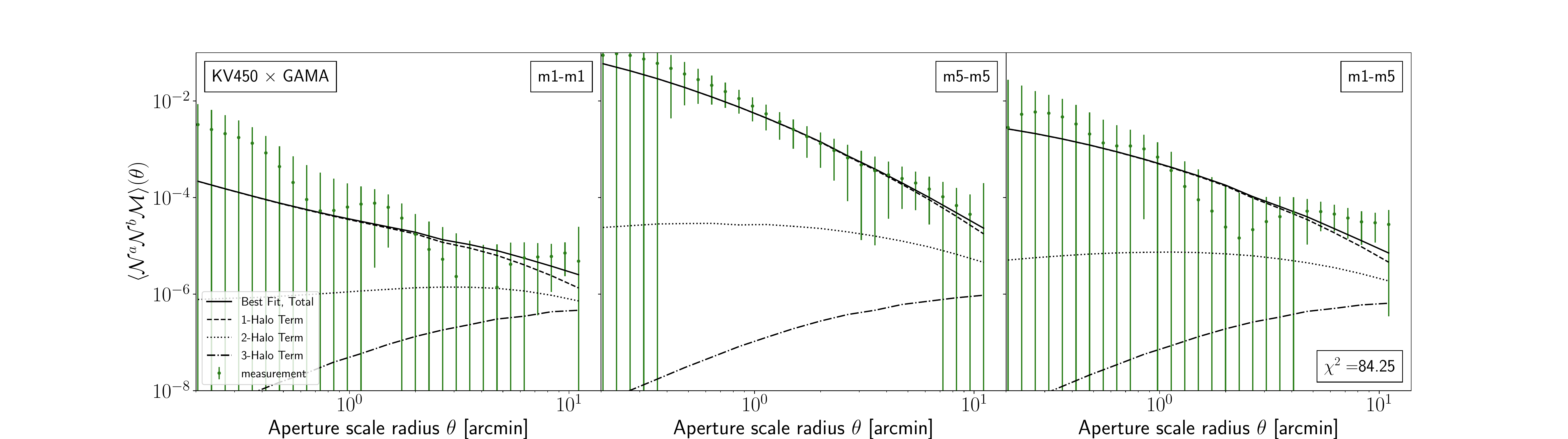}	
	\includegraphics[width=\textwidth, trim={5cm 0 5cm 2cm}, clip]{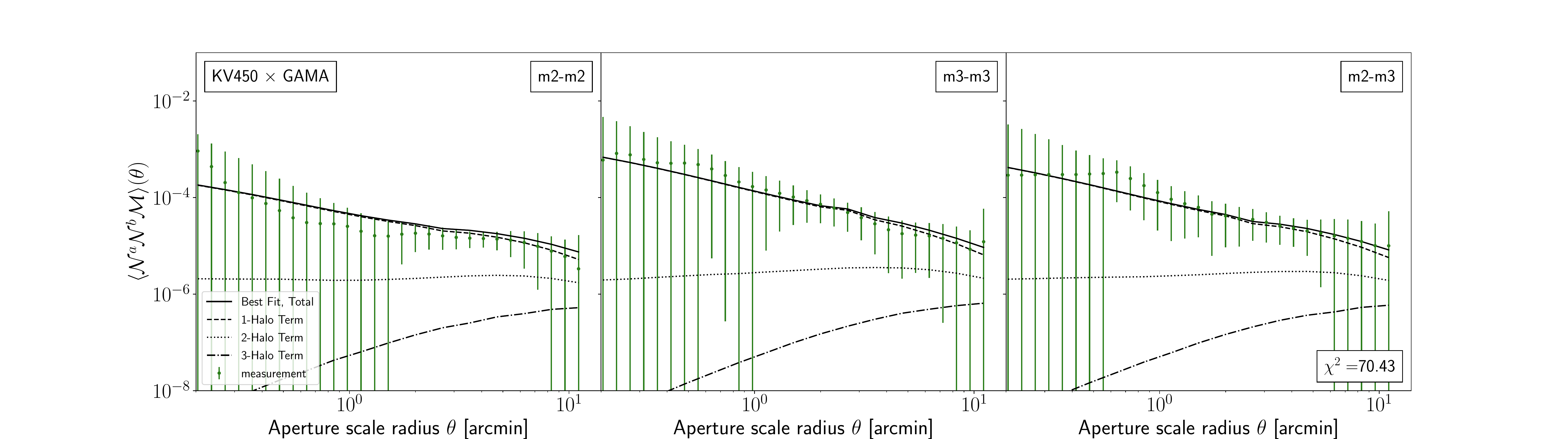}
	\includegraphics[width=\textwidth, trim={5cm 0 5cm 2cm}, clip]{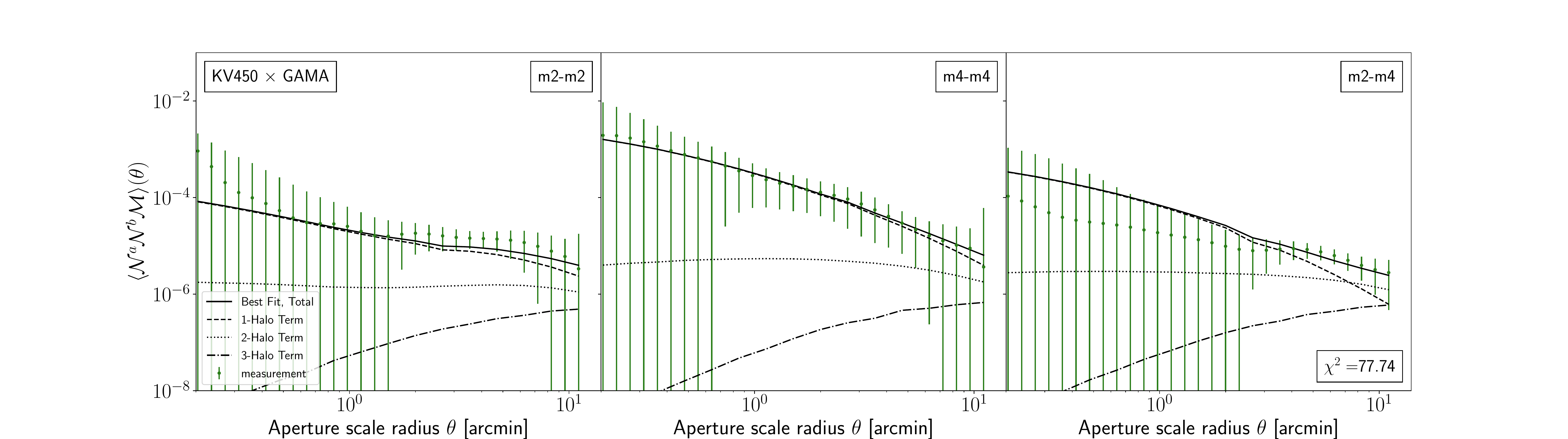}
	\includegraphics[width=\textwidth, trim={5cm 0 5cm 2cm}, clip]{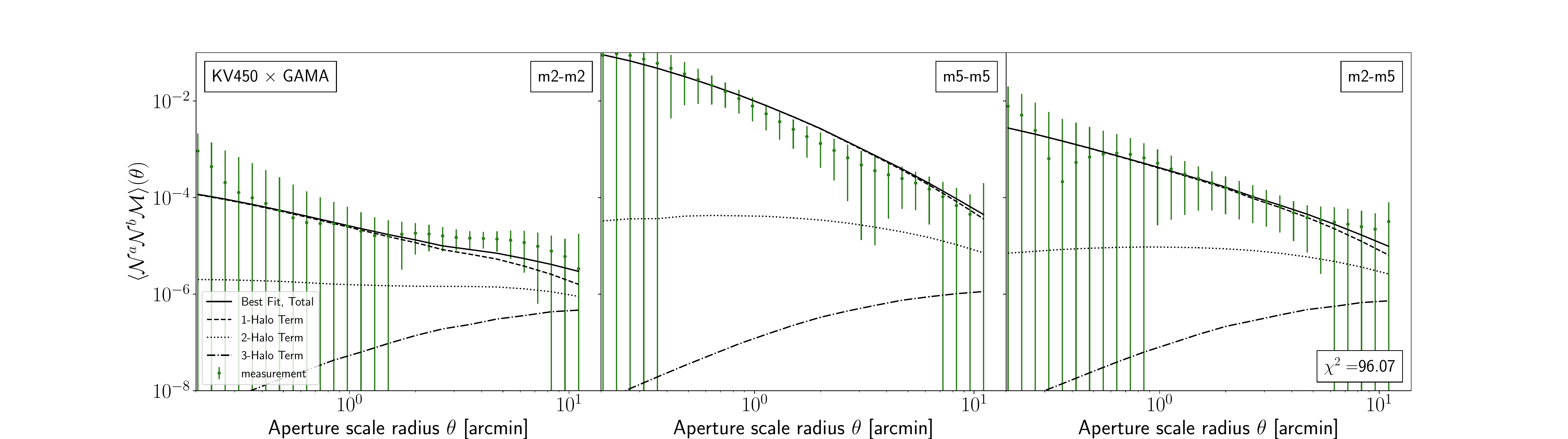}
	\includegraphics[width=\textwidth, trim={5cm 0 5cm 2cm}, clip]{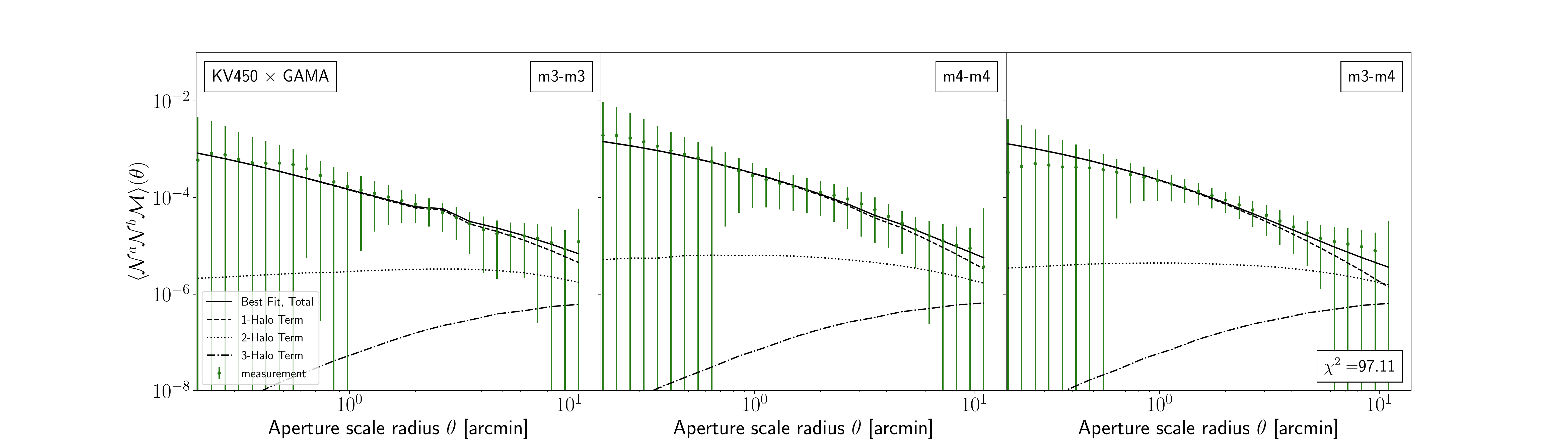}
	\includegraphics[width=\textwidth, trim={5cm 0 5cm 2cm}, clip]{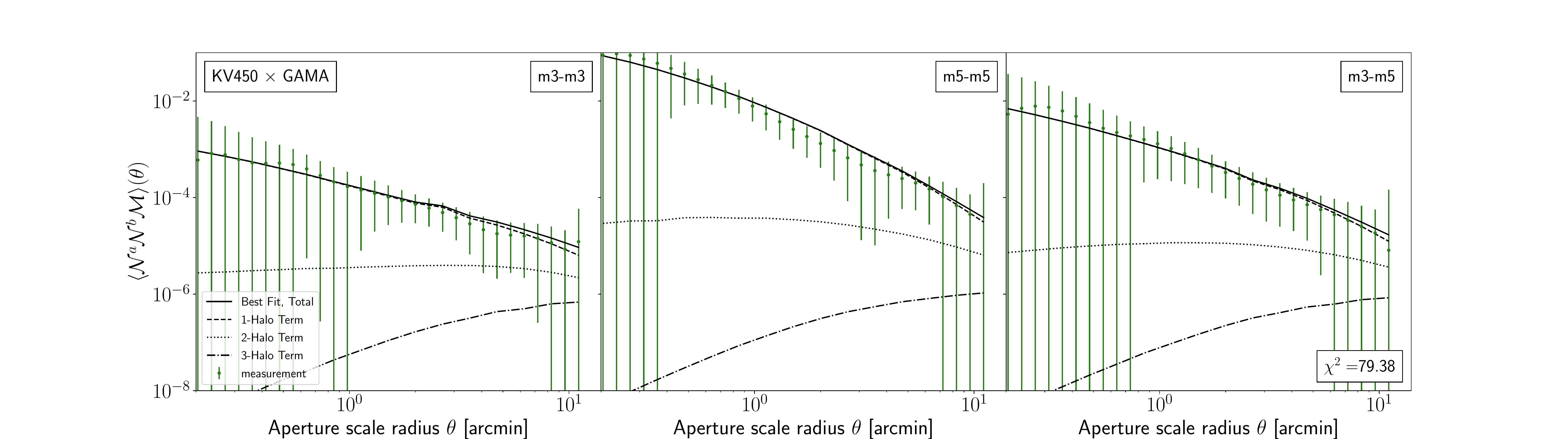}
	\includegraphics[width=\textwidth, trim={5cm 0 5cm 2cm}, clip]{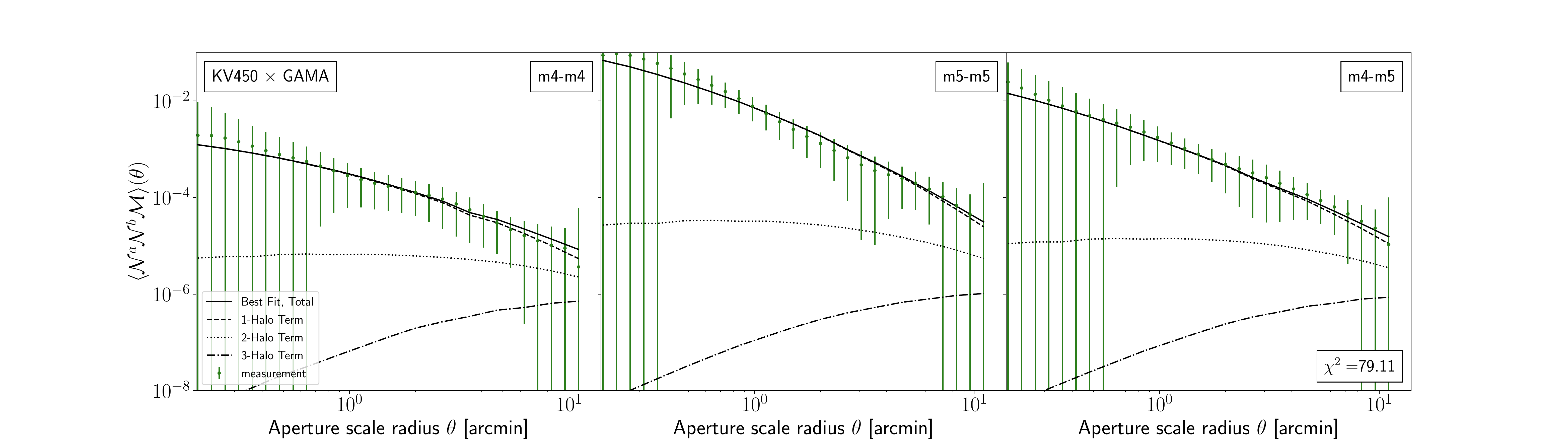}
		\captionof{figure}{{G3L} measurement in {KV450 $\times$ GAMA} (points) and best fitting halo model (lines) for stellar mass-selected lens samples, as defined in Table~\ref{tab:sub-samples}. Solid lines indicate the total aperture statistics, dashed lines the 1-halo, dotted lines the 2-halo, and dash-dotted lines the 3-halo term of the fit. Each row was fitted individually, leading to the $\chi^2$ values in the last panel. The corresponding halo model parameters are given in Table~\ref{tab:params_KV450_sm}}
	\label{fig:fitresults_masses_KV450}
\end{center}

\begin{table}[p]
		\caption{Best-fit values for halo model parameters for stellar-mass-selected lenses in KV450 $\times$ GAMA for each stellar mass sample $a$.}
\centering
	\label{tab:params_KV450_sm}
\begin{tabularx}{\textwidth}{llXXXXX}
	\hline
	                        &                                             & From correlation with m1     & From correlation with m2 & From correlation with m3    & From correlation with m4    & From correlation with m5               \\ \hline
	\multirow{9}{*}{$a=$m1} & $\alpha^a$                                  &                              & $0.18\pma{0.11}{0.11}$   & $0.07\pma{0.08}{0.19} $     & $ 0.03\pma{0.02}{0.12}    $ & $ 0.04\pma{0.01}{0.10}  $              \\
	                        & $\sigma^a$                                  &                              & $0.67\pma{0.37}{0.38} $  & $0.47\pma{0.26}{0.37}     $ & $ 0.51\pma{0.40}{0.49}   $  & $ 0.67\pma{0.33}{0.34}   $             \\
	                        & $M_\mathrm{th}^a [10^{11}\mathrm{M}_\odot]$ &                              & $2.2\pma{2.1}{2.6}     $ & $1.3\pma{1.2}{1.9}    $     & $ 0.9\pma{0.6}{2.1}       $ & $ 1.3\pma{1.9}{3.0}       $            \\
	                        & $\beta^a$                                   &                              & $0.90\pma{0.14}{0.09}$   & $0.87\pma{0.11}{0.19} $     & $ 0.81\pma{0.36}{0.29}  $   & $ 0.88\pma{0.27}{0.47}   $             \\
	                        & $M^{\prime a} [10^{13}\Msun]$               &                              & $16.9 \pma{2.0}{2.0}$    & $12.8\pma{2.6}{3.9}       $ & $ 17.0\pma{2.1}{7.6}   $    & $ 18.9\pma{1.5}{2.9}                $  \\
	                        & $f^a$                                       &                              & $0.99\pma{0.48}{0.48}  $ & $0.93\pma{0.41}{0.57}$      & $ 1.25\pma{0.49}{0.36}$     & $ 0.97\pma{0.43}{0.42}               $ \\
 \hline
	\multirow{9}{*}{$a=$m2} & $\alpha^a$                                  & $0.13 \pma{0.11}{0.12}$      &                          & $0.15 \pma{0.11}{0.13}$     & $0.12 \pma{0.10}{0.20}$     & $0.15 \pma{0.14}{0.17}$                \\
	                        & $\sigma^a$                                  & $0.47 \pma{0.35}{0.62} $     &                          & $0.46 \pma{0.44}{0.39}$     & $0.39 \pma{0.29}{0.62}$     & $0.43 \pma{0.33}{0.37}$                \\
	                        & $M_\mathrm{th}^a [10^{11}\mathrm{M}_\odot]$ & $5.3 \pma{2.4}{3.1}    $     &                          & $5.2 \pma{5.4}{5.3} $       & $5.4 \pma{3.9}{6.1}$        & $5.3 \pma{2.4}{7.3}$                   \\
	                        & $\beta^a$                                   & $1.11 \pma{0.16}{0.17}  $    &                          & $1.19 \pma{0.16}{0.11}$     & $1.07 \pma{0.16}{0.11}$     & $1.01 \pma{0.18}{0.14}$                \\
	                        & $M^{\prime a} [10^{13}\Msun]$               & $17.0 \pma{1.5}{2.1}      $  &                          & $16.2 \pma{2.5}{2.1} $      & $18.3 \pma{2.1}{2.4}$       & $15.1 \pma{0.9}{1.1}$                  \\
	                        & $f^a$                                       & $0.69 \pma{0.43}{0.43}     $ &                          & $0.68 \pma{0.47}{0.49}$     & $0.94 \pma{0.43}{0.49}$     & $0.71 \pma{0.39}{0.37}$                \\
	 \hline
	\multirow{9}{*}{$a=$m3} & $\alpha^a$                                  & $0.23 \pma{0.14}{0.14}   $   & $0.23 \pma{0.18}{0.18} $ &                             & $0.19 \pma{0.17}{0.32}$     & $0.23 \pma{0.15}{0.14}$                \\
	                        & $\sigma^a$                                  & $0.54 \pma{0.48}{0.36}$      & $0.60 \pma{0.44}{0.49}$  &                             & $0.59 \pma{0.50}{0.49}$     & $0.56 \pma{0.33}{0.45}$                \\
	                        & $M_\mathrm{th}^a [10^{11}\mathrm{M}_\odot]$ & $9.4 \pma{4.4}{1.1}$         & $9.3 \pma{5.4}{5.3}$     &                             & $10.8\pma{2.2}{3.6}$        & $9.1 \pma{4.5}{8.6}$                   \\
	                        & $\beta^a$                                   & $0.67 \pma{0.31}{0.31}$      & $0.69 \pma{0.10}{0.26}$  &                             & $0.81 \pma{0.19}{0.16}$     & $0.77 \pma{0.09}{0.10}$                \\
	                        & $M^{\prime a} [10^{13}\Msun]$               & $7.8 \pma{2.7}{2.1}$         & $8.0 \pma{1.9}{2.3}$     &                             & $7.9  \pma{1.5}{2.0}$       & $5.8 \pma{1.4}{1.3}$                   \\
	                        & $f^a$                                       & $0.86 \pma{0.36}{1.9}$       & $0.84 \pma{0.57}{0.89}$  &                             & $0.77 \pma{0.08}{0.13}$     & $0.69 \pma{0.48}{0.55}$                \\
	 \hline
	\multirow{9}{*}{$a=$m4} & $\alpha^a$                                  & $0.48 \pma{0.42}{0.30}$      & $0.46 \pma{0.20}{0.22} $ & $0.48 \pma{0.18}{0.16}$     &                             & $0.45 \pma{0.19}{0.18}$                \\
	                        & $\sigma^a$                                  & $0.51 \pma{0.38}{0.36}$      & $0.64 \pma{0.46}{0.46}$  & $0.59 \pma{0.49}{0.42}$     &                             & $0.55 \pma{0.41}{0.49}$                \\
	                        & $M_\mathrm{th}^a [10^{11}\mathrm{M}_\odot]$ & $46.1 \pma{5.3}{5.3}$        & $34.7 \pma{3.6}{7.8}$    & $41.2 \pma{2.3}{3.3}$       &                             & $44.9  \pma{3.1}{4.7}$                 \\
	                        & $\beta^a$                                   & $1.05 \pma{0.32}{0.39}$      & $0.66 \pma{0.28}{0.33}$  & $0.82 \pma{0.38}{0.39}$     &                             & $0.79 \pma{0.17}{0.21}$                \\
	                        & $M^{\prime a} [10^{13}\Msun]$               & $15.0 \pma{1.8}{3.9}$        & $9.0 \pma{2.5}{1.9}$     & $17.3 \pma{2.3}{2.0}$       &                             & $17.5\pma{2.1}{2.4} $                  \\
	                        & $f^a$                                       & $1.15 \pma{0.16}{0.14}$      & $1.38 \pma{0.47}{0.27}$  & $0.97 \pma{0.27}{0.35}$     &                             & $1.25 \pma{0.31}{0.12}$                \\
	 \hline
	\multirow{6}{*}{$a=$m5} & $\alpha^a$                                  & $0.54 \pma{0.51}{0.43}$      & $0.55 \pma{0.25}{0.39}$  & $0.51 \pma{0.36}{0.33}$     & $0.56 \pma{0.38}{0.32}$     &                                        \\
	                        & $\sigma^a$                                  & $0.61 \pma{0.45}{0.45}$      & $0.66 \pma{0.46}{0.50}$  & $0.54 \pma{0.47}{0.56}$     & $0.76 \pma{0.46}{0.44}$     &                                        \\
	                        & $M_\mathrm{th}^a [10^{11}\mathrm{M}_\odot]$ & $155 \pma{15}{17} $          & $146 \pma{13}{14}$       & $151 \pma{9}{11}$           & $147.6  \pma{5.7}{8.2}$     &                                        \\
	                        & $\beta^a$                                   & $0.80 \pma{0.26}{0.34}$      & $0.75 \pma{0.28}{0.24}$  & $0.88 \pma{0.22}{0.21}$     & $0.78 \pma{0.17}{0.21}$     &                                        \\
	                        & $M^{\prime a} [10^{13}\Msun]$               & $65.7 \pma{12}{10}$          & $70.9 \pma{9.8}{8.9}$    & $62.2 \pma{3.6}{3.6}$       & $64.0 \pma{5.2}{4.6}$       &                                        \\
	                        & $f^a$                                       & $1.7 \pma{1.5}{1.1}$         & $1.5 \pma{1.3}{1.9}$     & $1.7 \pma{1.2}{1.1}$        & $1.3  \pma{1.0}{1.8}$       &                                        \\ \hline
\end{tabularx}
\end{table}

\begin{table}[]
\caption{Best-fit values of HOD parameters describing satellite number cross-correlation, and $\chi^2$ and $p$-values for G3L Halomodel fit to KV450 $\times$ GAMA}
\label{tab:params_sm_kv450_chisq}
\begin{tabular}{llllll}
\hline
                    &                      & $b=$m2                   & $b=$m3                     & $b=$m4                     & $b=$m5                           \\ \hline
\multirow{4}{*}{$a=$m1} & $A^{ab} $[$10^{-2}$] & $19.4\pma{5.5}{4.2}$ & $11.3\pma{1.9}{1.9}$   & $0.86\pma{0.56}{0.65}$ & $0.093\pma{0.097}{0.083}$    \\
                    & $\epsilon^{ab}$      & $0.97\pma{0.2}{0.2}$ & $1.02\pma{0.21}{0.19}$ & $1.05\pma{0.18}{0.21}$ & $0.97\pma{0.17}{0.18}$       \\
                    & $\chi^2$             & $72.54$              & $94.37$                & $81.72$                & $84.25$                      \\
                    & $p$                  & $0.591$              & $0.075$                & $0.306$                & $0.242$                      \\\hline
\multirow{4}{*}{$a=$m2} & $A^{ab}$             &                      & $6.4\pma{2.0}{2.3}$    & $5.0\pma{1.4}{1.4}$                  & $1.0\pma{1.1}{1.3}$          \\
                    & $\epsilon^{ab}$      &                      & $1.02\pma{0.21}{0.19}$ & $1.2\pma{0.21}{0.19}$  & $1.11\pma{0.16}{0.17}$       \\
                    & $\chi^2$             &                      & $70.43$                & $77.74$                & $96.07$                      \\
                    & $p$                  &                      & $0.659$                & $0.423$                & $0.060$                      \\\hline
\multirow{4}{*}{$a=$m3} & $A^{ab}$             &                      &                        & $0.16\pma{0.12}{0.12}$ & $0.019\pma{0.021}{0.019}$    \\
                    & $\epsilon^{ab}$      &                      &                        & $1.10\pma{0.18}{0.17}$ & $1.02\pma{0.21}{0.22}$       \\
                    & $\chi^2$             &                      &                        & $97.11$                & $79.38$                      \\
                    & $p$                  &                      &                        & $0.052$                & $0.373$                      \\\hline
\multirow{4}{*}{$a=$m4} & $A^{ab}$             &                      &                        &                        & $0.0016\pma{0.0016}{0.0013}$ \\
                    & $\epsilon^{ab}$      &                      &                        &                        & $1.18\pma{0.23}{0.25}$       \\
                    & $\chi^2$             &                      &                        &                        & $79.11$                      \\
                    & $p$                  &                      &                        &                        & $0.381$                      \\ \hline
\end{tabular}
\end{table}

\end{appendix}

\end{document}